\documentclass[useAMS,usenatbib]{mn2e}

\usepackage{epsfig}
\usepackage{textcomp}
\usepackage{longtable}
\usepackage{lscape}

\newcommand{\xmm} {{\it XMM-Newton}}
\newcommand{\cmsq} {cm$^{-2}$}
\newcommand{\nh} {$N_{\rm{H}}$}
\newcommand{\chisq} {$\chi^2$}
\newcommand{\dchisq} {$\Delta\chi^2$}
\newcommand{\rchisq} {$\chi^2_r$}
\newcommand{\xspec}{{\sc xspec}}
\newcommand{\mic}{{${\umu}$m}}

\newcommand{\degree}{{$^\circ$}}
\newcommand{\ergs}{\mbox{\thinspace erg\thinspace s$^{-1}$}}

\title[XMM survey of the 12MGS]{An XMM-Newton spectral survey of 12 micron selected galaxies. I. X-ray data}

\author[M. Brightman and K. Nandra]{Murray Brightman$^{2,1}\thanks{E-mail:  mbright@mpe.mpg.de}$ and Kirpal Nandra$^{2,1}$\\
$^{1}$Astrophysics Group, Imperial College London, Blackett Laboratory, Prince Consort Road, London SW7 2AW\\
$^{2}$Max-Planck-Institut f\"{u}r extraterrestrische Physik, Giessenbachstrasse 1, D-85748, Garching bei M\"{u}nchen, Germany\\}

\begin{document}

\date{Accepted 0000 December 00. Received 0000 December 00; in original form 0000 October 00}

\pagerange{\pageref{firstpage}--\pageref{lastpage}} \pubyear{0000}

\maketitle

\label{firstpage}

\begin{abstract}

We present an X-ray spectral analysis of 126 galaxies of the 12 micron galaxy sample (12MGS). By studying this sample at X-ray wavelengths, we aim to determine the intrinsic power, continuum shape and obscuration level in these sources. We improve upon previous works by the use of superior data in the form of higher signal to noise spectra, finer spectral resolution and a broader band pass from \xmm. We pay particular attention to Compton thick AGN with the help of new spectral fitting models that we have produced, which are based on Monte-Carlo simulations of X-ray radiative transfer, using both a spherical and torus geometry, and taking into account Compton scattering and iron fluorescence.  We use this data to show that with a torus geometry, unobscured sight lines can achieve a maximum equivalent width (EW) of the Fe K$\alpha$ line of $\sim150$ eV, originally shown by \citet*{ghisellini94}. In order for this to be exceeded, the line of sight must be obscured with \nh$>10^{23}$ \cmsq, as we show for one case, NGC 3690. We also calculate flux suppression factors from the simulated data, the main conclusion from which is that for \nh$\ge10^{25}$ \cmsq, the X-ray flux is suppressed by a factor of at least 10 in all X-ray bands and at all redshifts, revealing the biases present against these extremely heavily obscured systems inherent in all X-ray surveys. Furthermore, we confirm previous results from \citet{murphy09} that show that the reflection fraction determined from slab geometries is underestimated with respect to toroidal geometries.  For the 12 micron selected galaxies, we investigate the distribution of X-ray power-law indices, finding that the mean  ($<\Gamma>=1.90_{-0.07}^{+0.05}$ and $\sigma_{\Gamma} = 0.31_{-0.05}^{+0.05}$) is consistent with previous works, and that the distribution of $\Gamma$ for obscured and unobscured sources is consistent with the source populations being the same, in general support of unification schemes. We determine a Compton thick fraction for the X-ray AGN in our sample to be 18 $\pm$ 5\%  which is higher than the hard X-ray ($>10$ keV) selected samples. Finally we find that the  obscured fraction for our sample is a strong function of X-ray luminosity, peaking at a luminosity of $\sim10^{42-43}$ \ergs.

\end{abstract}


\section{Introduction}
Studying AGN at X-ray wavelengths allows an assessment of the intrinsic source properties, such as the measurement of the continuum and the source power, due to the ability of X-rays to penetrate and measure large columns of obscuring material. This makes X-ray observations essential for testing AGN unification schemes \citep{awaki91} which invoke obscuration to explain the perceived difference between Seyfert 1s and Seyfert 2s \citep{antonucci93, urry95}. The \nh\ and $\Gamma$ distributions for AGN are also an important ingredients for synthesis models of the cosmic X-ray background \citep*[XRB,][]{gilli07}. 

Most of our knowledge of the X-ray properties of local AGN comes from X-ray selected samples, such as the \citet{piccinotti82} sample derived from {\it HEAO 1} observations \citep[e.g.][]{turner89, nandra94}. X-ray selection of AGN is in principle very effective as in this band the nucleus predominantly outshines the host galaxy, and can penetrate obscuration. The hard X-ray telescopes aboard {\it INTEGRAL} and {\it Swift} and their large sky coverage have allowed for the compilation of hard X-ray selected AGN catalogues. \citet{beckmann06} and \citet{beckmann09} have presented catalogues for {\it INTEGRAL} selected AGN, which total 199 after 5 years of observations, and \citet{tueller08} and \citet{tueller10} have presented 229 {\it Swift/BAT} selected Seyfert galaxies after 22 months of observations. However, the discovery from these surveys seems to be that the fraction of Compton thick AGN detected is smaller than expected from XRB studies \citep*{gilli07}. This points to the empirical fact that even the X-ray band is biased against the most obscured, Compton thick, AGN.  X-ray selection at energies greater than 10 keV is less affected by Compton thick obscuration than 2-10 keV selection though. 

An alternative to X-ray AGN selection is mid-infrared (MIR) selection where the primary radiation of the AGN is re-emitted after having been reprocessed by hot dust. The extended {\it IRAS} 12 micron galaxy sample \citep*[12MGS -][RMS]{rush93} is a sample of 893 MIR selected local galaxies which contains a high fraction of AGN (13\% at the time of publication, RMS). The sample is taken from the {\it IRAS Faint Source Catalogue, version 2 (FSC-2)} and imposes a flux limit of 0.22 Jy, including only sources with a rising flux density from 12 to 60 microns (to exclude stars) and with a galactic latitude of $|b| \geq$25 deg. Being selected in the mid-infrared (MIR) this sample is also relatively unbiased against absorption, low luminosity systems and `hidden AGN'. \citet{spinoglio89} showed that a wide variety of AGN types emit a constant fraction of their bolometric luminosities at this wavelength, and furthermore shown to be true for star forming galaxies as well by \citet{spinoglio95}. The 12MGS should be therefore also representative of the true number of different active galaxy types in relation to each other. 

It is possible, though, that even at 12 \mic, the AGN emission may be suppressed in the most heavily obscured nuclei. \citet{pier92} present radiative transfer modelling for the infrared emission from dust tori for varying torus orientations. The most significant result of this study was that throughout the infrared, the emission in the edge-on direction is weaker with respect to the face-on direction, and at 12 \mic, the difference can be up to several orders of magnitude for optically thick tori. These models suggest that AGN selection at 12 microns is biased against heavily obscured nuclei. However, recently this was tested observationally by \citet{horst07}. They used high resolution 12.3 \mic\ imaging of a sample of local Seyfert nuclei, with X-ray data to explore the ratio of the MIR to intrinsic X-ray nuclear luminosities in order to detect any difference between the obscured nuclei (Seyfert 2s) and unobscured nuclei (Seyfert 1s). As the X-ray luminosities have been corrected for absorption, $L_{\rm X}$ should represent the intrinsic emission from the nucleus. For smooth tori modelled by \citet{pier92}, obscured systems should then present lower $L_{\rm MIR}/L_{\rm X}$ ratios than unobscured systems due to the suppressed MIR flux seen in these systems. A difference of an order of magnitude for a difference in \nh\ of between $\sim10^{20}$ and $\sim10^{24}$ \cmsq\ is predicted by the torus models, and should be measurable. However, this is not seen observationally, as \citet{horst07} find no significant difference between Seyfert 1s and Seyfert 2s, or even Compton thick nuclei. They argue from this that the torus may not be smooth, and instead clumpy, where the torus is seen to be optically thin, but the clumps are optically thick, though with a small volume filling factor. Most importantly for this work though, it suggests that 12 micron selection is in fact not heavily biased against obscuration. This therefore makes the 12MGS the ideal parent sample for study of obscuration and unification using X-ray observations and is as such the subject of this paper.

The 12MGS has previously been studied at X-ray wavelengths by \citet{barcons95} who used hard X-ray observations by {\it HEAO 1}, and at soft X-rays by \citet{rush96x} with {\it ROSAT} data. \citet{barcons95} use the X-ray and IR data sets to show that most of the local X-ray emissivity originates from Seyfert galaxies and that most of the local 2-10 keV X-ray luminosity function between 10$^{42}$ and 10$^{46}$ \ergs\ can be accounted for by 12 micron emitting AGN. They also test the unification scheme and come to the conclusion that it must be modified to include the reduction of the covering fraction of the torus with increasing source luminosity, an observation first suggested by \citet{lawrence91} which has come to be known as the `receding torus model'. \citet{rush96x} conducted a spectral analysis of the 0.1-2.4 keV {\it ROSAT} data. For spectra with enough counts, they fit absorbed power-laws, finding a median soft X-ray spectral index of 2.3 for all Seyfert types.

In this paper, we present an X-ray spectral analysis of the galaxies of the 12MGS with an observation by \xmm, which consists of a heterogeneous mix of Seyferts, LINERs and star-forming galaxies. We focus on the X-ray spectral properties of the sources without regard for their optical spectroscopic types and use X-ray luminosity to select unambiguous AGN. In a companion paper though, we continue our investigation into this sample combining the X-ray data from this paper with optical spectroscopic and infrared data. In Section \ref{transsec} we present work on Monte-Carlo simulations of X-ray radiative transport which we use in our spectral analysis. Section \ref{specfit} describes our spectral fitting methods and Section \ref{specfitresults} gives our spectral fitting results. Finally we discuss all of these results in Section \ref{discussion} and present our conclusions in Section \ref{conclusions}.



\section{New X-ray spectral models for heavily obscured sources}\label{transsec}

In our X-ray spectral analysis of the 12MGS, we pay particular attention to whether sources are Compton thick, as the Compton thick fraction is particularly crucial for the XRB models and knowledge of the accretion history of the universe. Compton thick AGN are most commonly identified in the 2-10 keV X-ray band by the measurement of a flat spectrum, where $\Gamma<1.0$, and the equivalent width (EW) of the iron K$\alpha$ line has been measured to be high \citep[EW$>$1 keV,][]{matt96_2}. This is typical of a reflection spectrum, produced by X-ray scattering from an optically thick surface, such as the accretion disk or wall of the torus in AGN. It is indicative of heavy obscuration of the central source as the reflection is dominating over the transmission of the intrinsic power-law through the obscuring material. 

In some cases, the transmission of the intrinsic power-law can be detected in the 2-10 keV band for these heavily obscured sources. However, due to the extreme suppression of flux in this band at these high column densities, this is very difficult. Also at the high optical depths in these Compton thick sources the calculations for modelling the transmission spectrum become non-linear due to multiple scatterings. Thus, models describing simple attenuation of the spectrum by absorption and scattering become invalid. 

Therefore, accurate models that correctly account for Compton scattering, especially multiple scatterings, are necessary when fitting the spectra of these sources. Currently, the best established model within {\sc xspec} for correctly describing X-ray reprocessing by photoelectric absorption and Compton scattering is {\tt plcabs}. This model describes the X-ray transmission from an isotropic source at the centre of a uniform spherical distribution of matter. It is valid up to energies between 10 and 18.5 keV and up to column densities of $5 \times 10^{24}$ cm$^{-2}$ \citep{yaqoob97}. Unfortunately {\tt plcabs} has a limited range of validity and does not include line emission, modelling the continuum only. Recently, two works by \citet{ikeda09} and \citet{murphy09} have produced improved Monte-Carlo models based on a toroidal obscuring medium similar to calculations by \citet{ghisellini94} which include iron line emission.

\begin{figure}
\begin{center}
\includegraphics[width=90mm]{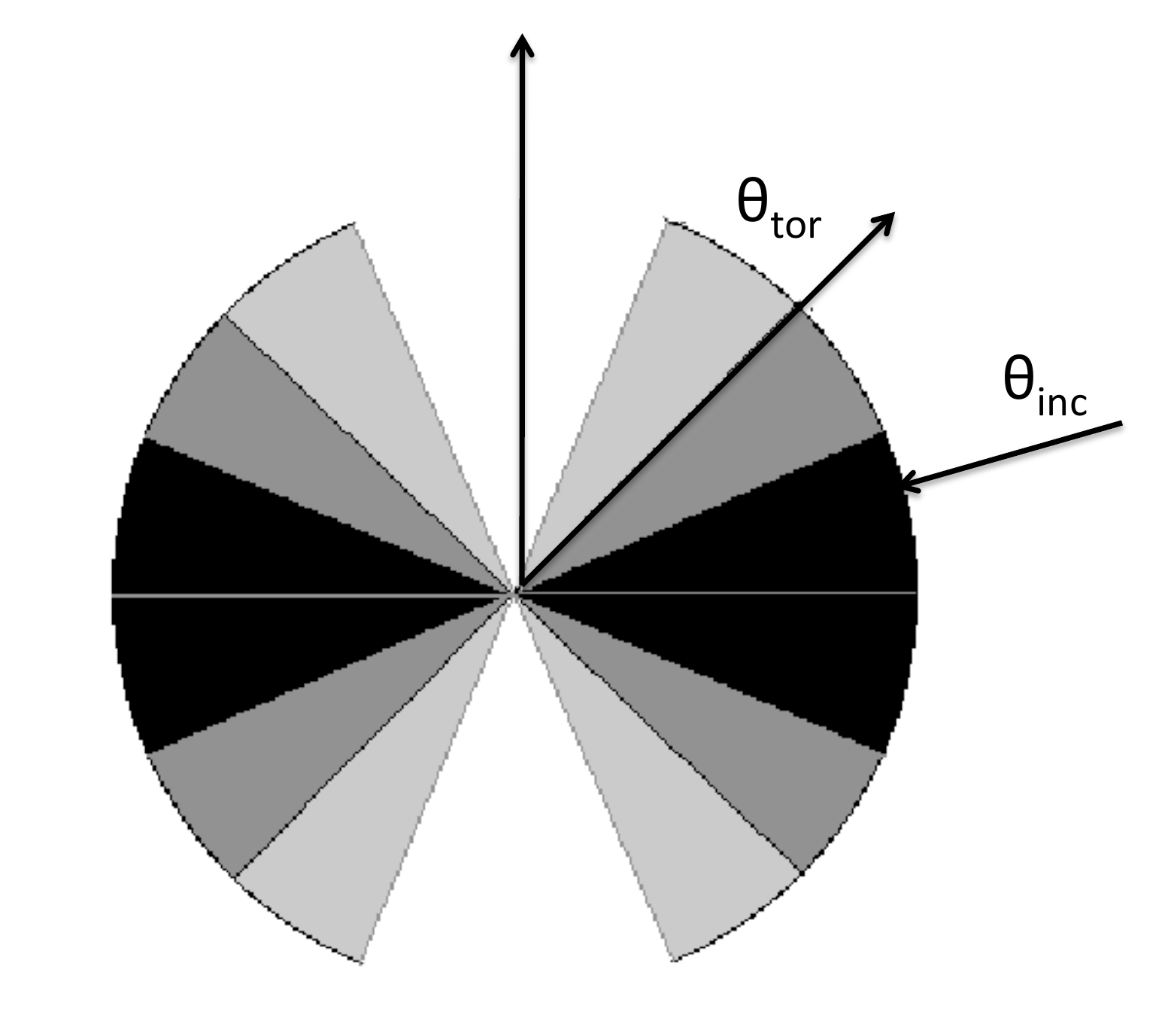}
\caption{Geometry of the biconical torus distribution of matter used. The shading only serves to differentiate the different opening angles used, and does not correspond to density.}
\label{fig_torgeom}
\end{center}
\end{figure}

Here we have carried out our own Monte-Carlo simulations based mainly on the methods of \citet[][NG94]{nandra94monte}, but with a point-source of X-rays located at the centre of both a spherical and toroidal distribution of gas emitting a power-law spectrum. These simulations follow the propagation of X-ray photons which interact with the medium via Compton scattering or photo-electric absorption, which can then result in fluorescent emission. The type of interaction is determined by the energy dependent interaction cross sections. We calculate the Compton scattering cross section described in \citet{rybicki86} and calculate the photo-electric absorption cross sections using data from \citet{verner96}, using abundances from \citet{anders89} and fluorescent yields from \citet{bambynek72}. The medium is assumed to be of constant density and there is no dependence of the line of sight column density on the inclination angle of the observer, with the exception of the unobscured sight lines in the torus geometry which have a zero column density. Fig. \ref{fig_torgeom} illustrates the geometry used in the torus simulations, where the shading only serves to differentiate the different opening angles used, and does not correspond to density.

The models importantly also include line emission, most significantly iron K$\alpha$ (6.4 keV), but also iron K$\beta$ (7.06 keV) and K$\alpha$ emission from several other elements described in Table \ref{elements}.  As mentioned before, the EW of the iron line is strongly dependent on absorption, as is the shape of the line's Compton shoulder.  The Compton shoulder feature is seen on the red side of the line, and is a result of Compton down scattering of the first order line and depends on the scattering optical depth.  In these models, the flux of the line and its spectral shape are inextricably linked to \nh\ and therefore it provides additional constraints on this parameter when spectral fitting.  \citet{yaqoob10} present a detailed examination of the Compton shoulder's properties, showing that it is also dependent on spectral shape and geometry of the absorber, and hence contains valuable information given that it is well resolved.

\begin{table*}
\begin{center}
\begin{tabular}{l l c c l }
\hline
Element 		& Abundance			& Line energy	& Edge energy		& Fluorescent 	\\
			& w.r.t Hydrogen		& (keV)		& (keV)		& yield, Y \\		
			& ($\times 10^{-4}$)	& & & \\			
\hline
Carbon		& 3.63	& 0.28		& 0.29	& 0.0025	\\
Oxygen		& 8.51	& 0.53		& 0.54	& 0.0086	\\
Neon		& 1.23 	& 0.85		& 0.87	& 0.0183	\\
Magnesium	& 0.38 	& 1.25		& 1.31	& 0.0303	\\
Silicon		& 0.355 	& 1.74		& 1.85	& 0.043	\\
Argon		& 0.0363 & 2.96		& 3.20	& 0.122	\\
Calcium		& 0.0229 & 3.69		& 4.04	& 0.168	\\
Chromium	& 0.00484 & 5.41		& 6.00	& 0.283	\\
Iron			& 0.468 	& 6.40, 7.06	& 7.12	& 0.342	\\
Nickel		& 0.0178	& 7.48		& 8.35	& 0.432	\\
\hline
\end{tabular}
\end{center}
\caption{Elemental data used in the Monte-Carlo simulations. Abundances are from \citet{anders89}, and fluorescent yields are from \citet{bambynek72} }
\label{elements}
\end{table*}%

We have made several improvements on the previous work by NG94, many with the aim to improve computational times and spectral quality.  These include the use of a technique described by \citet{lucy02}, in which radiation in Monte-Carlo simulations is represented by indivisible monochromatic `packets' of energy which propagate through the system in the same way as the photons were simulated previously. This technique allowed us to inject any number of quanta at each energy without sacrificing the power-law relation. Each packet is assigned a `weight' dependent on the width of the energy bin from which is it injected, the number of packets being injected at that energy and the energy-flux relation taken from the power-law.

\begin{displaymath}
weight = \frac{\Delta E \times F(E)}{N_{packets}}
\end{displaymath}

The weights of the emergent packets are then binned at the end in the same way. This technique has the advantage that it is possible to vary the number of quanta released depending on how detailed the spectrum needs to be at that energy. It is especially useful for increasing the signal to noise of the output spectrum around the iron K-shell regime (6-8 keV).  Valuable computational time may also be saved by decreasing the signal to noise in the featureless parts of the spectrum, above 10 keV for example.  We also made an improvement to the original method by introducing variable elemental abundances (with respect to the abundance of Hydrogen).  The abundance of iron is a separate parameter also, as the modelling of the iron lines are particularly sensitive to this.

We have constructed table models for use within {\sc xspec}, which we call {\tt trans} for the spherical distribution of matter, and {\tt torus} for the toroidal distribution of matter. The parameters of the models are the line of sight column density of the neutral material (10$^{20}$ \cmsq\ $\le$ \nh\ $\le$ 10$^{26}$ \cmsq) and the photon index of the intrinsic power-law ($1 \le \Gamma \le 3$). For {\tt trans} the abundance of iron (0.1 solar $\le$ Fe$_{abund} \le$ 10 solar) and the total elemental abundance (0.1 solar $\le$ abund. $\le$ 10 solar) are included as parameters, whereas for {\tt torus}, the opening angle of the torus ($25.8^{\circ} \le \theta_{tor} \le 84.3^{\circ}$) and the inclination angle of the torus ($18.2^{\circ} \le \theta_{i} \le 87.1^{\circ}$) are included. Both models have validity in an energy range of 0.1 keV to 320 keV.  

\subsubsection{Fe K$\alpha$ Line Equivalent Width Calculations}

We also use the results from our Monte-Carlo simulations to make predictions of the equivalent width of the iron K$\alpha$ line for both geometries given a set of model parameters. In the simulations the photon packets produced by a simulated iron K-shell fluorescence are flagged and summed as they exit the simulation.  In this way, we can calculate the equivalent width of the line produced, by the sum of iron K$\alpha$ photon flux divided by the continuum level at 6.4 keV and the width of the energy bin at this energy. The calculation represents the EW of the entire line, including the Compton shoulder.

\begin{figure*}
\begin{minipage}{150mm}

 \includegraphics[width=150mm]{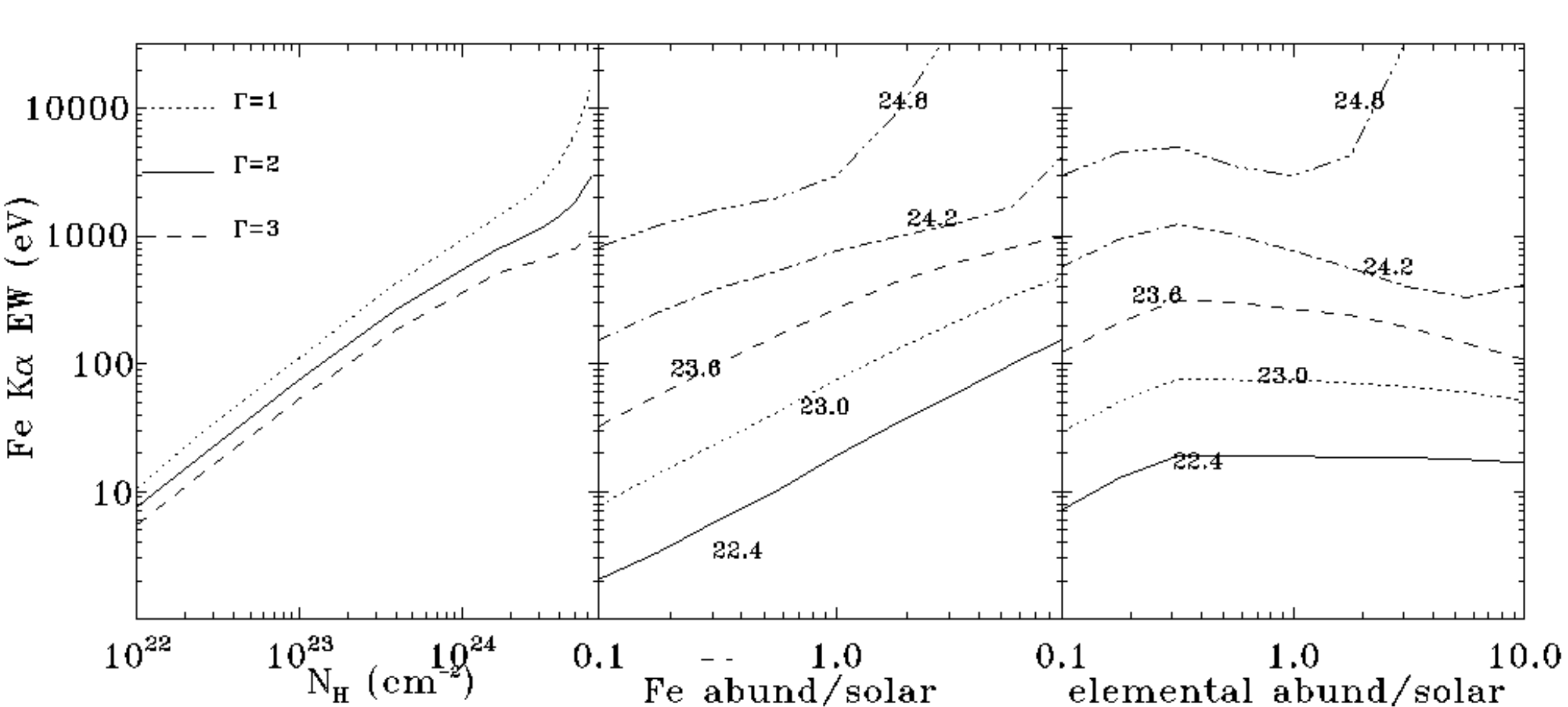}
 \caption{Predictions of the equivalent width of the iron K$\alpha$ line for the spherical distribution and varying parameters. Left panel - EW vs \nh\ with the three lines representing $\Gamma$ = 1,2 and 3 in ascending order for solar abundances. Middle panel - EW vs iron abundance (in units of solar abundance) with different lines for varying \nh\ labelled in the figure as log$_{10}$(\nh/\cmsq) and fixed $\Gamma$=2. Right panel EW vs total elemental abundance (in units of solar abundance) with different lines for varying \nh\ labelled in the figure as log$_{10}$(\nh/\cmsq) and fixed $\Gamma$=2. }
 \label{sphereews}
  \end{minipage}

\end{figure*}

Figure \ref{sphereews} shows how the EW varies with \nh, $\Gamma$ and the iron and elemental abundances for the spherical geometry. The EW of the line increases linearly versus \nh\ up to $\tau \sim 0.1$, where the relationship increases faster than linearly beyond that. This is due to the suppression of the continuum at 6.4 keV at high column densities, against which the EW is measured, but possibly also due to the greater number of photons at and above the 7.1 keV iron K shell threshold having been down-scattered from higher energies. Increasing the iron abundance will as expected increase the EW of the line due to the greater number of iron atoms per hydrogen atom to produce the transition. 
The abundance of the other elements also effects the EW of the iron line, though not as significantly. The EW is seen to increase from low elemental abundance to $\sim0.3\times$ solar abundance. This is also expected as an increase in the elemental abundances will cause some continuum suppression at 6.4 keV, leading to an increase in the EW of the iron line. This trend flattens off for \nh$<10^{23}$ as the elemental abundance increases beyond solar though, but for higher \nh\ values, the EW decreases above $\sim0.3\times$ solar abundance. Again though, at these extreme \nh\ values the continuum against which the EW is measured is very suppressed so it is difficult to know if this behaviour is real.

\begin{figure*}
\begin{minipage}{150mm}

 \includegraphics[width=150mm]{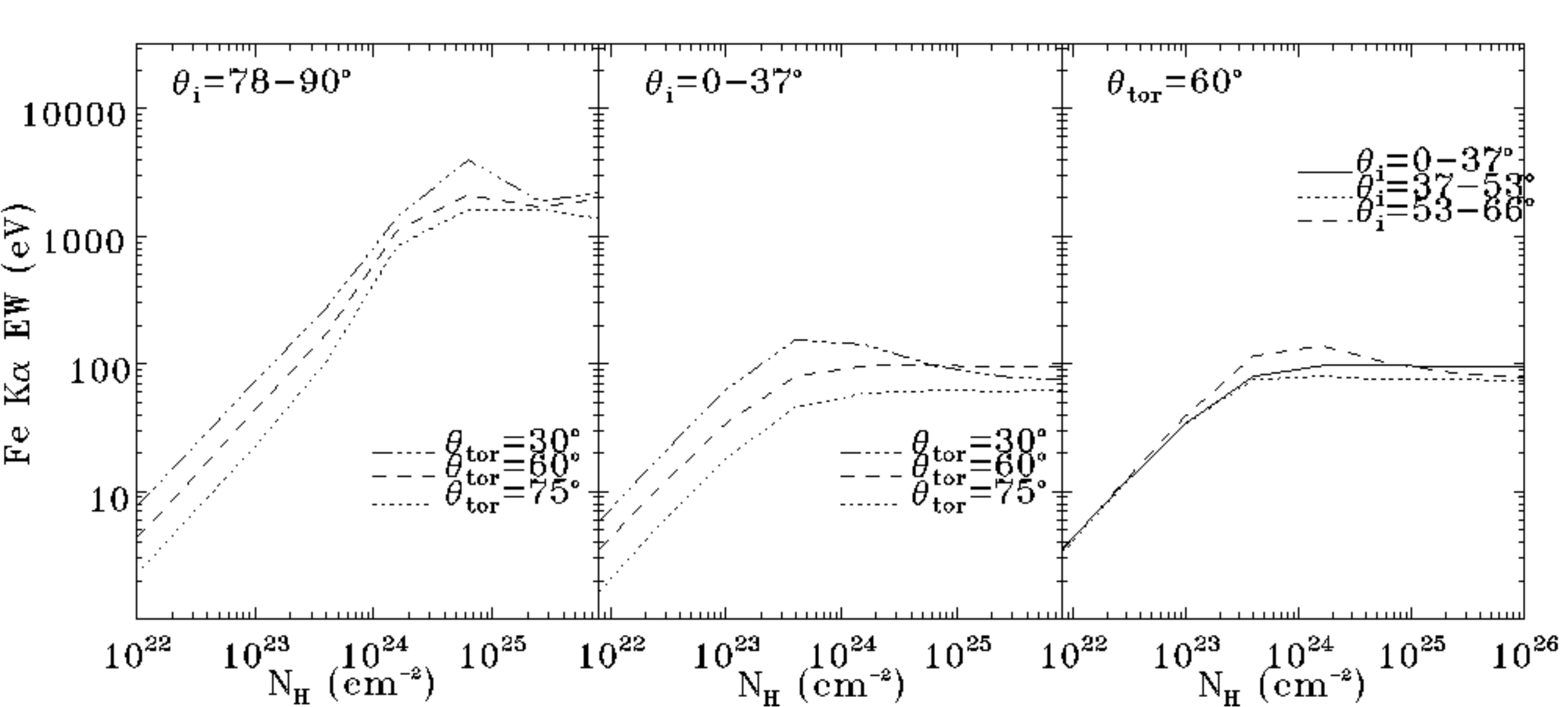}
 \caption{Predictions of the equivalent width of the iron K$\alpha$ line for the torus distribution with varying torus opening angles and viewing angles. Left panel -  EW vs \nh\ for equatorial viewing angles (78-90\degree), and varying torus opening angles, $\theta_{tor}$. Middle panel -  EW vs \nh\ for polar viewing angles (0-37\degree), and varying torus opening angles, $\theta_{tor}$; Right panel -  EW vs \nh\ for different viewing angles, $\theta_i$, with a fixed torus opening angle, $\theta_{tor}$=60\degree; All with fixed $\Gamma$=2.}
\label{torusews}
\end{minipage}

\end{figure*}

Figure \ref{torusews} shows how the EW of the iron line changes with \nh\ in the torus model for different torus parameters. The left plot shows this relationship for equatorial sight-lines, where the different lines show different opening angles. The smaller the opening angle, and hence the larger solid angle of the torus, the larger the EW of the iron line. The EW reaches a maximum here of 4 keV, 2.1 keV and 1.6 keV for torus opening angles of $\theta_{tor}$=30\degree, 60\degree and 75\degree. The middle panel shows the same relationship, but for the polar sight-line. The EWs here are systematically lower due to the greater, unabsorbed continuum. The EWs reach a maximum of 150 eV, 96 eV and 63 eV for 30\degree, 60\degree and 75\degree opening angles. Due to this dependance on column density and the opening angle of the torus, it may be possible to make inferences about the circum-nuclear matter distribution in AGN from the EW of the iron line, even for unobscured sources. For example, unobscured sources with an EW greater than $\sim150$ eV must have a very narrow opening in the torus, or else super-solar iron abundance. Alternatively, the source may not be unobscured at all. It may be very heavily obscured with the intrinsic emission extremely suppressed below 10 keV, and only a weak scattered or galactic component visible. We use these data later when determining the Compton thick nature of our sources. Finally the right panel in this figure shows the variation of EW for a fixed torus opening angle of 60\degree, and varying viewing angles, and shows that for unobscured sources, the viewing angle does not affect the EW of the iron line greatly.

\subsubsection{X-ray flux suppression}

\begin{figure}
\begin{center}
\includegraphics[width=90mm]{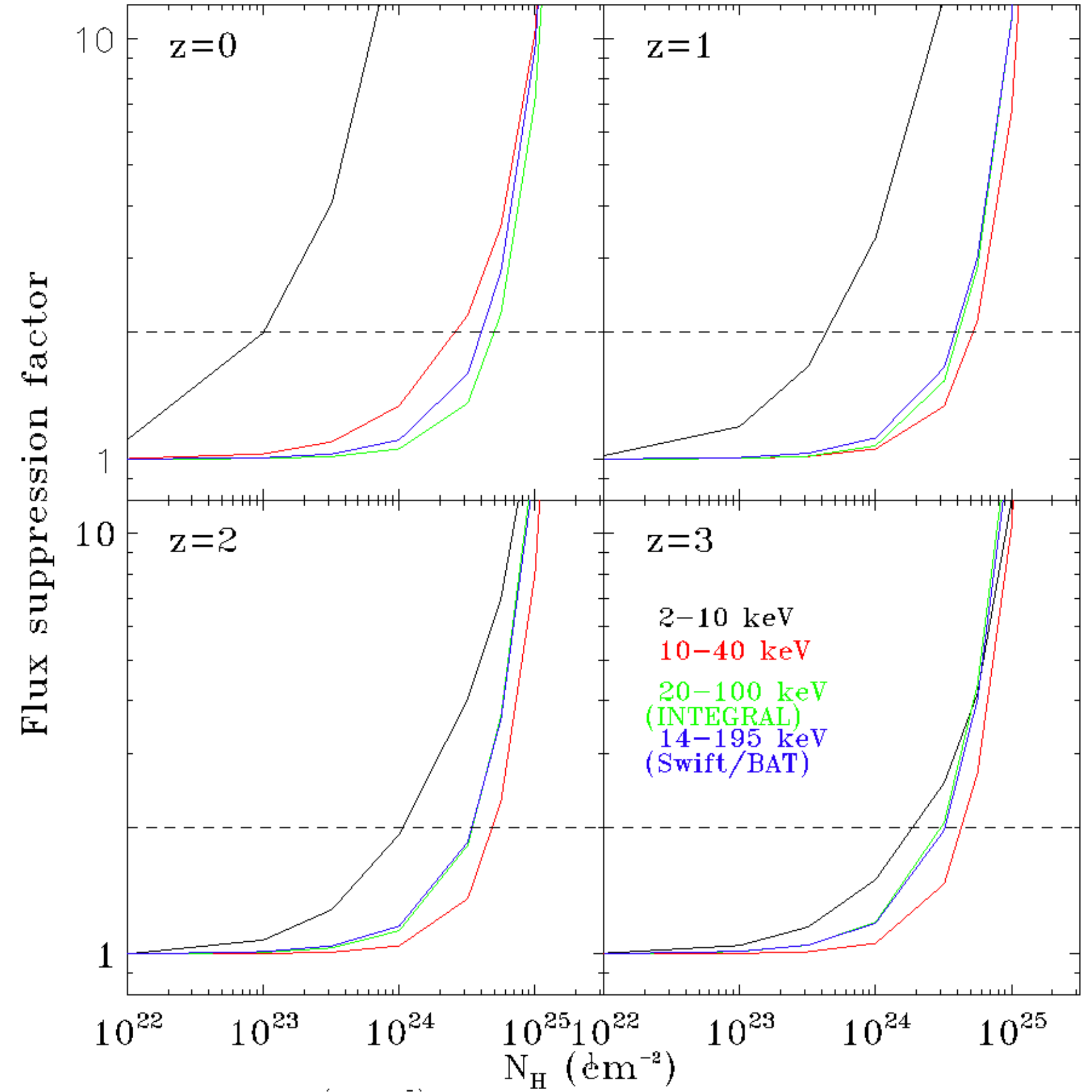}
\includegraphics[width=90mm]{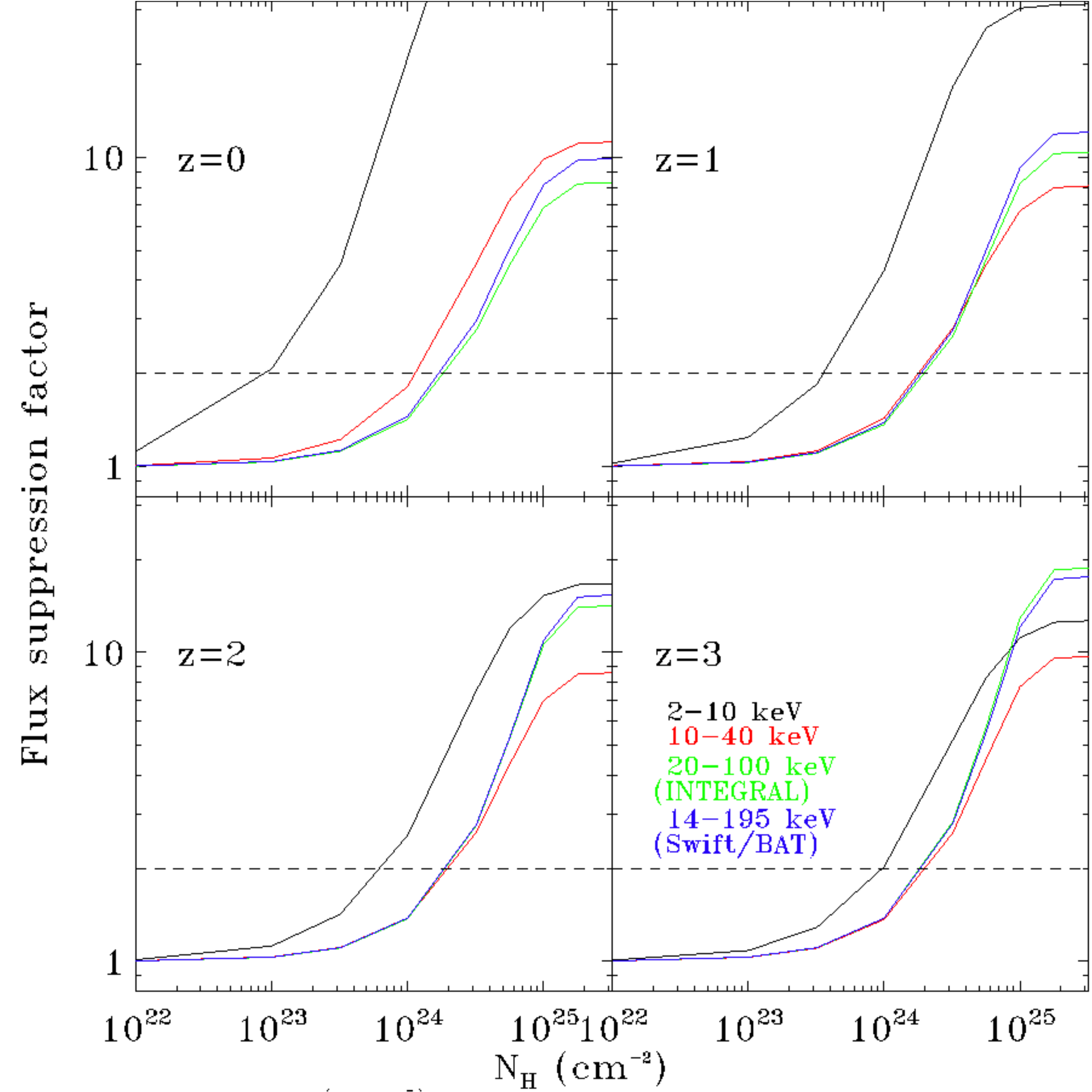}
\caption{The flux suppression factors for the spherical model (top) and the torus model (bottom) for the observed frame 2-10 keV, 10-40 keV, 20-100 keV (INTEGRAL) and 14-195 keV (Swift/BAT) bands as a function of \nh\ at redshifts of 0, 1, 2 and 3. The flux suppression factor is defined as the unabsorbed flux in that band divided by the observed flux, for $\Gamma = 2$.}
\label{fig_fsuppress}
\end{center}
\end{figure}

We also use the results from our new models to calculate expected flux suppression factors for different X-ray bands, as a function of absorption and for different redshifts. This information is useful when estimating the \nh, given only an observed flux and a predicted intrinsic flux, or when considering the biases against heavily obscured sources in X-ray selected AGN samples. We do this for both spherical and toroidal distributions of matter with solar abundances, $\Gamma=2$ intrinsic power-laws and for the case of the torus, a 60$^\circ$ opening angle and 90$^\circ$ viewing angle. Figure \ref{fig_fsuppress}  gives the flux suppression factor at different column densities and in different bands. The flux suppression factor is calculated as the intrinsic, unabsorbed energy flux in a particular band divided by the flux observed in that band. 

The attenuation of the intrinsic emission by the absorbing material is evident as the \nh\ increases. For the spherical case, at a redshift of zero and for energy bands above 10 keV, the effect of this attenuation is negligible ($<2$) for \nh\ up to $\sim4\times 10^{24}$ \cmsq, however, at \nh=10$^{25}$\cmsq, the flux in these bands is suppressed by a factor of 10. This relationship is roughly independent of redshift for these bands. For the 2-10 keV band, locally, the flux is suppressed by a factor of 10 at  \nh=$6\times 10^{23}$ \cmsq. As redshift increases, the \nh\ at which the flux suppression factor reaches 10 increases due to the photo-electric absorption cut-off being redshifted through the band. The \nh\ at which the flux suppression factor is 10 for redshifts one, two and three is $3\times10^{24}$ , $7\times10^{24}$ and $10^{25}$ \cmsq\ respectively for the 2-10 keV band.

Flux suppression factors calculated from the torus geometry behave similarly to the spherical case until $\sim4\times10^{24}$ \cmsq\ where they flatten off, rather than continuing to increase as they do for the spherical geometry. This is due to reflected X-ray flux from the torus which is present regardless of the level of obscuration seen directly through the torus. The flux suppression factor flattens off at $\sim100$ for the 2-10 keV band at z=0, however, this improves to $\sim30$, $\sim15$ and $\sim12$ for z=1,2 and 3 due to this band observing higher energy rest-frame X-rays which are less effected by photo-electric absorption that dominate in the 2-10 keV band at rest-frame energies.  For the $>10$ keV bands, the flux suppression factor levels off at $\sim10$ at all redshifts. At z=3, flux suppression in the 2-10 keV band is less than that seen in the 20-100 keV (INTEGRAL) and 14-195 keV (Swift/BAT) bands. This is most likely due to the Compton hump, produced by the 'down-scattering' of higher energy X-rays and peaks at rest-frame $\sim30$ keV. At z=3 it appears directly in the 2-10 keV band, whereas the higher energy bands only view the tail of the hump.

For both geometries, all redshifts and X-ray bands, flux suppression is $>10$ at \nh=10$^{25}$\cmsq. This shows that for sources with extreme obscuration, even X-rays are strongly attenuated.

\section{X-ray spectral analysis}
\label{specfit}

The X-ray spectra of galaxies and AGN are very complicated and as such it is necessary to utilise high signal to noise spectra with good spectral resolution and a wide band pass in order to correctly determine the source properties. These have been provided by \xmm\ better than ever before, so in using data obtained from this observatory, we aim to gain new and improved insights into the X-ray spectral properties of the 12MGS. \citet{rush96x} studied the soft X-ray spectra of the 12MGS from {\it ROSAT} All-sky survey observations, but in doing so will have missed any moderate to heavily obscured emission, and furthermore, the signal to noise and resolution of {\it ROSAT} was incomparable to that of \xmm. \citet{barcons95} used hard X-ray data from {\it HEAO 1} and will therefore have been sensitive to obscured sources. However, their data also suffered from poor signal to noise and low spectral resolution, meaning they could only determine the source intensity with no meaningful investigation of the absorption or continuum slope. 

We develop a systematic approach to the spectral fitting of our sample in order to determine the intrinsic X-ray continuum slope, $\Gamma$, the neutral absorption column density, \nh, and the 2-10 keV intrinsic source power, $L_{\rm X}$. Characterising $\Gamma$ is necessary for insights into the primary X-ray generation process in AGN, thought to be the Comptonisation of accretion disk photons by a hot corona \citep{sunyaev80, haardt91}. This intrinsic source continuum parameter is also useful for comparing different source populations, such as Seyfert 1s and 2s, to determine if their central engines are the same, as predicted by unified schemes. Determining the \nh\ distribution in AGN is a key ingredient to XRB synthesis models, and measuring the intrinsic power allows for comparison to the power at other wavelengths. We will discuss most of these issues in a subsequent publication however, and we focus on an accurate determination of these parameters here.

\subsection{Sample selection and data reduction}

\begin{figure}
\begin{center}
\includegraphics[width=80mm]{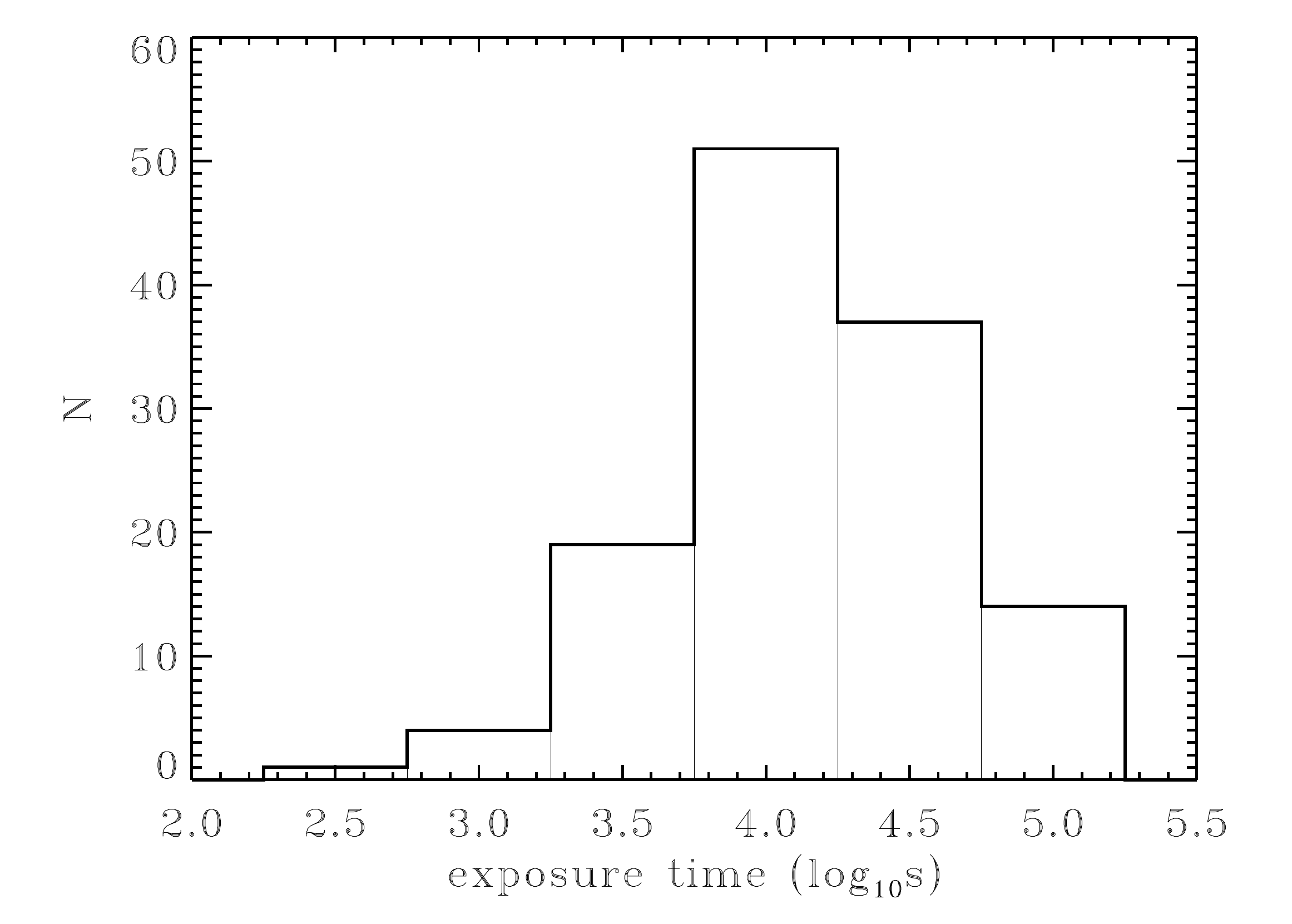}
\includegraphics[width=80mm]{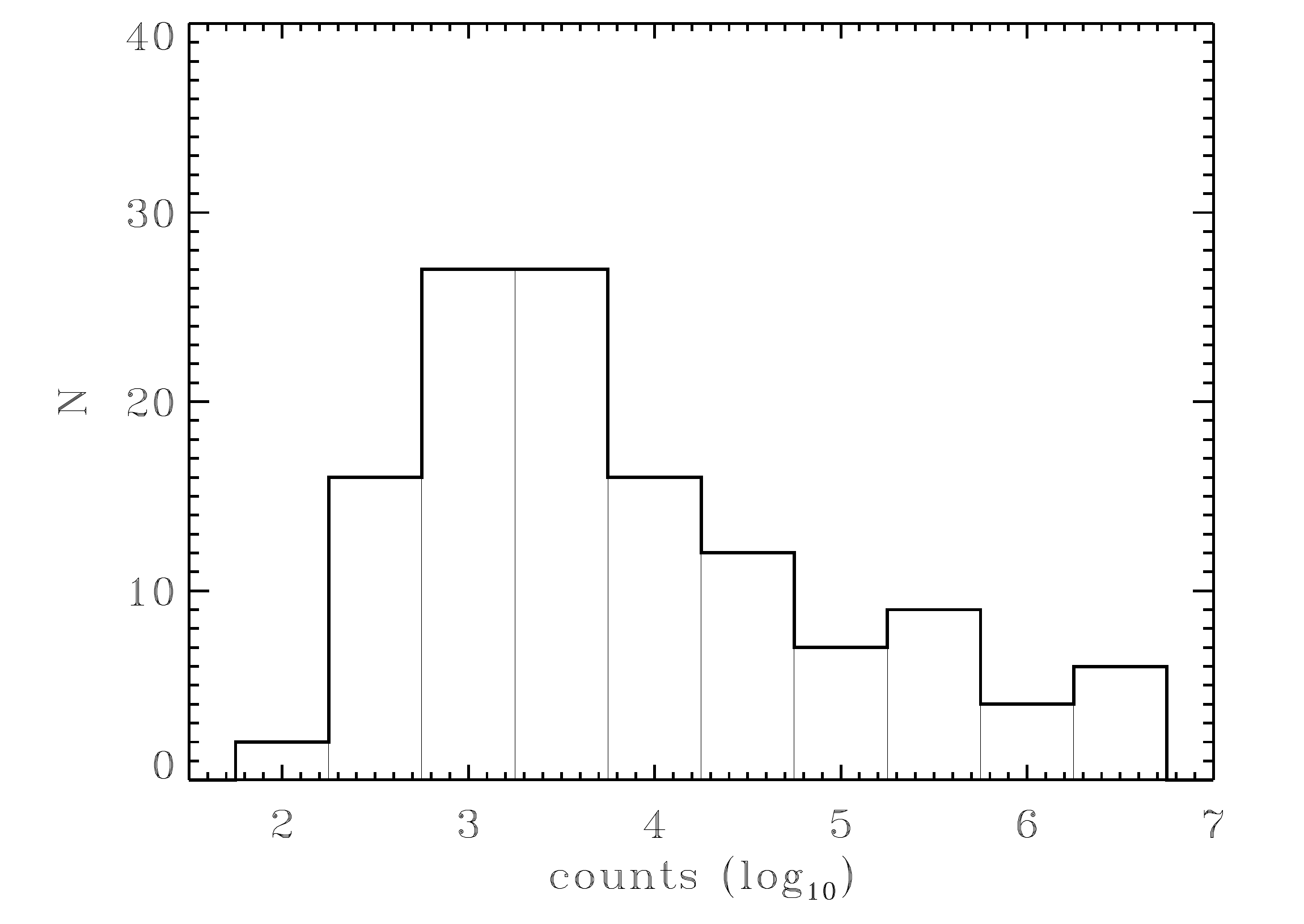}
\includegraphics[width=80mm]{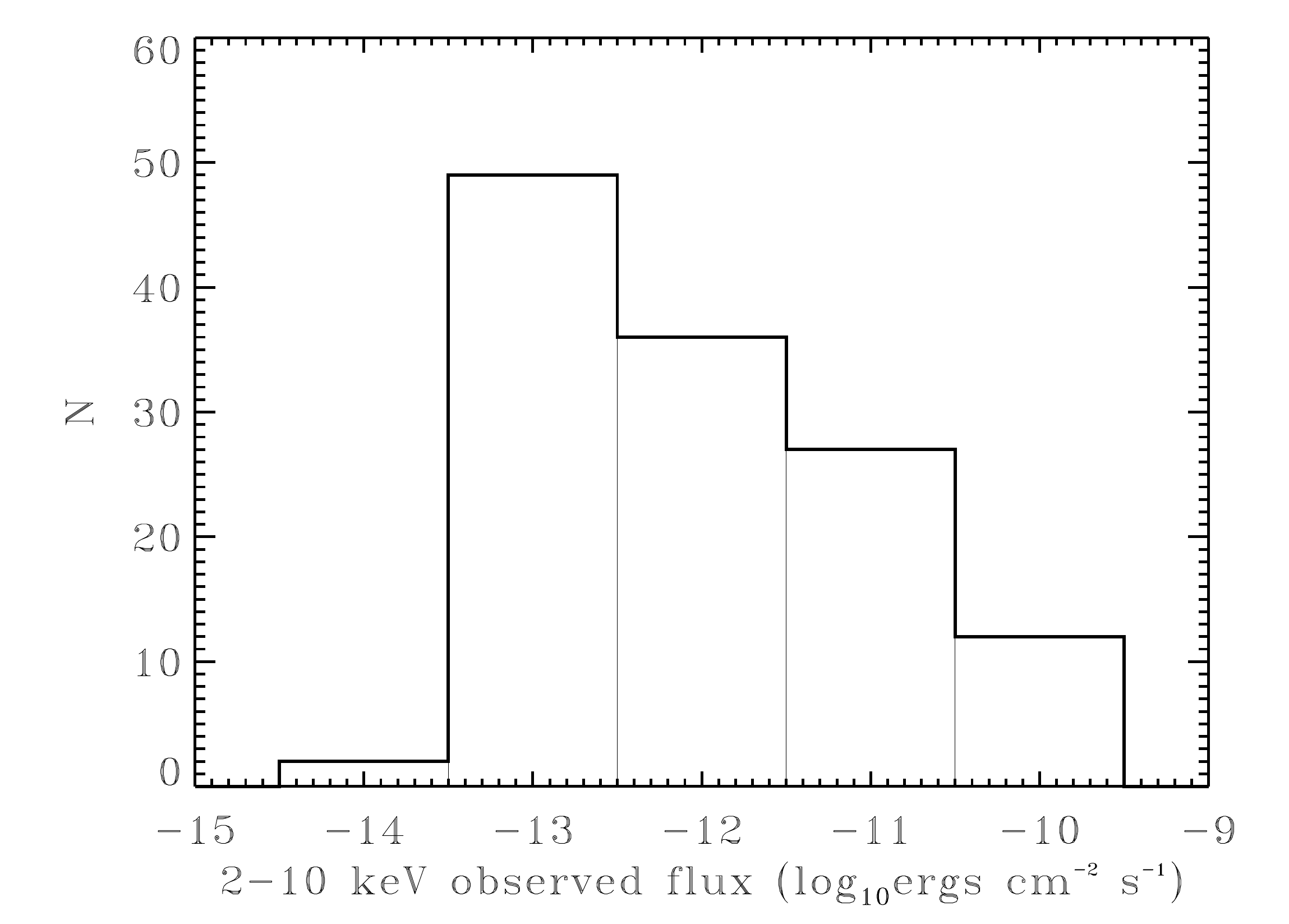}

\caption{X-ray observation statistics for the sample, with exposure time distribution (top), source count distribution (middle) and observed source 2-10 keV flux (bottom)}
\label{fig_obsstats}
\end{center}
\end{figure}

The sample of observations presented here consists of all EPIC-pn \xmm\ observations of the extended 12MGS \citep{rush93} that were public in the \xmm\ data archive as of December 2008, including sources which were observed serendipitously.  In the case of multiple observations of the same source, we use the longest observation available.  The data were reduced homogeneously and systematically using the method described by \citet{nandra07} using {\sc sas} v7.0 tasks.  This involves extraction of source events from a circular region, radius 35 arcsecs, centred on the source position, and extraction of background events from rectangular regions, ensuring no contamination from interloping sources. We search for pileup effects in the observations by comparing the distribution of single and double pixel events, which are sensitive to these effects, to model distributions.  Pileup effects are reduced if and when detected, by using only CCD pattern 0 (single) events \citep{ballet99}.  Background flares are filtered out by calculating the level of the background count rate at which the excess variance is determined to be zero.  A background count rate of twice this level is used to filter the observation of flares.  The subsequent background-subtracted spectra extracted are then binned with a minimum of 20 counts per bin, appropriate for $\chi^2$ fitting.  We include for analysis only those sources for which a meaningful spectrum has been produced by this process. This excludes those observations for which the flaring of the background has rendered the observation useless, where the source has not been detected above the background, or the detection is at low significance. We determine a meaningful spectrum to be one where the normalisation of a simple power-law model can be constrained to be greater than zero ($\Delta\chi^2>2.71$) with a fit to the 0.2-10 keV spectrum. After these considerations, the sample consists of 126 observations. The names, positions and redshifts of these galaxies along with the details of the \xmm\ observations are given in Table \ref{table:xrayanalysis_obsdat}. This set of \xmm\ observations is quite varied, with exposures varying from hundreds of seconds to tens of kilo-seconds (Fig. \ref{fig_obsstats}), with differing observation modes and filters used. We do however, treat each spectral fit the same, regardless of the number and distribution of counts in the spectrum (see Fig. \ref{fig_obsstats} for distribution of counts in these observations and the observed flux distribution). 

\subsection{Spectral fitting method: 2.5-10 keV}
We use {\sc xspec} v11.3 to carry out the spectral fitting of the 12 micron selected galaxies. As each spectrum has been grouped with a minimum of 20 counts per bin, the errors on the data points can be considered as Gaussian, and therefore $\chi^2$ fitting can be used to find the best fit parameters for a given model. 

Factors complicating the determination of the intrinsic parameters and the \nh\ are the soft excess components, ubiquitous in the soft X-ray spectra of AGN, which may arise from the accretion disk, thermal emission, scattered intrinsic emission and/or unresolved emission lines. For this reason, where possible, we determine the intrinsic source parameters from a fit to the 2.5-10 keV spectra for these galaxies, where the soft excess components are negligible. The following is a description of the spectral fitting methods we use.

We begin by fitting the 2.5-10 keV spectrum with a simple power-law model ({\tt powerlaw} in \xspec). We then systematically include further model components in order to fit the more complex features that the spectrum presents.  We note the change in the \chisq\ value of the fit after the addition of each added component and the change in the number of degrees of freedom (DOF).  For a model combination to be allowed, we stipulate that \dchisq $\geq 4.0$ for each DOF lost in adding the additional component (based loosely on 95\% confidence levels quantified using an F-test).   The combination of model components which meets these criteria and has the smallest reduced \chisq\ (\rchisq\ = \chisq/DOF) is presented as our best systematic fit for that spectrum.  

We use the following models in combination to acquire a fit to the 2.5-10 keV spectrum: 
\begin{itemize}

\item Neutral absorption model, {\tt zwabs} \citep{morrison83}, for absorption intrinsic to the source.
\item Neutral reflection from Compton thick material, {\tt pexmon} \citep{nandra07}, which is derived from the {\tt pexrav} model of \citet{magdziarz95} and includes self consistent Fe K$\alpha$, with its Compton shoulder, and Fe K$\beta$ line emission \citep{george91, matt02monte}.
\item Our new model describing a heavily absorbed power-law, {\tt trans}, based on Monte-Carlo calculations, with a spherical geometry, which take into account Compton scattering and includes line emission (as described in Section \ref{transsec}). We use this model in addition to the {\tt zwabs} model as it is more likely to pick out a heavily obscured source due to the inclusion of line emission.
\item Additional line emission, modelled by a gaussian: 

\end{itemize}

The {\tt pexmon} and {\tt trans} models include Fe K$\alpha$ line emission, produced by cold distant matter, and therefore, if a narrow emission feature exists at 6.4 keV, these models should be preferentially included in the fit. We also check to see if any emission is present from ionised iron by adding narrow gaussians at 6.7 and 6.96 which correspond to K$\alpha$ emission from He-like and H-like iron respectively. We also add a third gaussian component with energy constrained to be $6.2<$E$_{line}<6.6$ keV and with a free width parameter, to check for broad Fe K$\alpha$. This represents $\Delta$E=0.2 keV from the rest-frame energy of neutral Fe K$\alpha$ so as to avoid fitting ionised Fe lines at higher energy. In reality, broad Fe K$\alpha$ lines have been found with a lower energy than the energy we permit \citep[e.g.][]{nandra07}, but we use this conservative range as broad Fe K$\alpha$ lines are not the focus of this investigation. We keep these gaussian components if the normalisation of each line is constrained to be greater than zero ($\Delta\chi^2>2.71$).   

We keep the power-law indices for the {\tt pexmon} and {\tt trans} components equal to the primary power-law for each fit as it is presumed that each component in this band has the same origin. For spectra where the power-law index, $\Gamma$, is badly constrained we fix the power-law index to $\Gamma=1.9$, the canonical value for Seyfert galaxies \citep{nandra94}, which make up the majority of our sample, but also roughly appropriate for galaxies \citep{ptak99}. We define fits where $\Gamma$ cannot be constrained within $\Delta\Gamma \leq 0.7$ to be unconstrained as such an uncertainty introduces a large overall uncertainty into the fit and other fit parameters as well as the source luminosity.

We fit the 2.5-10 keV spectra initially as this band is least affected by galactic and other complex soft X-ray emission, and is therefore best suited for determining the intrinsic parameters of the nuclear source. Furthermore, the higher signal-to-noise ratio at soft energies often results in spectral fits being driven by the soft features. For example, strong soft emission may cause the power-law index to steepen and as a result the measured \nh\ may increase to compensate for this.  However, the lower signal-to-noise ratio in the hard band can sometimes lead to unconstrained fits. We do not use fits where the normalisation of the power-law is not constrained to be greater than zero and instead use the full band fit method, described below. For fits that produce a low $\Gamma$, a hidden transmission or reflection component is likely to be present and not recognised by the 2.5-10 keV fit. Where $\Gamma<1.4$, this presents a scenario where the photon index lies greater than 3-$\sigma$ away from the mean measured by \citet{nandra94}. In this case we also refer to the full band fit, where a reflection or heavily obscured transmission component is often then detected.

\subsection{Spectral fitting method: 0.2-10 keV}

Ignoring data below 2.5 keV leaves us insensitive to neutral columns less than $\sim 10^{22}$ \cmsq.  For satisfactory hard band fits, we then freeze the hard band fit and continue fitting the 0.2-10 keV spectrum with these parameters frozen, with the exception of the \nh\ parameter. The following models are then systematically added, again requiring that \dchisq $\geq 4.0$ for each degree of freedom lost in adding the additional component. Again, the combination of model components which meets these criteria and has the smallest reduced \chisq\ (\rchisq\ = \chisq/DOF) is adopted as our best systematic fit for the 0.2-10 spectrum.  

\begin{itemize}

\item Neutral absorption ({\tt wabs}) associated with the Galaxy, frozen to the value given in Table \ref{table:xrayanalysis_specfit} and applied to the whole spectrum. This component is assumed to be present regardless of the \dchisq.
\item Absorption applied to the whole spectrum and intrinsic to the host galaxy, be it neutral or ionised (see below). If no neutral absorption is detected in the hard band, we also check to see if a neutral absorption component is required for the primary power-law in the soft band ({\tt zwabs}).
\item A warm absorber ({\tt cwa18}, based on {\tt XSTAR} \citep{kallman04}. See \citet{nandra07} and references therein for description) to model cases where ionised absorption features are present in the spectrum. If evidence for neutral absorption is present in the hard band fit, we also check whether an improved fit results by adding a warm absorber on top of this or indeed, replacing this with a warm absorber.
\item A thermal plasma component \citep[{\tt apec\rm,}][]{smith01}. The soft X-ray spectra of obscured AGN have been shown to consist partly of emission lines photo-ionised by the AGN \citep{guainazzi07}. Models such as {\tt apec} are often used to account for this emission when these lines are not resolved. When using this model, the abundances are initially frozen to the solar values, however, once the best fit is obtained, we check if thawing the abundances produces a better fit.
\item A second power-law, where the power-law index, $\Gamma_2$ is constrained such that $\Gamma_2 = \Gamma_1$. This component is used to fit any `scattered power-law' often seen in Seyfert 2 type spectra \citep{turner97}. It can only be seen when moderate or high absorption is occurring to the primary power-law, and therefore we only add it if absorption is detected in the hard band fit. 
\item We also check to see whether $\Gamma_2 \geq \Gamma_1$ produces a better fit.

\end{itemize}

\subsubsection{Modelling the soft-excess seen in Seyfert 1 like spectra}

Modelling the soft-excess seen in quasar and Seyfert 1 like spectra is complicated, especially as we do not fully understand the nature of this feature. The models used to fit the soft excess are additive (often a blackbody or Comptonised component), thus accounting for flux in the soft X-ray band. However, if an absorption component is used in conjunction, adding the soft excess component can lead to an over-estimation of the \nh\ parameter due to `balancing' of emission and absorption. Thus, we apply strict criteria in order to include a soft excess model in the spectral fit. 

\begin{figure}
\begin{center}
\includegraphics[width=80mm]{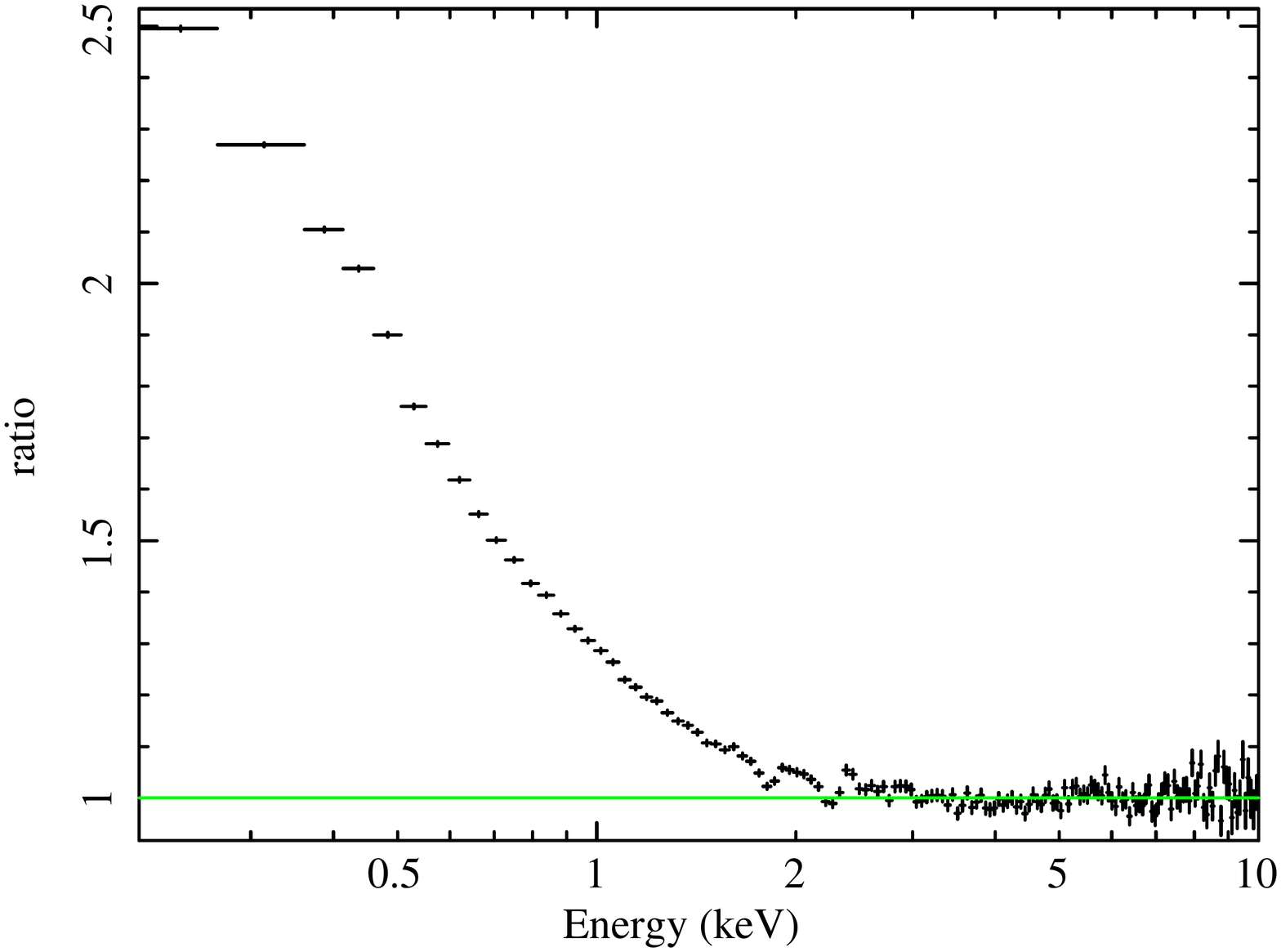}
\includegraphics[width=80mm]{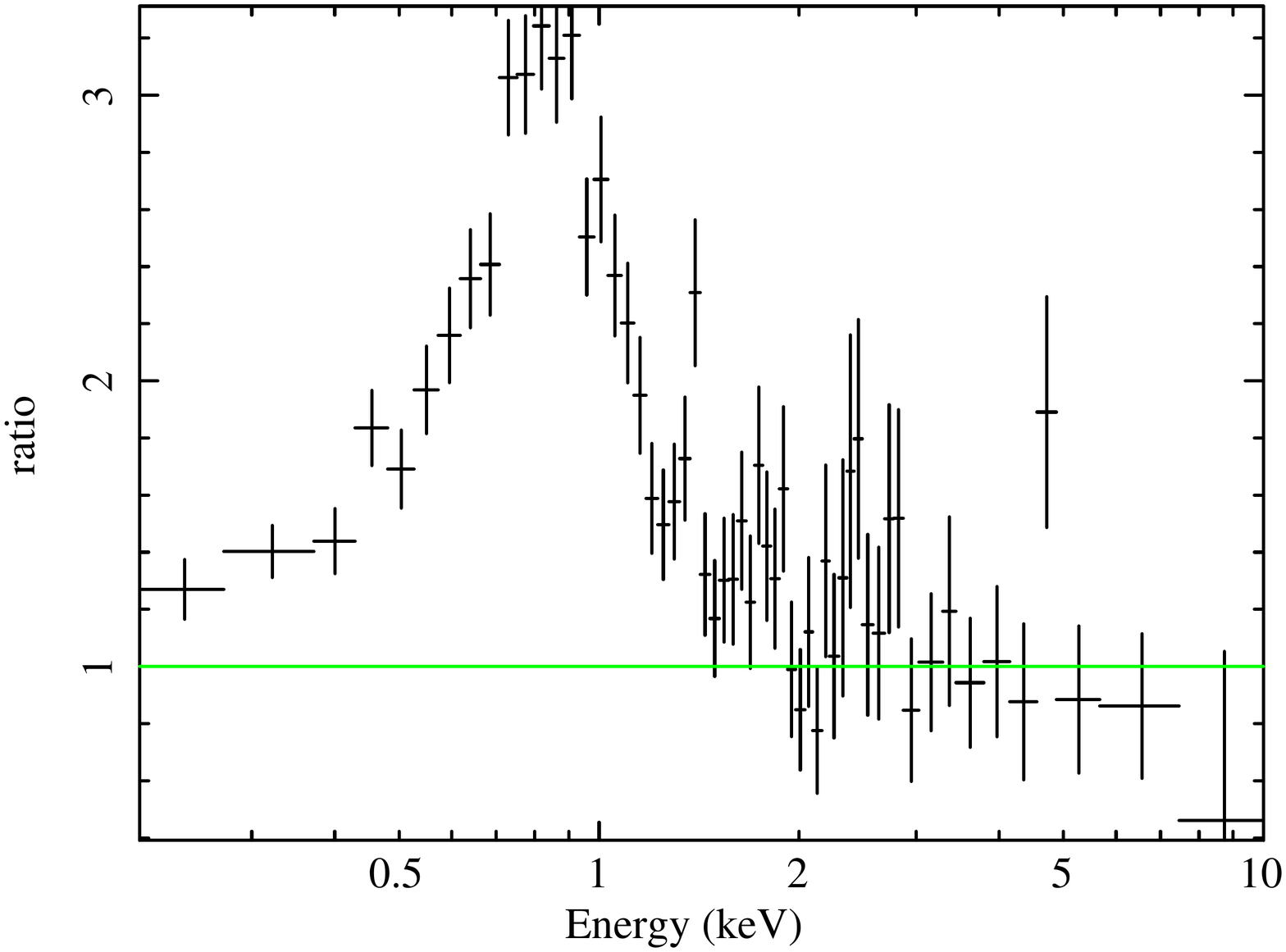}
\caption{The ratio of the data to the extrapolated 2.5-10 keV power-law for 3C273 (top) and Messier 100 (bottom). 3C273 has a classical soft excess at up to more than twice the level of the extrapolated power-law. Messier 100 has strong emission from a  thermal plasma, but this peaks at $\sim1$ keV.}
\label{soft_rat}
\end{center}
\end{figure}

This soft excess is thought to originate from close to the central engine, and thus should be an excess above the intrinsic, {\it unabsorbed} power-law spectrum. To test for the soft excess, we extrapolate our fit of the 2.5-10 keV spectrum into the soft band, removing any absorption components. Only then, if there is emission in excess of this extrapolated fit in the soft band, do we include soft-excess components in the soft band fit. To detect excess emission in the soft band, we use a `soft ratio', defined as the data-to-model ratio at 0.4 keV, done  similarly by \citet{sobolewska07}. An energy of 0.4 keV is used to avoid atomic emission from oxygen at $\sim0.7$ keV. We choose a soft ratio lower cut off of 1.2 (20\%), which is high enough to exclude uncertainties in the spectral calibration (see below). We also require this ratio to be rising towards softer energies, as emission from thermal plasmas can cause an excess at 0.4 keV, but that peaks at higher energy. Figure \ref{soft_rat} gives two examples where this soft ratio is greater than 20\%, but one shows a classical soft excess which rises from 1 keV to 0.5 keV (3C273), whereas the other shows an excess due to the emission from a thermal plasma, which falls from 1 keV to 0.5 keV (Messier 100). If the soft ratio exceeds 1.2 and the ratio is rising from 1 to 0.5 keV, we then proceed to include soft excess model components in the spectral fit. The model components we use to fit the soft excess are either a black body ({\tt bbody}) or a Comptonised  ({\tt compTT}) component, whichever provides the best fit.

\subsubsection{Bad spectral fits}

We define fits which give \rchisq\ $>2.0$ as `bad'. Although this definition is arbitrary, it is useful for identifying cases where the models used are not a good representation of the spectrum. For the 2.5-10 keV fits, only NGC 1068, NGC 1365 and M82 have bad fits. These sources have notoriously complex spectra, which are high signal to noise in this case, with several emission and absorption features. We do not attempt to fit these additional complexities, however, and present only the basic hard band parameters for these fits. No errors are quoted because the standard formalism does not apply unless the model fits the data \citep{lampton76}. 

When we fit the full band spectrum, we sometimes find that keeping the hard band parameters fixed will cause a bad full band fit, so we then thaw these parameters for an improvement. After this procedure, 17/126 fits are still bad. These bad fits usually result from high signal-to-noise spectra which reveal highly structured soft X-ray features which are too complex for our models. However, the nature of the soft X-ray emission is not the focus of this investigation,  as we seek mainly the intrinsic hard X-ray power of the source, along with any neutral absorption present. In these cases, we do not present the results from the full band fit, rather only the hard band fit. These are denoted by an $^{(H)}$ in the spectral fit table. Furthermore, we note that the calibration uncertainties on \xmm\ are at the level of $\sim$5\% \citep{kirsch04}. If the ratio of the data to the model is less than this in the soft band, the bad fit may be partially due to calibration uncertainties. In this case we present the soft band fit results, despite a bad \chisq. We present the parameters $\Gamma$ and the power-law normalisation from the hard band fits here as these fits had `good' \rchisq\ values and thus the uncertainties could be estimated. These are denoted by a $^{(D)}$ in the spectral fit table. 3C120 is an example of a source with a bad spectral fit, but a data/model ratio of less than 5\%, which we show as an example in Fig. \ref{xrayanalysis_specfig}. Our procedure should not result in grossly inaccurate estimates of $\Gamma$, $L_{\rm X}$ or \nh\ in such cases.

For cases where the hard band fit is unconstrained (i.e. the power-law normalisation is not constrained to be greater than zero) or the hard band fit is bad (\rchisq\ $>2.0$), then we cannot use the method described above, whereby the hard band fit is extended into the soft band. In these cases, we carry out simultaneous full, 0.2-10 keV band fits with all model components. Although this is less desirable, it is useful in the case where the hard band fit is unsatisfactory and for cases where the hard band measures a flat power-law index. These are denoted by an $^{(F)}$ in the spectral fit table.


\subsection{Intrinsic power and reflection fraction}

\begin{figure}
\begin{center}
\includegraphics[width=80mm]{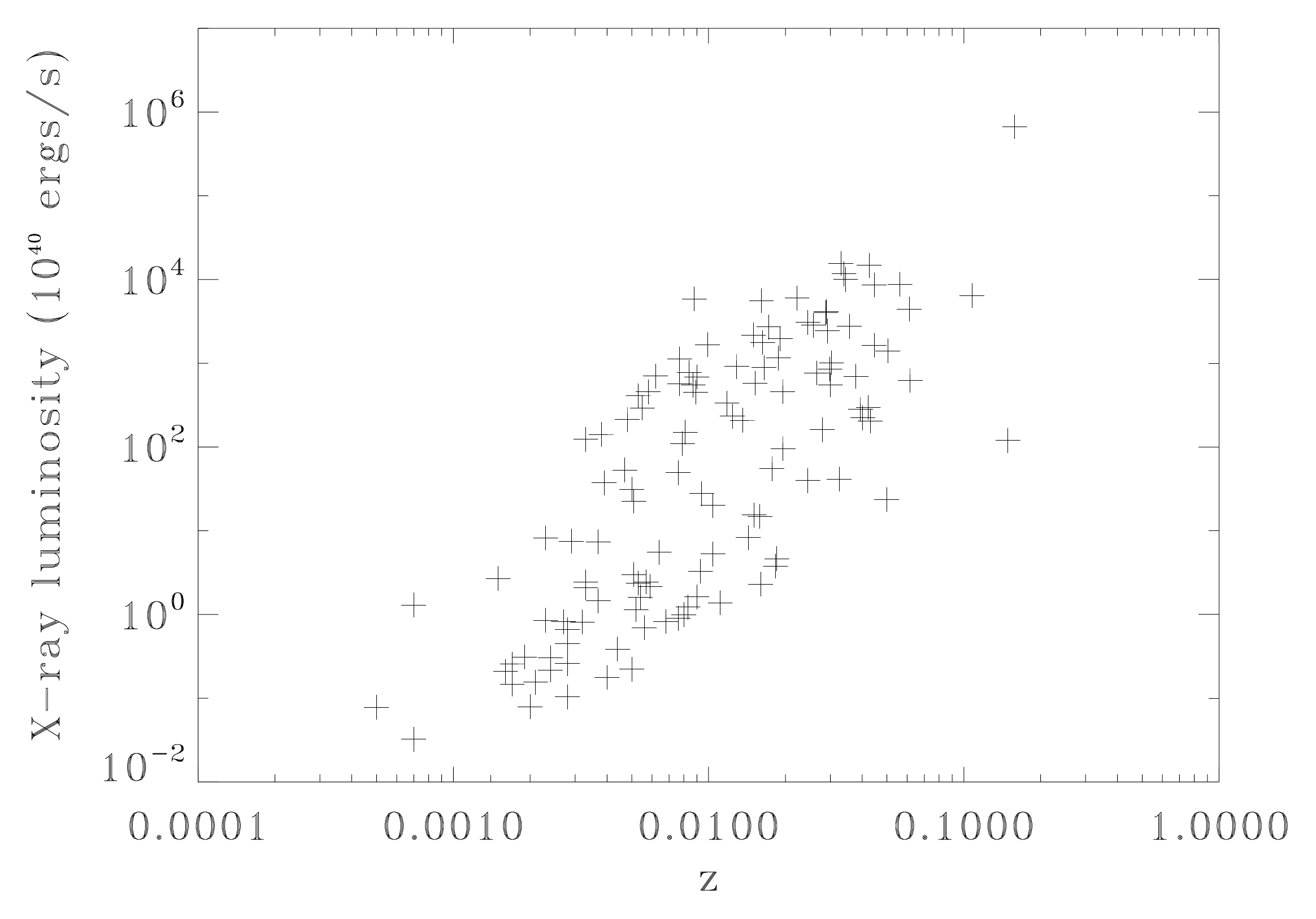}
\includegraphics[width=80mm]{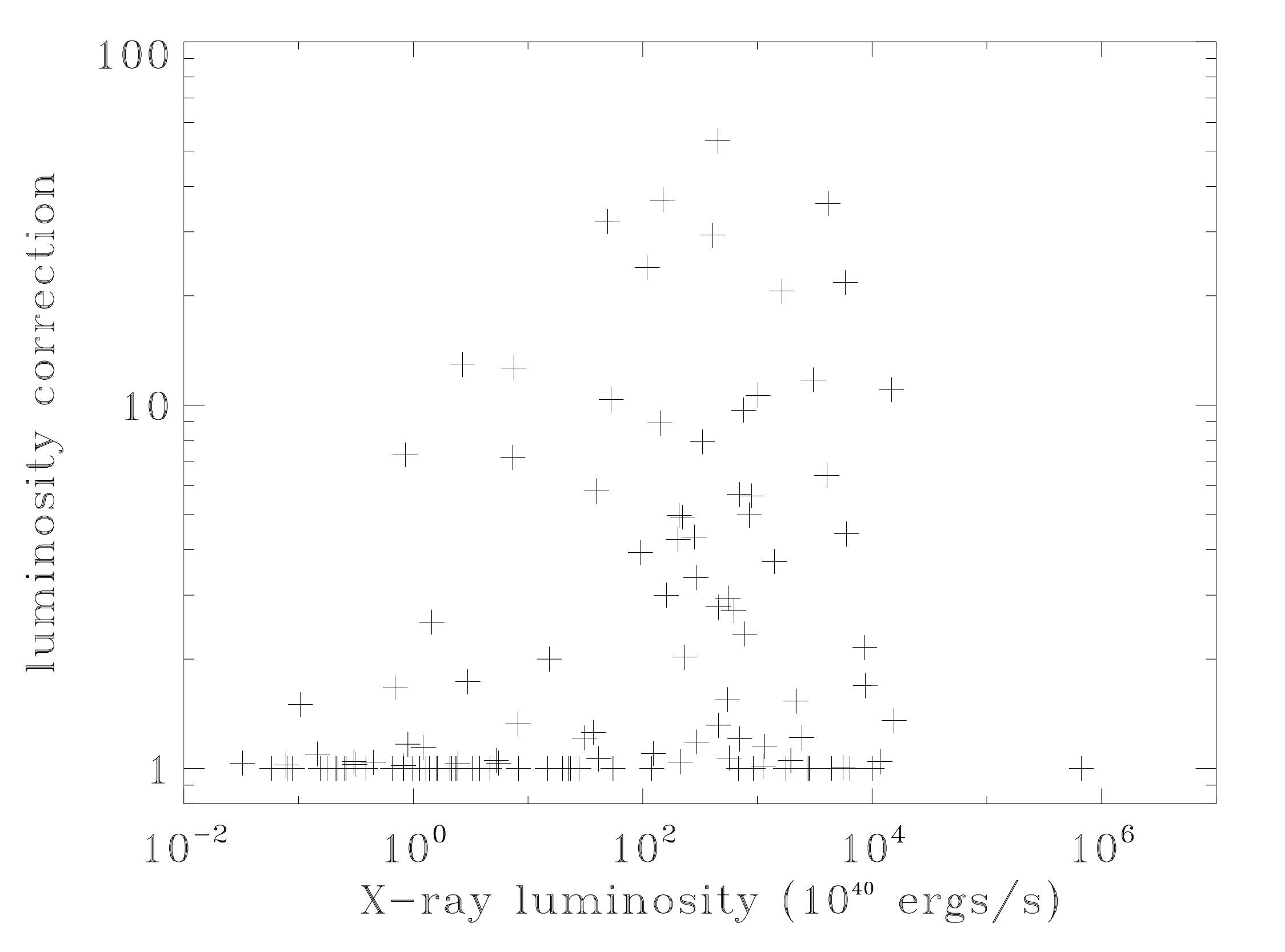}
\caption{(top) 2-10 keV intrinsic (absorption corrected) luminosity plotted against redshift for all objects. (bottom) The luminosity correction ($L_{\rm X} (int)/L_{\rm X} (obs)$) against the intrinsic X-ray luminosity.}
\label{xrayanalysis_zlum}
\end{center}
\end{figure}

We calculate the intrinsic power emitted in the 2-10 keV band for each galaxy based on the absorption corrected flux in the 2-10 keV band from the spectral fits to the primary power-law. We define the primary power-law as the power-law component with the greatest normalisation, i.e. the intrinsic flux at 1 keV. This may include the {\tt trans} component, if this is present, which will typically model a heavily obscured hard component visible only at hard energies.  Standard cosmological parameters are assumed throughout ($\Omega_m=0.3$, $\Lambda = 0.7$, H$_0 = 70$ km s$^{-1}$ Mpc$^{-1}$). In the top panel of Fig. \ref{xrayanalysis_zlum} the intrinsic, absorption corrected 2-10 keV luminosity of each galaxy is plotted against its redshift. This figure illustrates that the 12MGS is a low redshift galaxy sample, with a wide range of X-ray luminosities from 10$^{38}$ \ergs\  through to 10$^{44}$ \ergs (though with one very luminous source, 3C273 at $\sim10^{46}$ \ergs). The lower panel shows the size of the correction made to the observed luminosity against the intrinsic luminosity, which can be up to a factor of $\sim$50.

For reflection dominated spectra, we derive the intrinsic power of the source from the intrinsic power-law required to produce the reflection component.  The {\tt pexmon} model calculates the reflection from a semi-infinte slab, subtending $2\pi$ steradians on the sky. The estimate of the intrinsic power in the reflection dominated sources is uncertain as it depends on the precise geometry of the reflector. This was shown by \citet{murphy09} who modelled reflection from a toroidal distribution of matter, finding that the reflection fraction can be underestimated in this way by a factor of $\sim6$. We also investigate the reflection fraction and intrinsic luminosity derived using the {\tt pexmon} model and compare these to those derived from our own {\tt torus} model.

\subsection{What can we determine using the torus spectral model?}

For spectra in which we find reflection is strong (R$_{pexmon}>1$, where R$_{pexmon}$ is the ratio of the normalisation of the {\tt pexmon} component to the normalisation of the primary power-law), we fit the same spectrum with model spectra from the torus simulations with two aims. Firstly, as mentioned above, we aim to investigate the reflection fraction and intrinsic source luminosity as determined using the torus model with respect to the {\tt pexmon} slab model, and secondly, we investigate if it is possible to determine and constrain the torus geometric parameters, $\theta_{tor}$ and $\theta_i$ from the 2-10 keV spectra.  For determining the reflection fraction, we fix $\theta_{tor}=60^{\circ}$, leaving $\theta_i$ free to vary. We then determine if the inclination angle can be constrained. Following this, we fix  $\theta_i=90^{\circ}$, and allow  $\theta_{tor}$ to vary, to see if this can be constrained.

\subsection{X-ray AGN in our sample}

The main goal of this work is to ascertain the properties of the primary emission, as well as the absorption characteristics of our sample. We do this initially for our whole sample, and we then go on to do so for AGN only. We define a set of unambiguous X-ray AGN as galaxies with an observed 2-10 keV luminosity greater than 10$^{42}$ \ergs\ as these sources are unambiguously powered by the accretion of material onto a super-massive black hole. No local pure star-forming galaxy has ever presented a 2-10 keV luminosity above this limit. One of the most X-ray luminous star forming galaxies is NGC 3256, which presents a 2-10 keV luminosity of 2.5 $\times 10^{41}$ \ergs, but shows no evidence for AGN activity \citep{moran99}.  Therefore, galaxies with an observed 2-10 keV luminosity greater than 10$^{42}$ \ergs\ are very likely to host an AGN. However, this will not select all AGN in this sample, so the remaining sources cannot be thought of as non-AGN. In this way 60/126=48\% of the sources in our sample are unambiguous AGN taking into account their X-ray luminosities only.

\section{Spectral fitting results}\label{specfitresults}

\subsection{Spectral fit components and parameters}

From our spectral fitting method, we have determined a best fit model for each source, which includes all the absorption and soft excess components, giving us an accurate estimate of $\Gamma$, \nh\ and the intrinsic power accounting for the complexities in the spectra. For 126 spectra, we find that in 62 of them, we can constrain $\Gamma$ on the primary power-law to $\Delta\Gamma<0.7$ (we fix $\Gamma=1.9$ otherwise). Furthermore, 96 spectral fits require an absorption component, 39 of which warm absorption has been detected. A thermal plasma component is ubiquitous in the fits as well, needed to fit 93 spectra. Fe K$\alpha$ emission is detected in 71 spectra, where 36 of these are broad lines ($\sigma>0.1$ keV). A secondary power-law component is required in 44 of our fits, and reflection is also required in 44 of the fits.

The spectral parameters of all fits are listed in Tables \ref{table:xrayanalysis_specfit} to \ref{table:xrayanalysis_specfit_feka}.  Table \ref{table:xrayanalysis_specfit} presents the basic spectral fit parameters such as the fit statistics, the parameters of the primary power-law and the thermal plasma component, where present, and the observed 2-10 keV flux and intrinsic 2-10 keV luminosity.  Table \ref{table:xrayanalysis_specfit_pl2} gives details of any secondary power-law component required in the fit; Table \ref{table:xrayanalysis_specfit_sx} gives details of the sources which have a soft excess detected and of the models used to fit this component; Table \ref{table:xrayanalysis_specfit_pex} presents details of the reflection component needed in each fit, including data obtained from the {\tt torus} model; Table \ref{table:xrayanalysis_specfit_cwa} presents the parameters of the warm absorber model if ionised absorption has been detected in the spectrum; and Table \ref{table:xrayanalysis_specfit_feka} gives details of the line emission detected in the iron K regime. Included on all fit parameters in which a satisfactory fit has been acquired are the 90\% confidence errors for one interesting parameter (\dchisq=2.706).

\begin{figure*}
\begin{center}
\begin{minipage}{160mm}

\includegraphics[width=80mm]{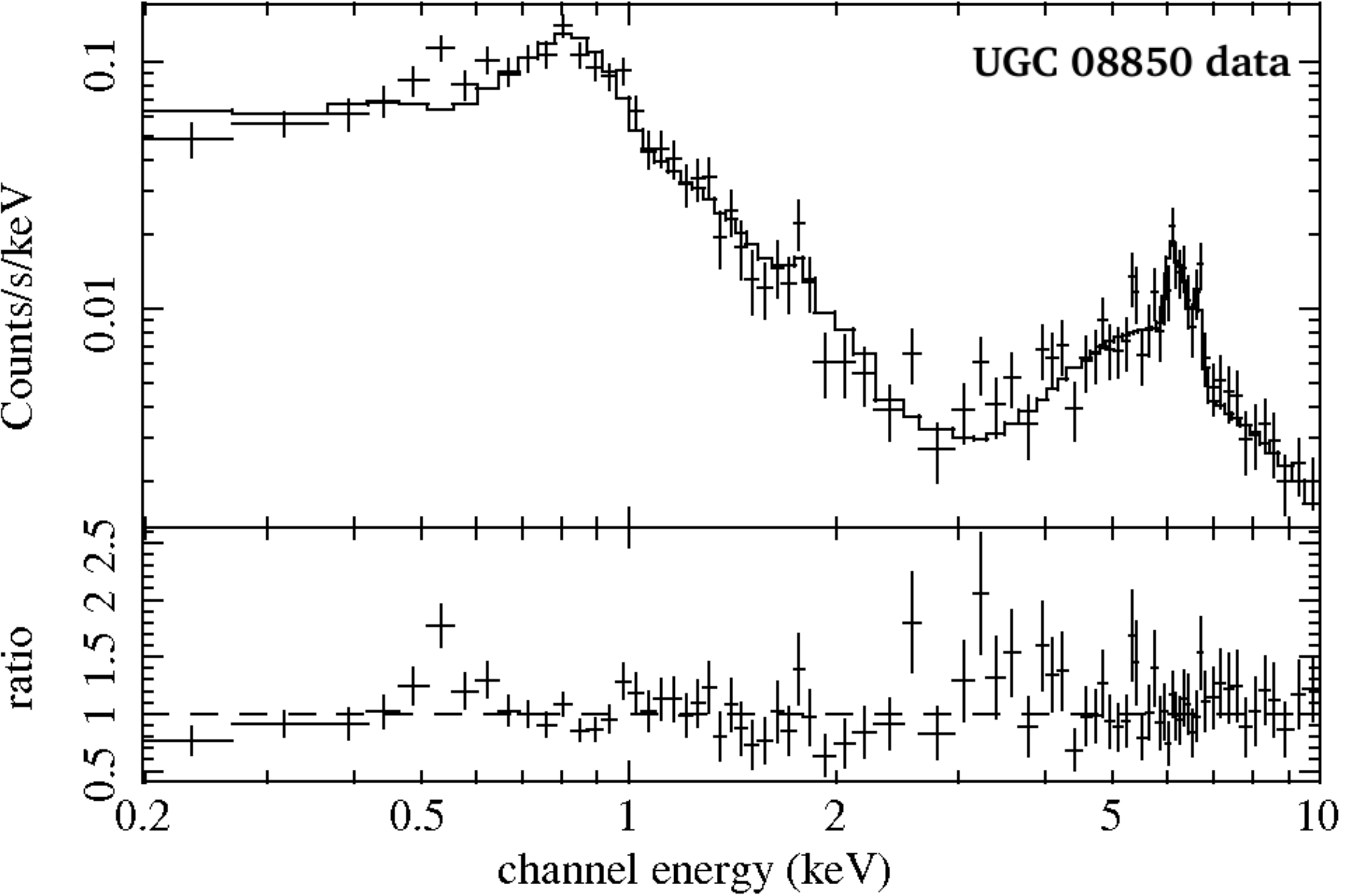}  \includegraphics[width=80mm]{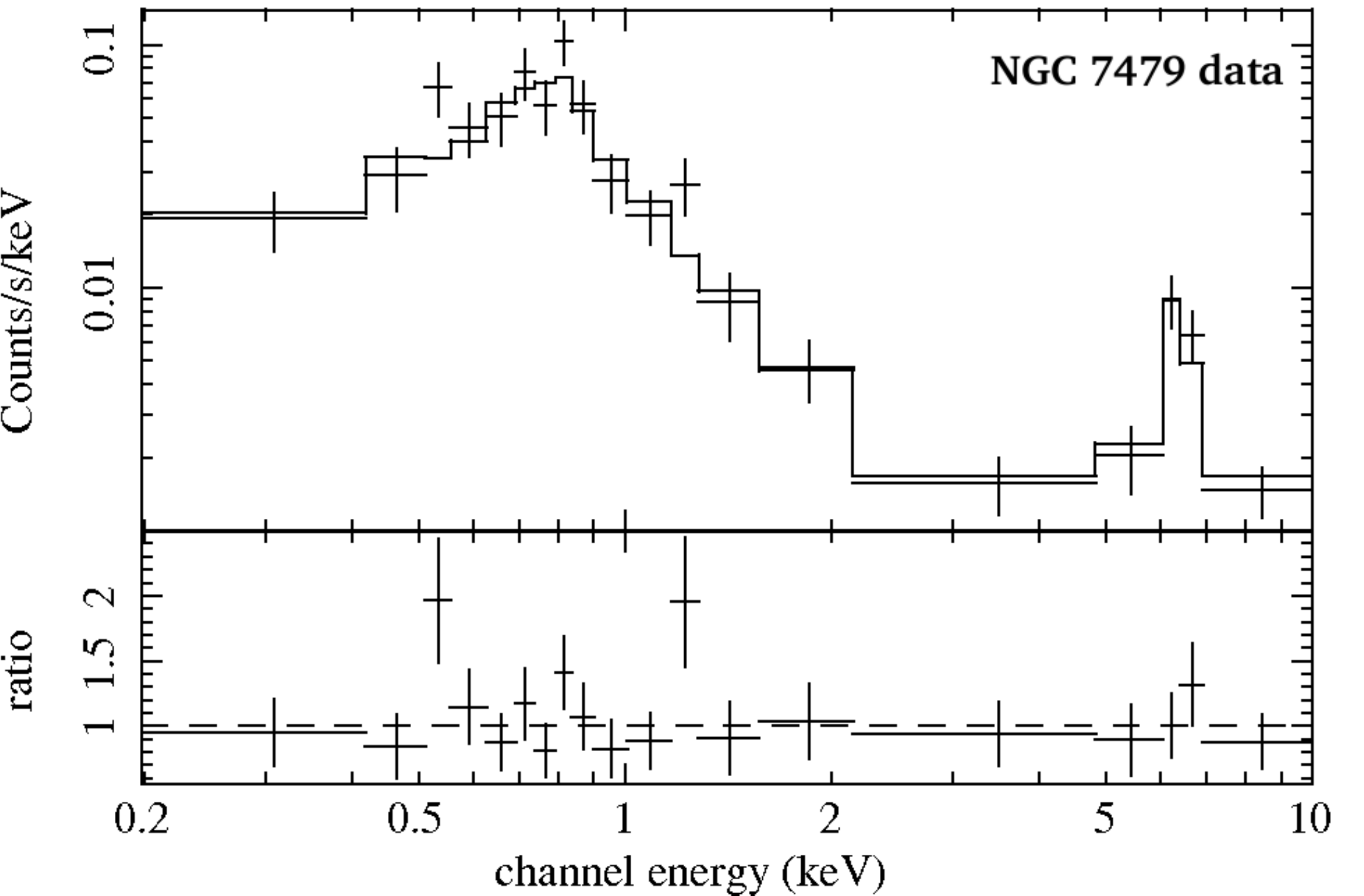}
\includegraphics[width=80mm]{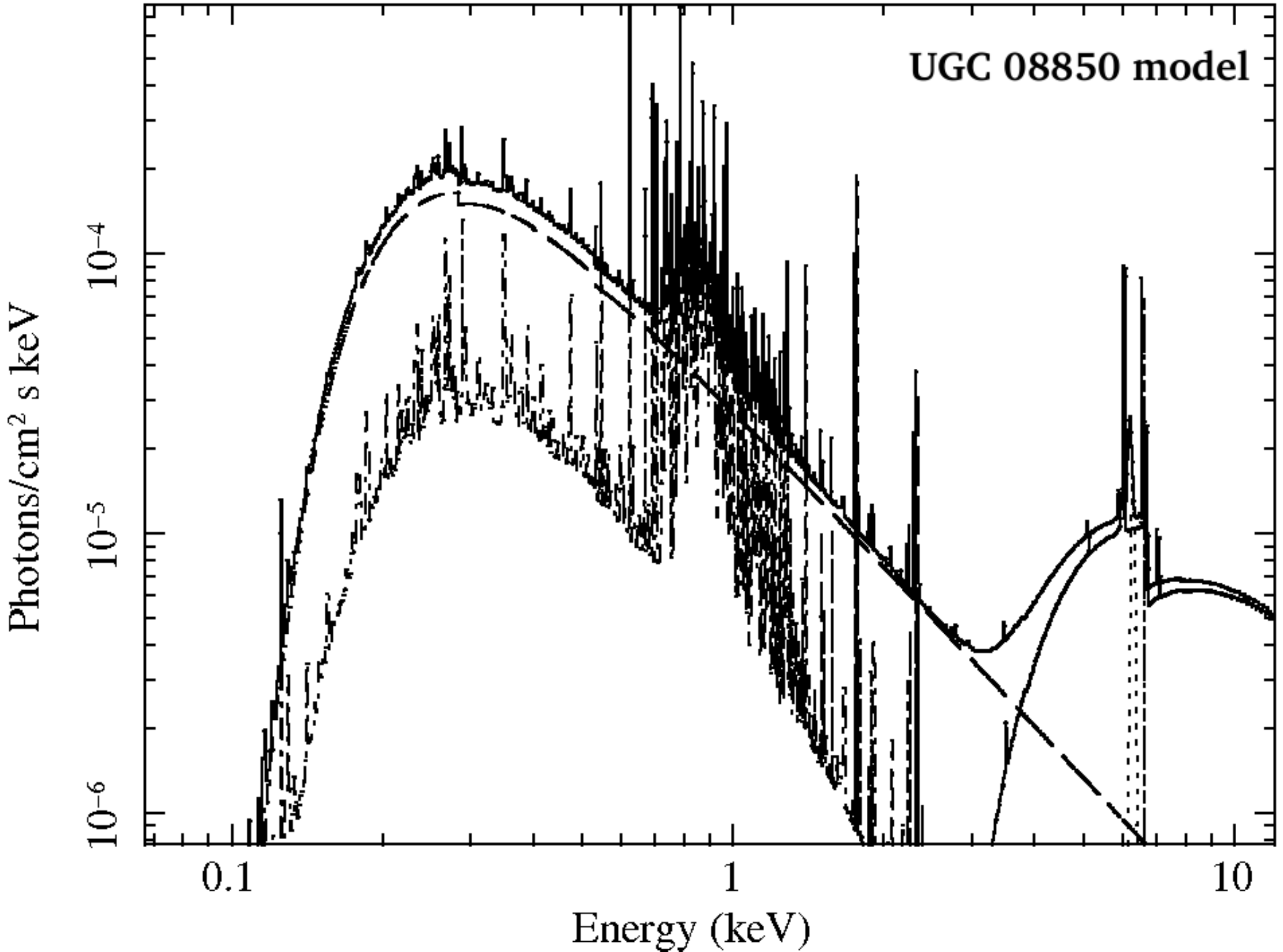}  \includegraphics[width=80mm]{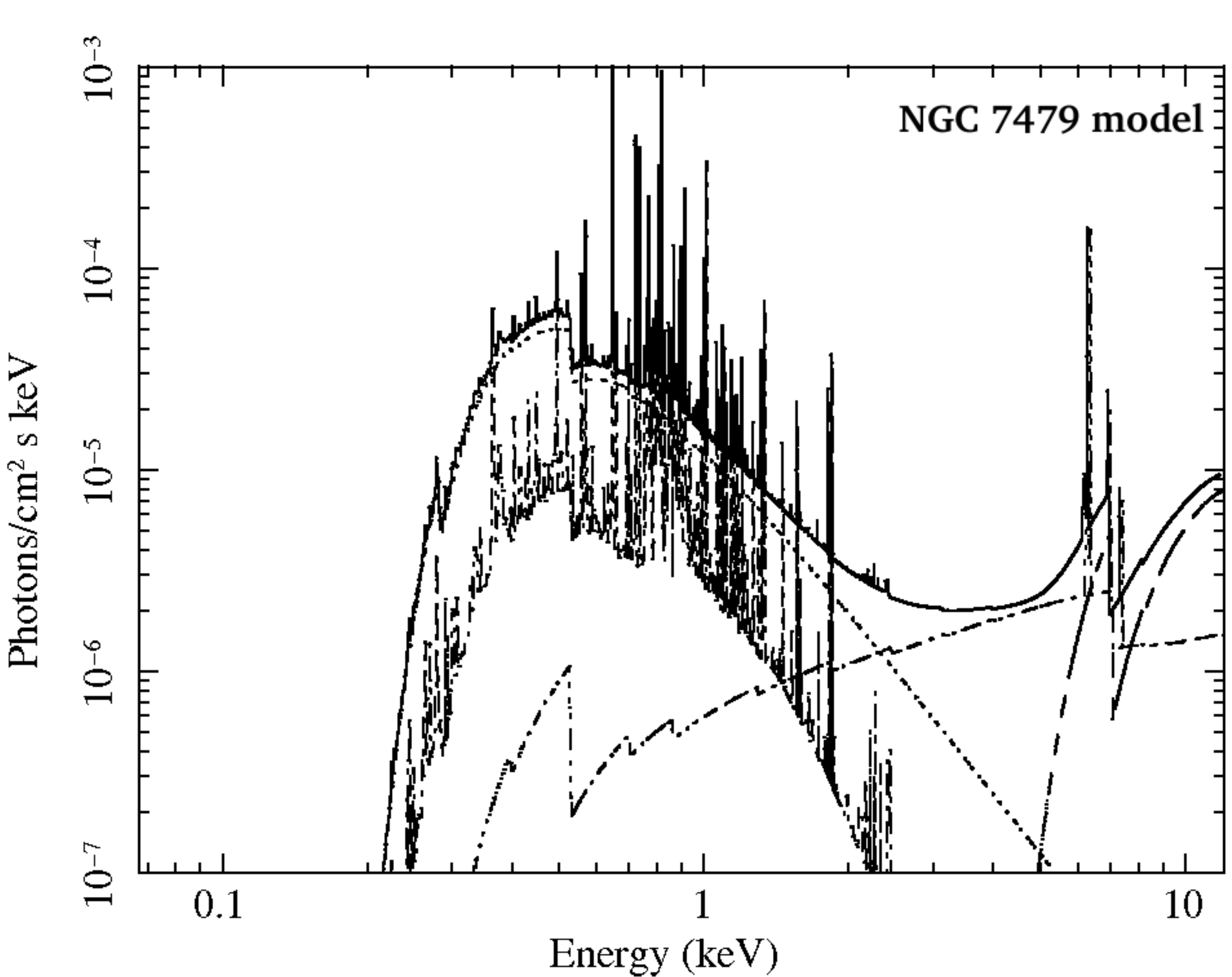}

\caption{0.2-10 keV spectra of 3C120, NGC 3079, NGC 4593, NGC 5170, UGC08850 and NGC7479, shown with best fit model from Table \ref{table:xrayanalysis_specfit}.}
\label{xrayanalysis_specfig}
\end{minipage}
\end{center}
\end{figure*}

A selection of example spectra are presented in Fig. \ref{xrayanalysis_specfig}. These are the spectra of 3C120, NGC 3079, NGC 4593, NGC 5170, UGC08850 and NGC7479, chosen to display a range in signal-to-noise ratio as well as model components. 3C120 and NGC 4593 have a soft excess present in their spectra as determined by the ratio of the data at 0.4 keV to the extrapolated 2.5-10 keV power-law, and fitted by the {\tt compTT} model. Reflection from neutral material is included in the fit, where in the case of NGC 4593, this component accounts for the Fe K$\alpha$ emission seen. 3C120 requires an additional gaussian component to model what seems to be broad Fe K$\alpha$ emission. These fits also include a thermal plasma component, however, it is possible that in these cases this model component is not representative of real thermal plasma emission as this component is not likely to be detected above the emission from the AGN, and is rather being used to fit the soft complexities of the spectra. A warm absorber has also been detected in the spectrum of NGC 4593. 

The spectrum of NGC 3079 is typical of a reflection dominated spectrum (R$_{pexmon}$ = 16.3) with an intense Fe K$\alpha$ component (940 eV) and is certainly Compton thick. UGC 08850 presents a transmission dominated spectrum, which is well fitted by our new model {\tt trans} measuring \nh\ $= 3.8 \times 10^{23}$ \cmsq, plus a scattered power-law. In the spectrum of NGC 7479 we have detected a hidden transmission component (\nh\ $=  2 \times 10^{24}$ \cmsq) beneath a reflection dominated spectrum, though it is not well constrained. The dominance of the reflection component and high EW of the Fe K$\alpha$ line (768 eV) imply a high column density in support the measured \nh\ in the transmission component.

The spectrum of NGC 5170 has a low signal to noise ratio and in this case we have had to fix the power-law index to 1.9. The only added model component here is low level neutral absorption.

\begin{figure*}
\begin{center}

\includegraphics[width=80mm]{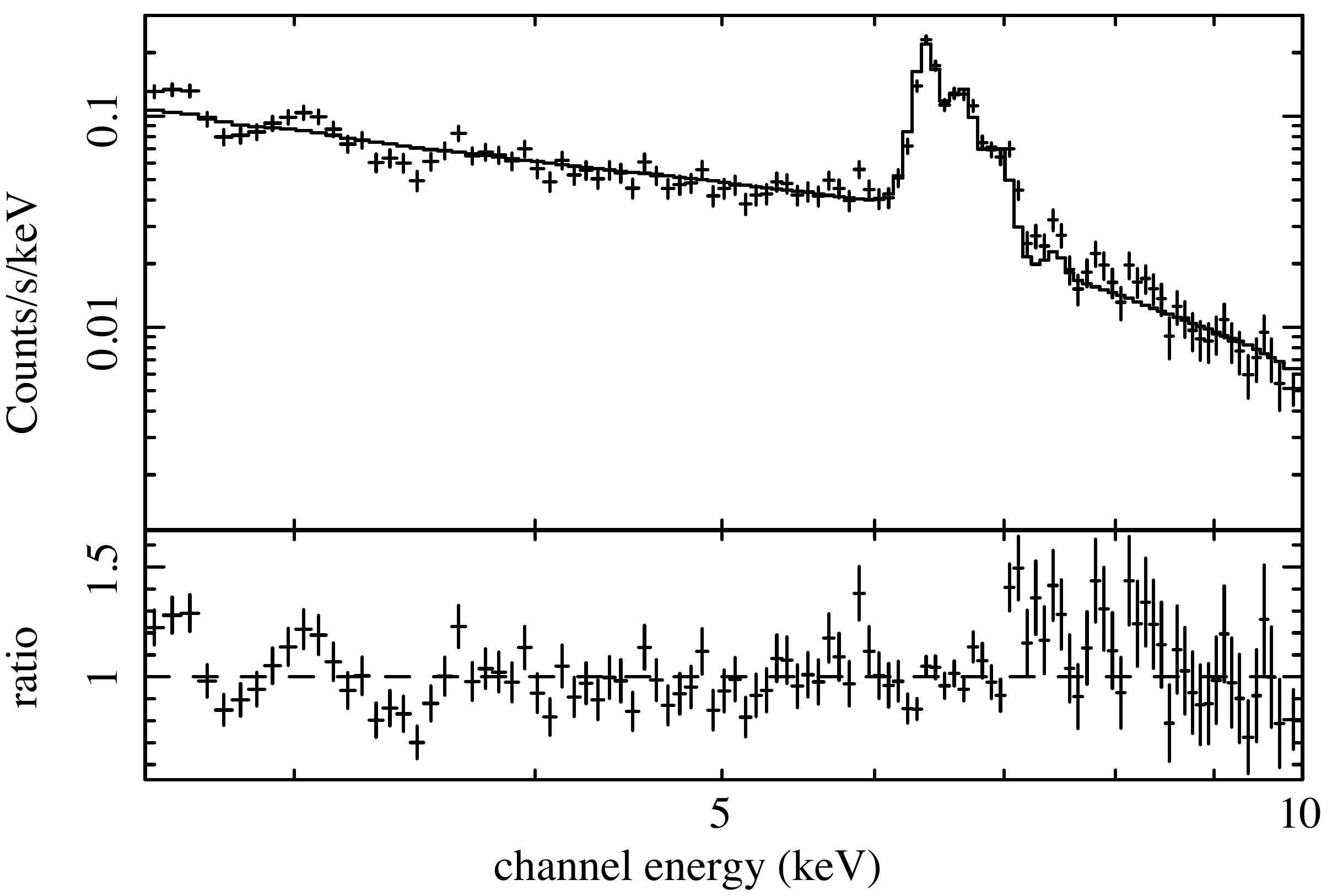}
\includegraphics[width=80mm]{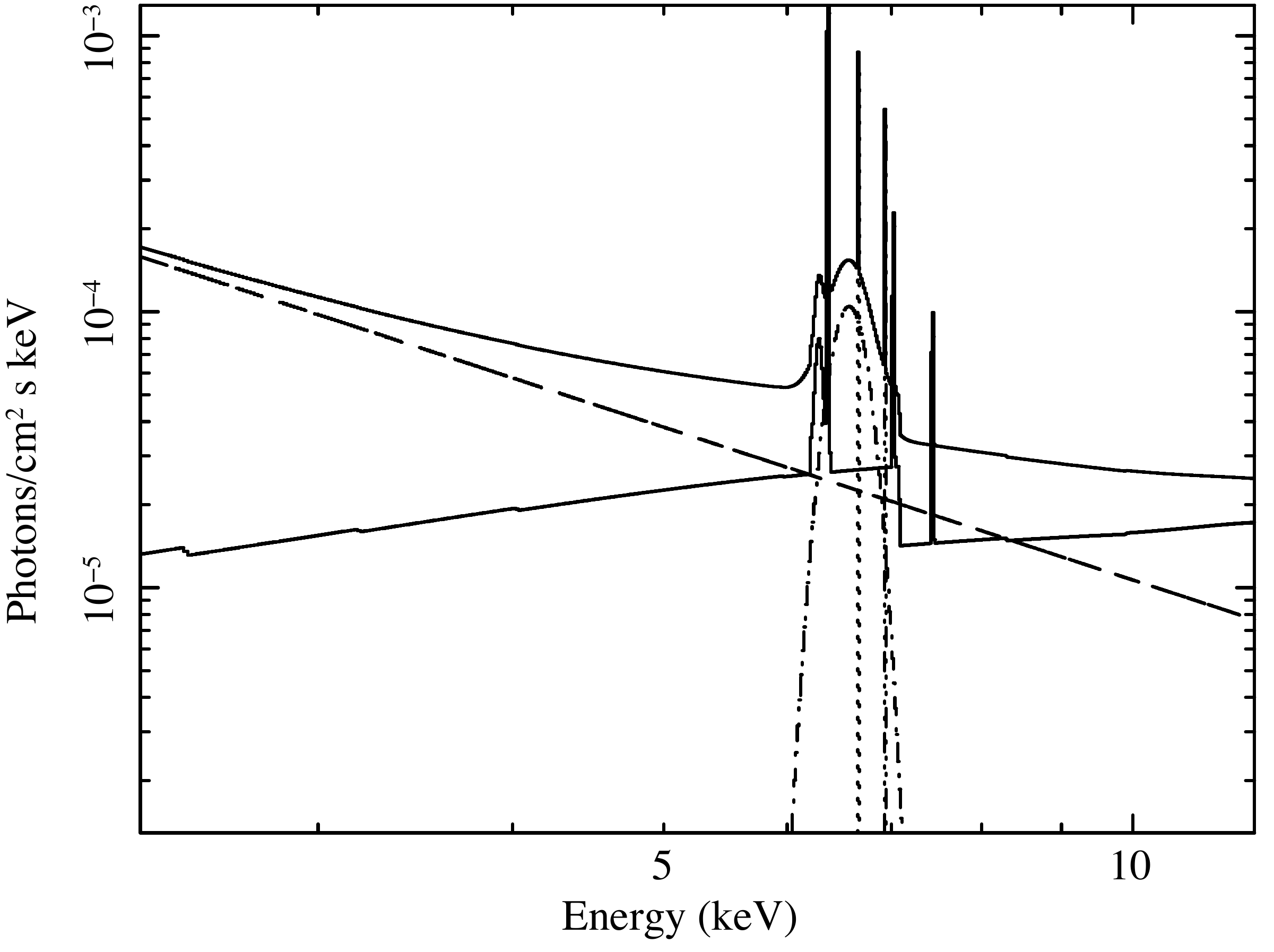}

\caption{2.5-10 keV spectrum of NGC 1068 (left), shown with best fit model (right) from Table \ref{table:xrayanalysis_specfit}. This is an example of a spectrum where the fit is bad due to spectral features too complex for the models we use.}
\label{NGC1068specfig}
\end{center}
\end{figure*}

In Fig. \ref{NGC1068specfig}, we present the 2.5-10 keV spectrum of NGC 1068, which is an example of where the 2.5-10 keV fit is bad. This is a high signal-to-noise spectrum which reveals multiple discrete features, possibly due to absorption and emission, or even calibration uncertainties.

\begin{figure*}
\begin{center}
\begin{minipage}{160mm}

 \includegraphics[width=80mm]{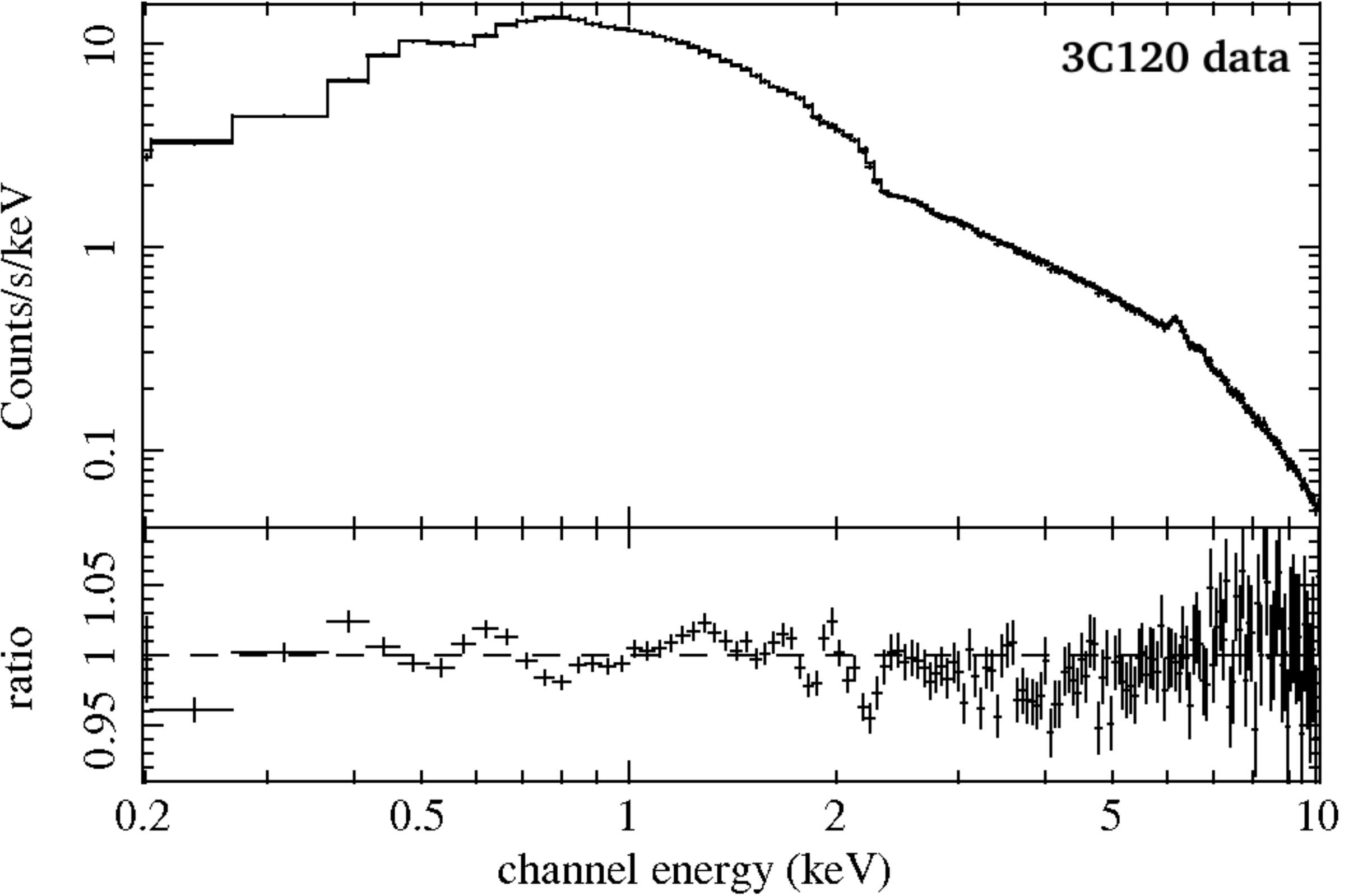}  \includegraphics[width=80mm]{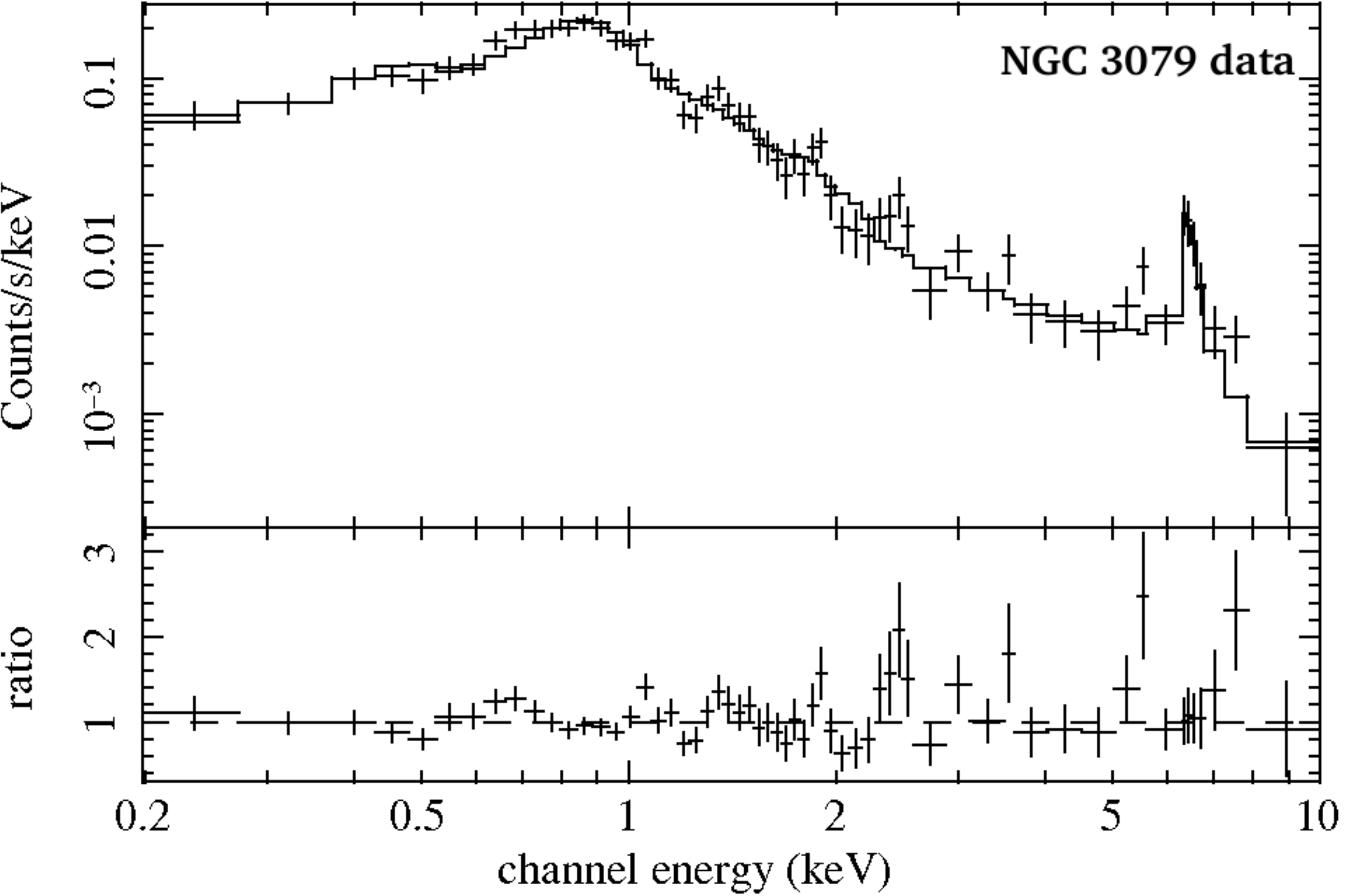}
 \includegraphics[width=80mm]{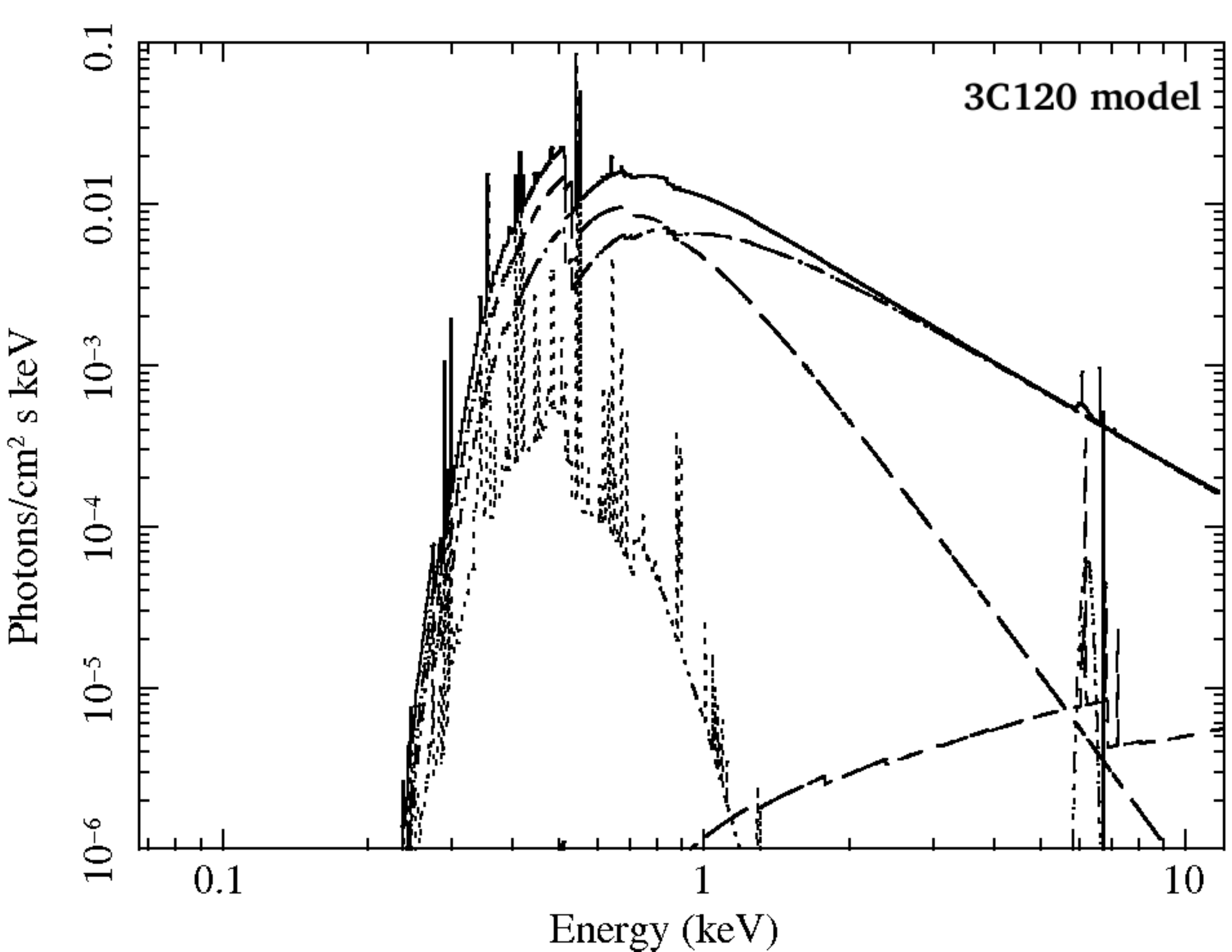}  \includegraphics[width=80mm]{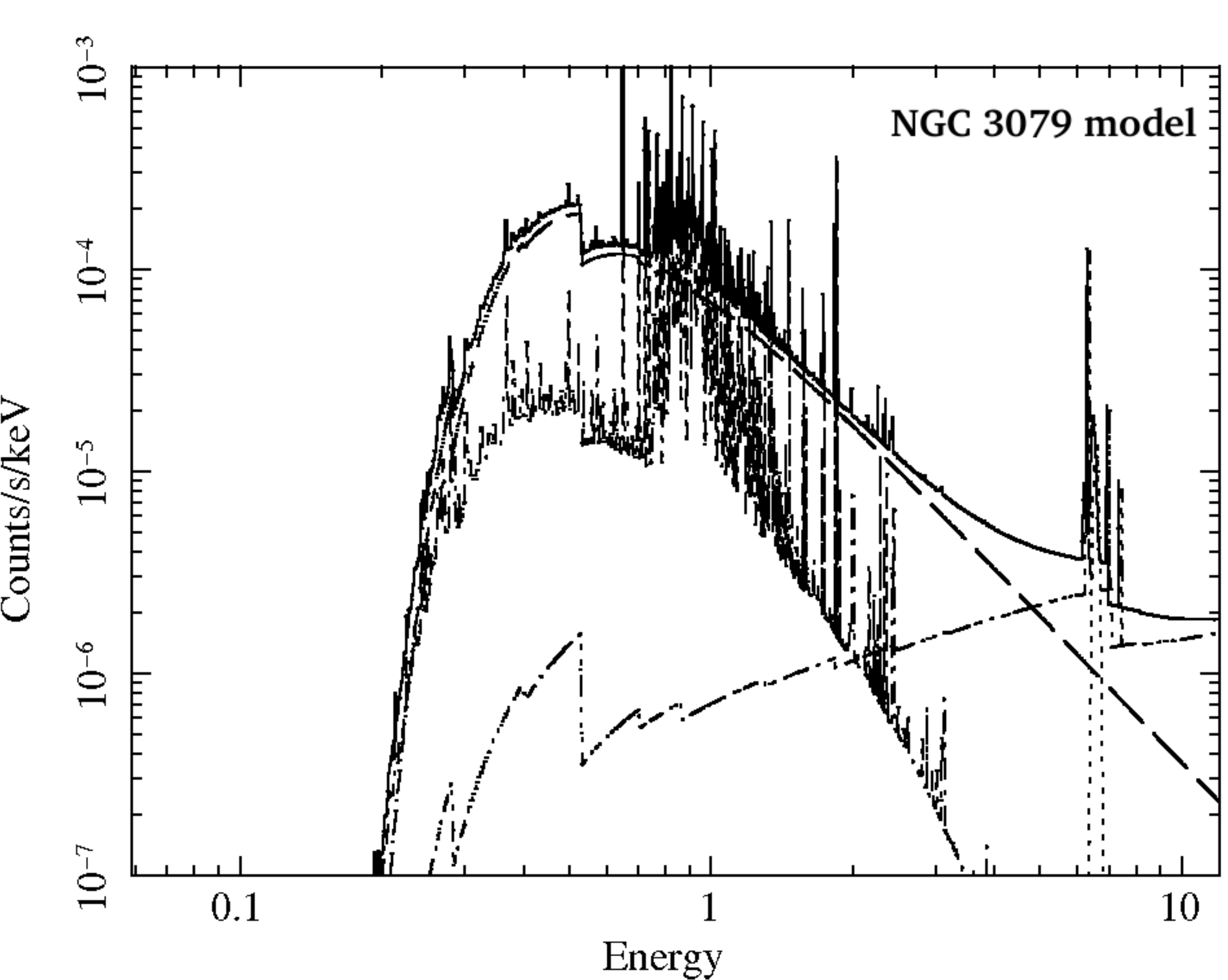}

\includegraphics[width=80mm]{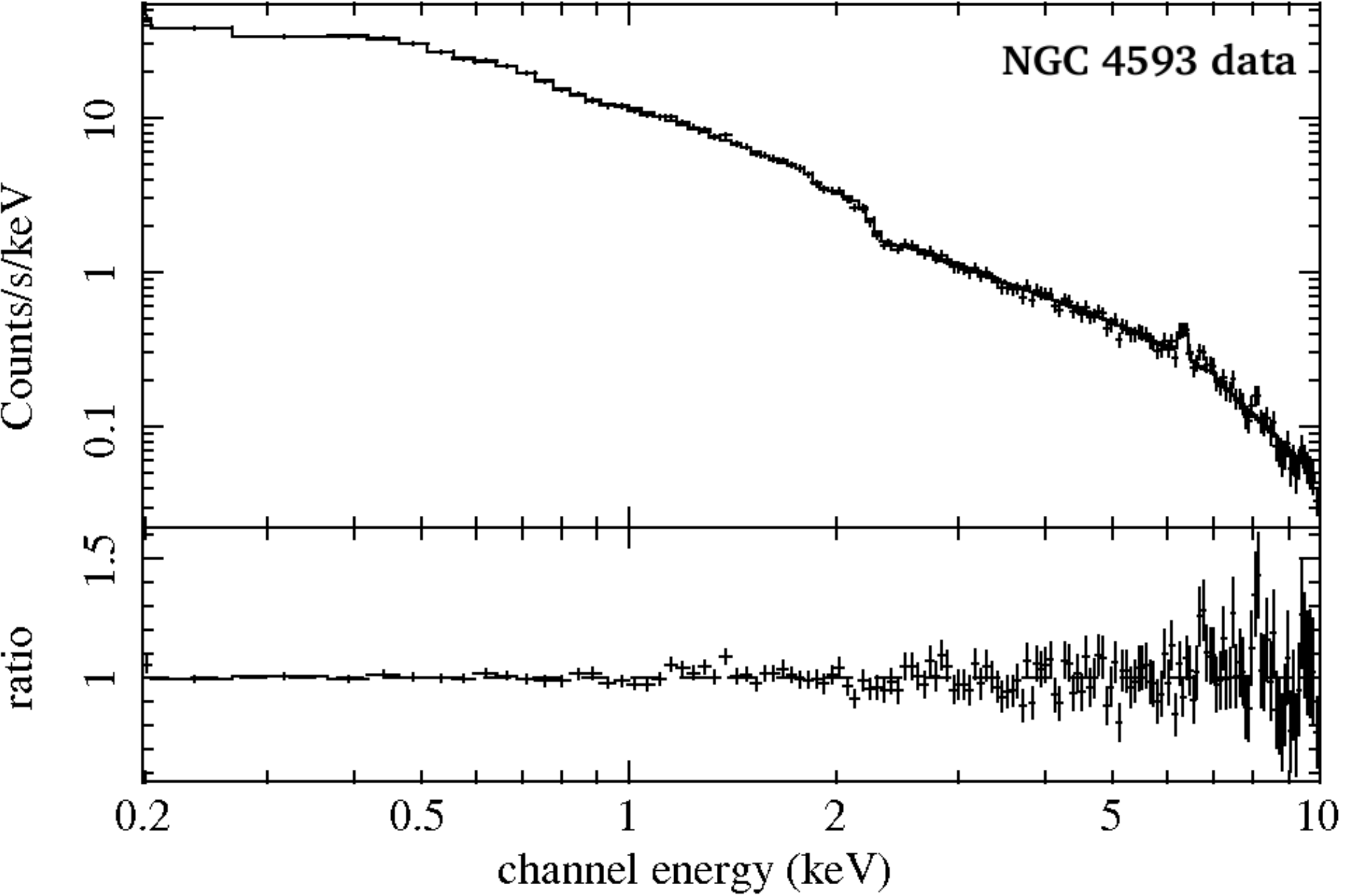}  \includegraphics[width=80mm]{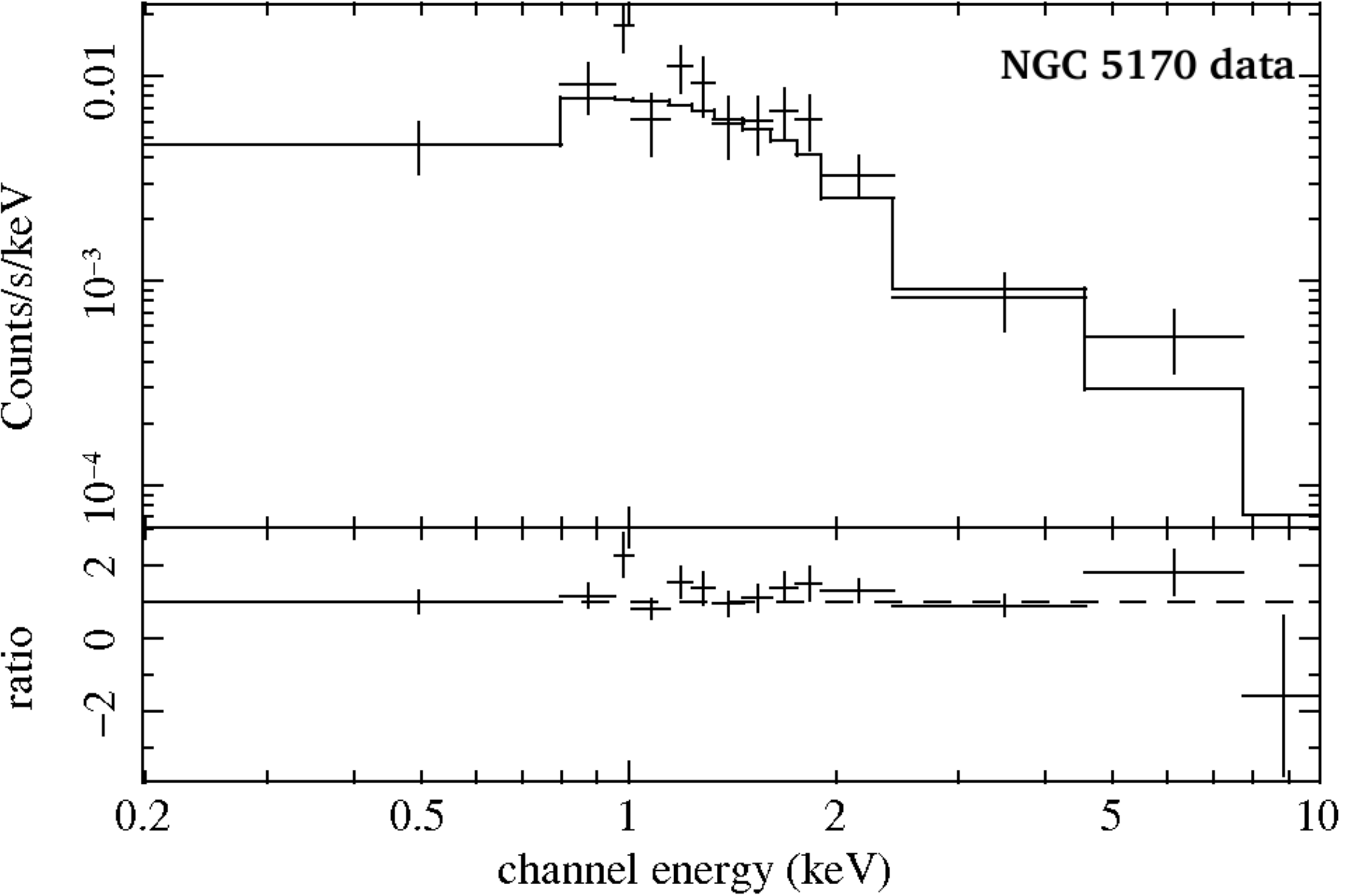}
\includegraphics[width=80mm]{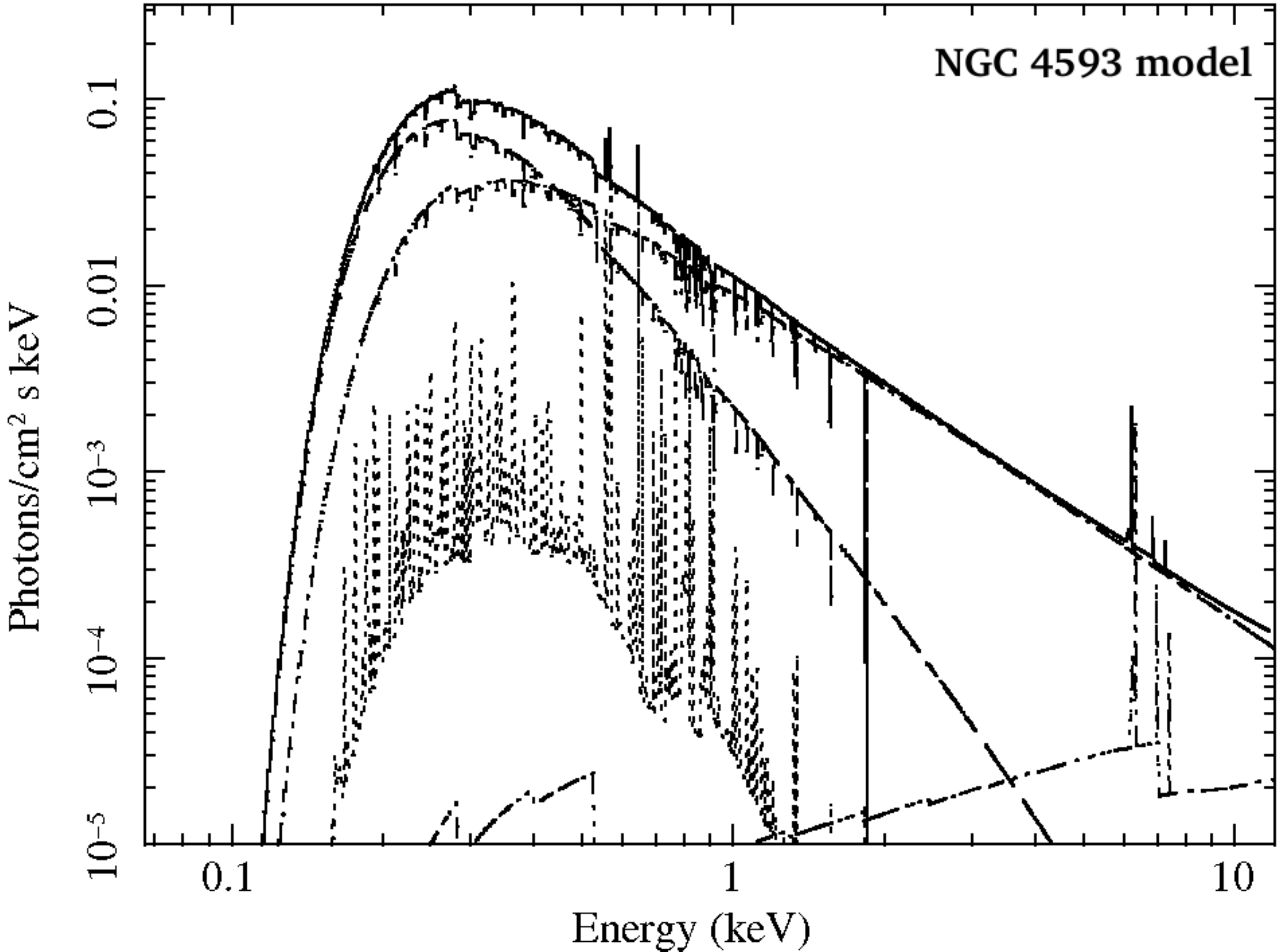}  \includegraphics[width=80mm]{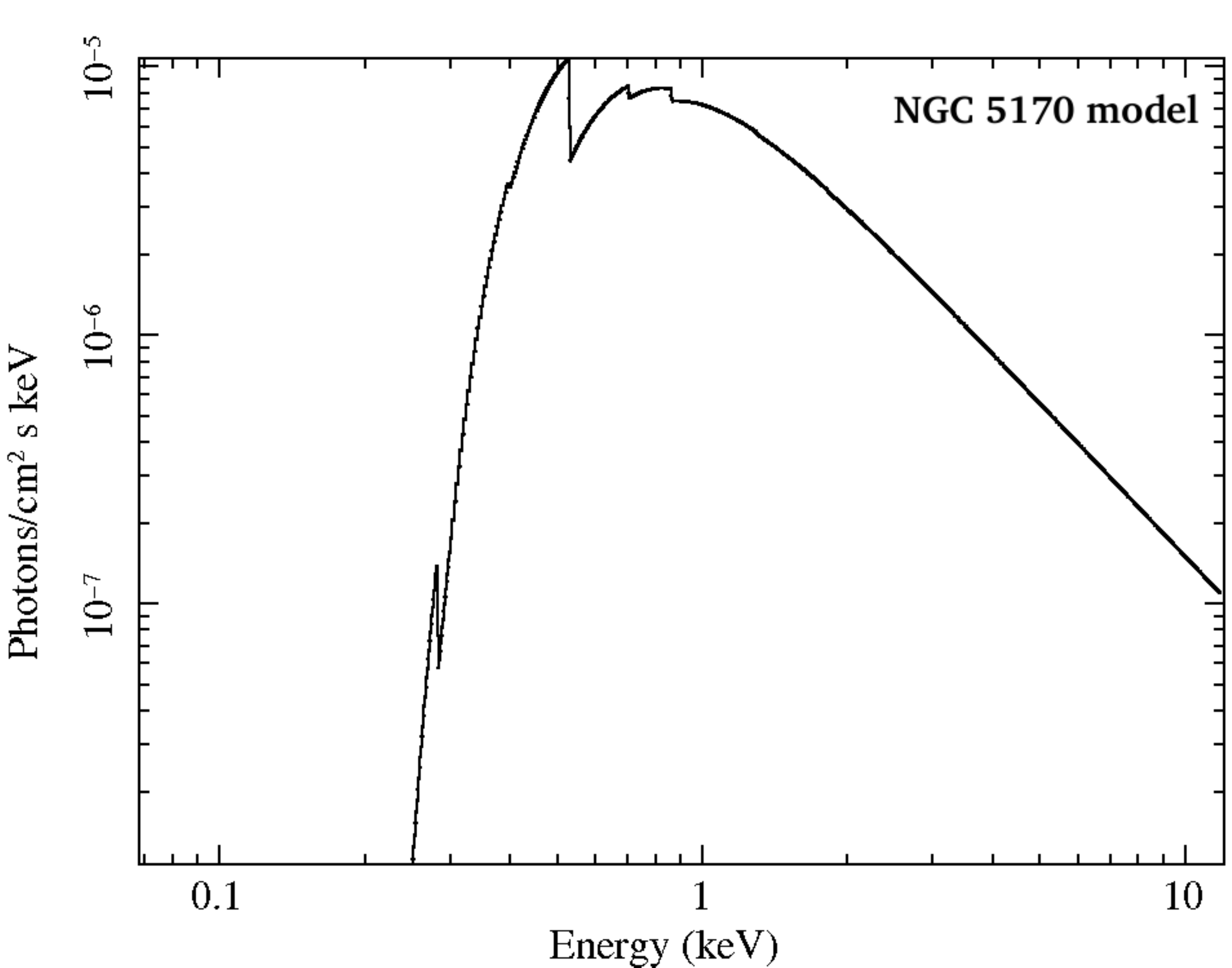}

\end{minipage}
\end{center}
\end{figure*}

\subsection{Properties of the primary emission}
\label{subsec:primaryx}

The primary power-law is found to have a mean power-law index, $<\Gamma>=1.90_{-0.07}^{+0.05}$ for the whole sample, and an intrinsic spread of $\sigma = 0.31_{-0.05}^{+0.05}$. We use the maximum likelihood method of \citet{maccacaro88}, which takes into account the measurement errors on $\Gamma$, to determine both the mean and the intrinsic standard deviation of this distribution. 

\begin{figure}
\begin{center}
\includegraphics[width=90mm]{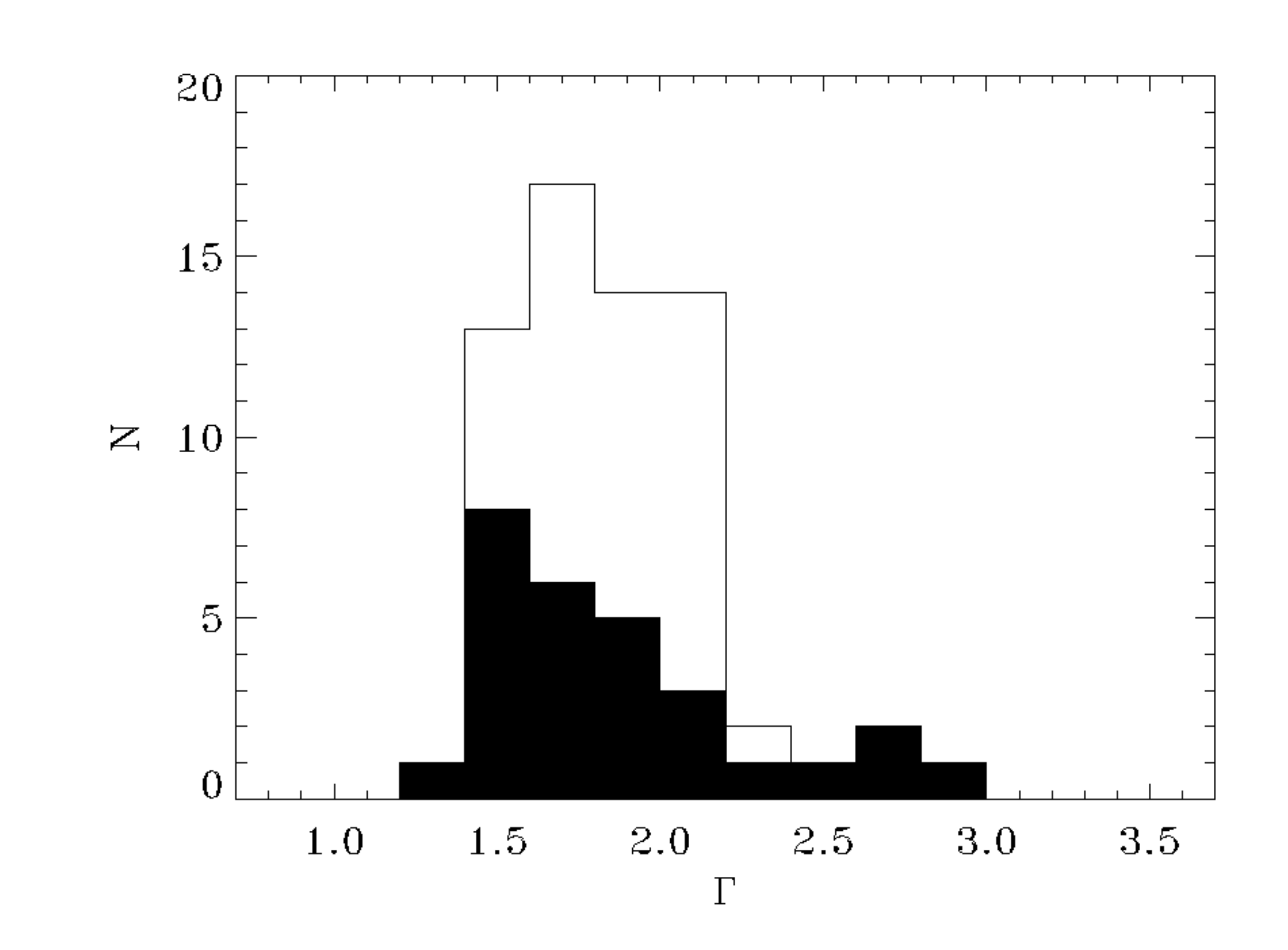}
\includegraphics[width=90mm]{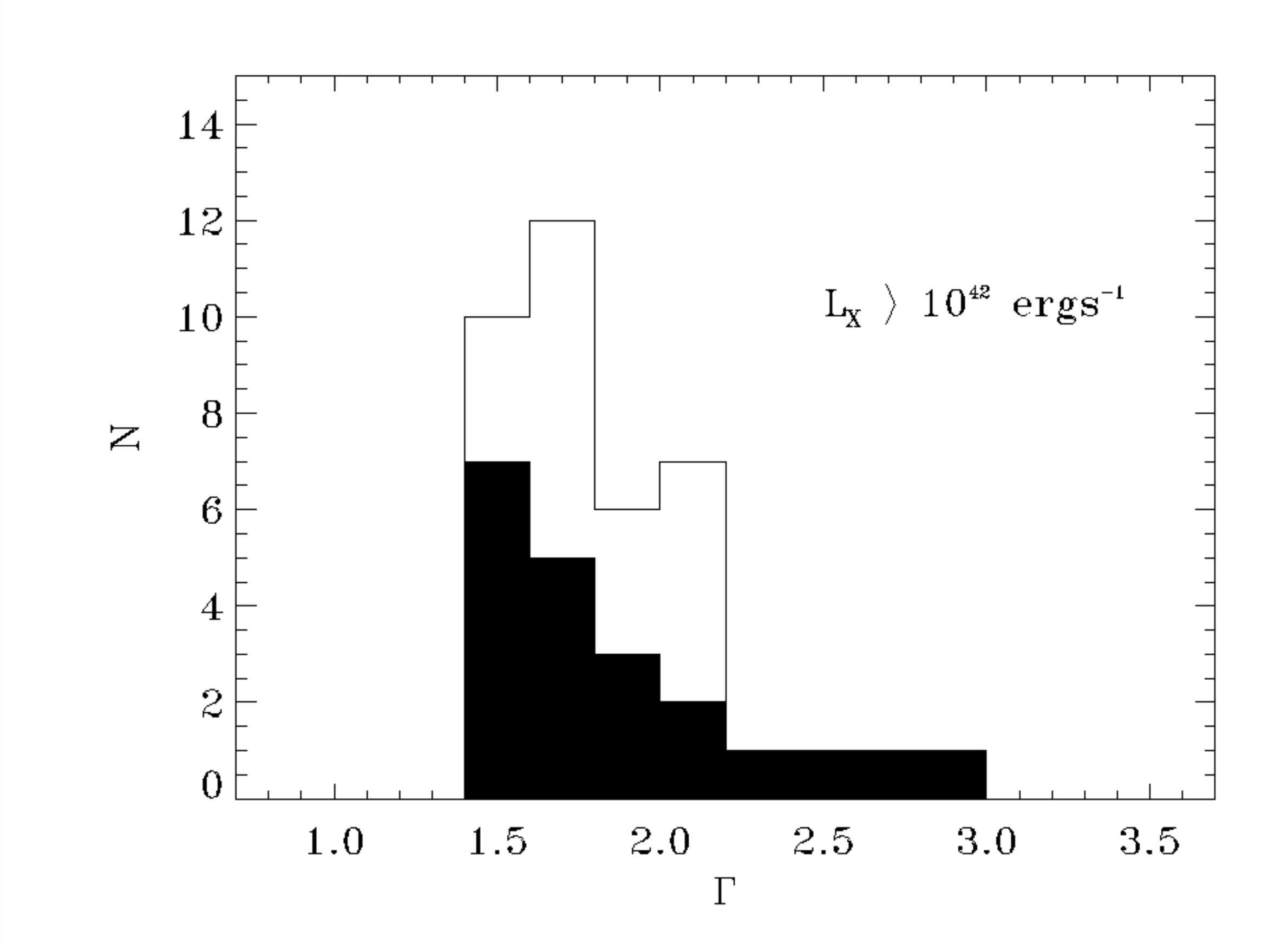}

\caption{Top: $\Gamma$ distribution of all sources (empty histogram), and obscured sources (\nh\ $>10^{22}$ \cmsq, black histogram). Bottom: $\Gamma$ distribution of all unambiguous AGN, defined by $L_{\rm X}>10^{42}$ ergs s$^{-1}$ (empty histogram), and obscured AGN (\nh\ $>10^{22}$ \cmsq, black histogram). The $\Gamma$ presented is that from the primary power-law only.}
\label{xrayanalysis_gammadist}
\end{center}
\end{figure}

We also investigate the $\Gamma$ distributions for obscured and unobscured sources, which is an important test for AGN unification. We do this for the entire sample, and also for unambiguous AGN with $L_{\rm X}>10^{42}$ \ergs. In Fig. \ref{xrayanalysis_gammadist} we show how the distribution of $\Gamma$ differs for obscured sources in these two cases. We again use the maximum likelihood method of \citet{maccacaro88} to determine both the mean and the intrinsic standard deviation for each of these distributions. We find that for all unobscured sources, $<\Gamma>=1.90_{-0.09}^{+0.09}$ and $\sigma = 0.33_{-0.06}^{+0.07}$ and for all obscured sources, $<\Gamma>=1.97_{-0.12}^{+0.14}$ and $\sigma = 0.44_{-0.09}^{+0.12}$. This is statistically consistent with the intrinsic power-law indices of these two populations being the same. We also find that for unobscured AGN, $<\Gamma>=1.84_{-0.14}^{+0.12}$ and $\sigma = 0.32_{-0.08}^{+0.11}$ and for obscured AGN, $<\Gamma>=1.90_{-0.14}^{+0.16}$ and $\sigma = 0.38_{-0.07}^{+0.13}$. This is again statistically consistent with the intrinsic power-law indices and their intrinsic spreads being the same for both obscured and unobscured AGN. A Kolmogorov-Smirnov (K-S) test confirms that the probability that these distributions are drawn from different parent samples is low (42\%) for AGN. 

\begin{figure}
\begin{center}
\includegraphics[width=90mm]{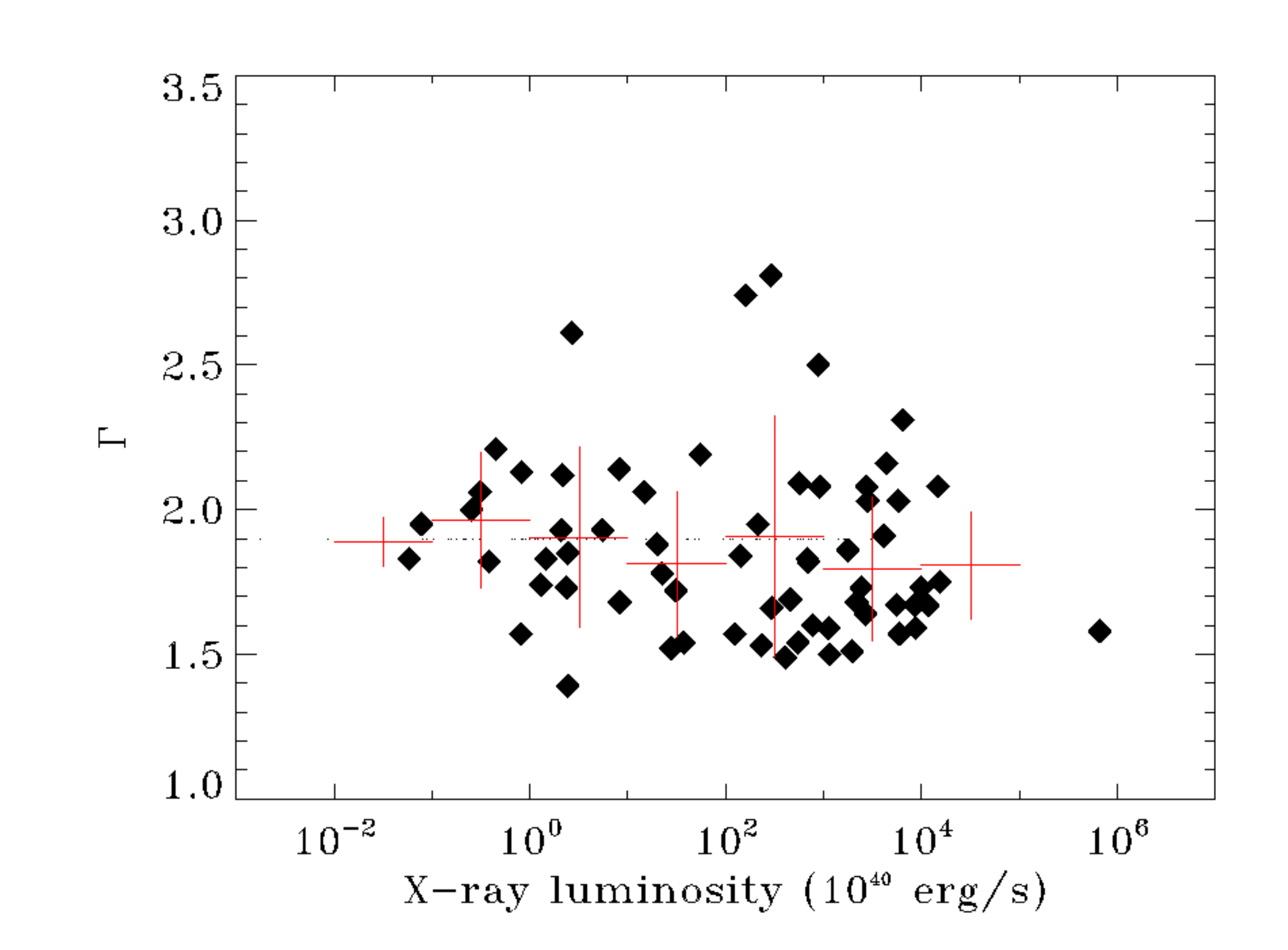}
\caption{Plot of $\Gamma$ vs. 2-10 keV intrinsic luminosity. Red crosses give the mean $\Gamma$ for each luminosity bin. The vertical error bars show one standard deviation spreads in the values.}
\label{xrayanalysis_lxgamm}
\end{center}
\end{figure}

Furthermore, we find no dependence of $\Gamma$ on X-ray luminosity. We show this in Fig. \ref{xrayanalysis_lxgamm} in the range of 10$^{38}$ to 10$^{45}$ \ergs, where the scatter of $\Gamma$ versus X-ray luminosity is plotted, as well as the mean values for each logarithmically spaced luminosity bins. The vertical error bars show one standard deviation spreads in the values.

\subsubsection{Soft-excess modelling}

\begin{figure}
\begin{center}
\includegraphics[width=90mm]{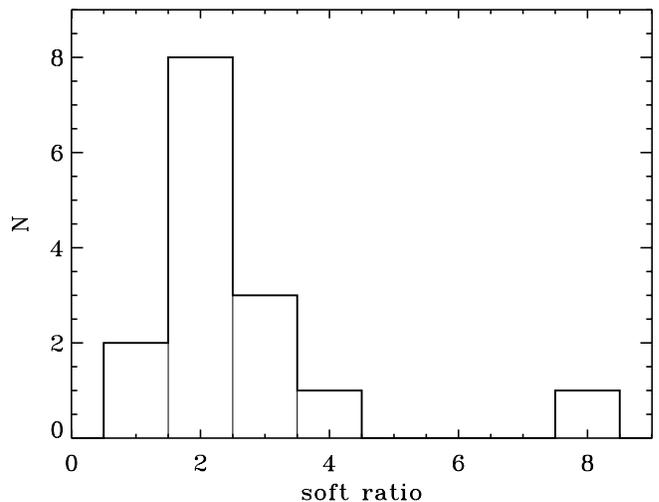}
\caption{Distribution of the soft ratio, the data-to-model ratio at 0.4 keV to the extrapolated 2.5-10 keV power-law, for 15 sources found to have a soft excess.}
\label{xrayanalysis_sx}
\end{center}
\end{figure}

We find that a total of 15 spectra in our sample require a soft-excess model component in their spectral fit. We have determined this from two soft ratio criteria, that the soft ratio must be $>1.2$ and that is must be rising from 1 keV to 0.5 keV.  3C120 has a soft excess which does not fit our criteria, as it does not have a data/model ratio greater than 20\% at 0.4 keV. However, this source has a clear soft excess at 0.7 keV not well modelled by thermal plasma emission. We find that a ratio of 3.2 at 0.4 keV is recovered when an additional intrinsic absorption component is used. In Fig. \ref{xrayanalysis_sx}, we plot the distribution of the soft ratio, which ranges from $\sim$1.3-7.8 for our sample, though with only one spectrum with a ratio greater than 4.  \citet{sobolewska07} note that this soft ratio ranges from $\sim$2-3 for their sample of quasars. The details of these fits are presented in Table \ref{table:xrayanalysis_specfit_sx}. All but three of the fits prefer a Comptonised component over a black-body component.

\subsection{Obscuration and reflection}

\subsubsection{Compton thin sources}

We find that 28/126 (22\%) of our spectra show no evidence for any absorption at all, whereas 77 (61\%) spectra present Compton thin (neutral) absorption, showing that absorption is commonplace in the X-ray spectra of all galaxies.

\subsubsection{Compton thick sources}

As mentioned previously, for 0.2-10 keV X-ray spectra, a direct measurement of a column density of Compton thick source is difficult due to the extreme suppression of the X-ray flux in this band. We have, however, successfully used our new model, {\tt trans}, to directly measure the \nh\ to be greater than $1.5 \times 10^{24}$ \cmsq\ in four sources (NGC 1320, NGC 1667, NGC 4968 and NGC 7479). The inclusion of a self-consistently modelled iron K$\alpha$ line is likely to have provided greater constraint on the \nh\ than models which include this component separately. 

\begin{figure}
\begin{center}
\includegraphics[width=90mm]{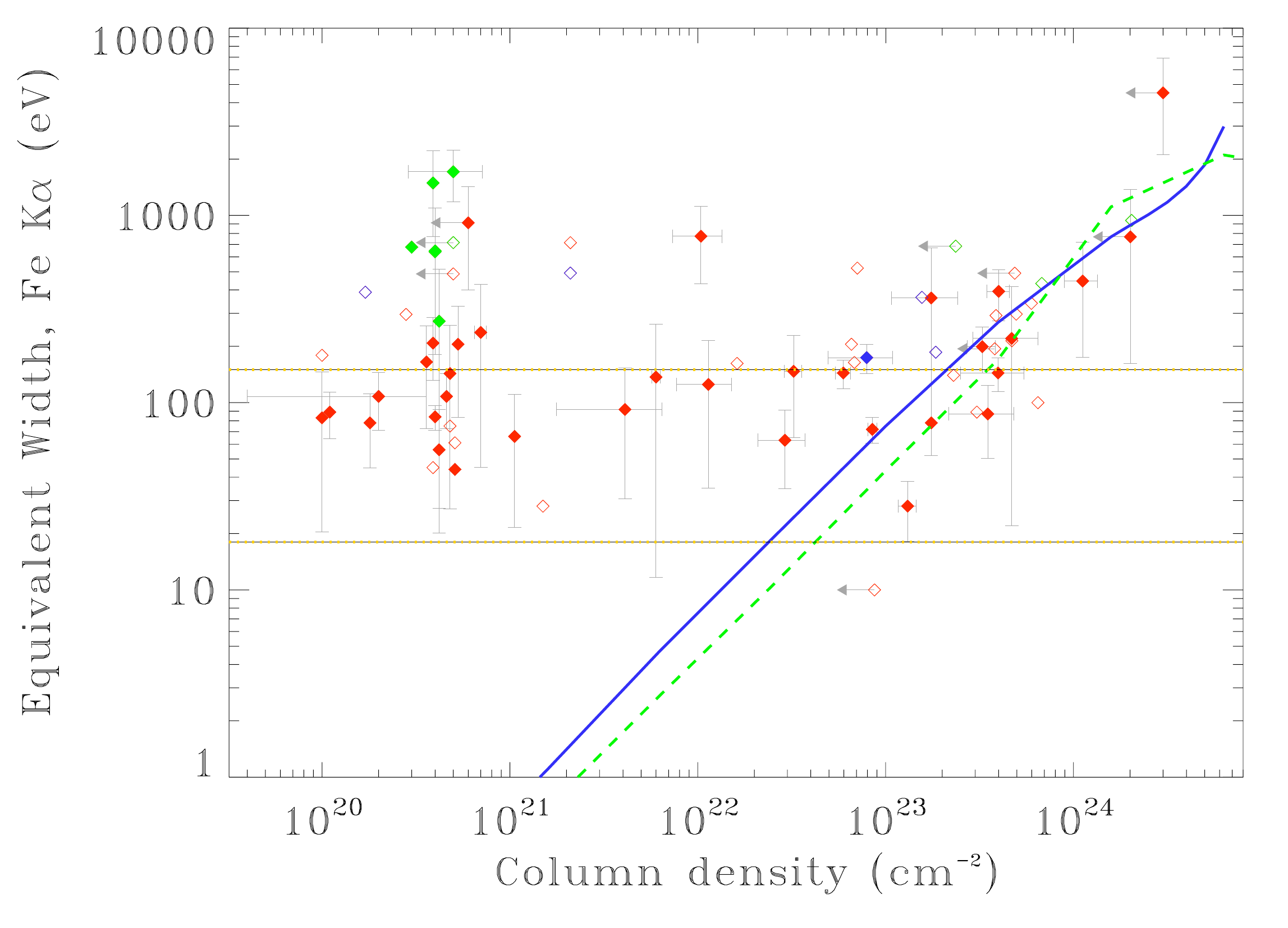}
\caption{The equivalent width of iron K$\alpha$ line versus directly measured \nh\ for our sample.  The lines drawn show the theoretical relationship between the EW of the Fe K$\alpha$ line and the \nh, calculated from the Monte-Carlo simulations, described in section \ref{transsec} for a power-law with $\Gamma$=2 and solar abundances. The solid blue line shows the calc8lations for the spherical distribution, the dashed green line shows the calculations from the torus distribution for the equatorial line of sight, and a torus opening angle of $\theta_{tor}$=60\degree. The dotted yellow lines show maximum and minimum EWs observed from unobscured sight lines in the torus distribution, taken from Fig \ref{torusews}, where the line of sight \nh\ through the torus is greater than 10$^{23}$ \cmsq. Thus, all truly unobscured sources must have an EW less than the maximum shown. Sources which are reflection dominated (R$>$8) are shown in green. Sources with $1<$R$<8$ are shown in blue for comparison. All grey lines are error bars. These are only shown where the EW is consistent with being non-zero (plotted as filled diamonds, otherwise plotted as open diamonds). Arrows indicate upper limits on the \nh.}
\label{xrayanalysis_nhew}
\end{center}
\end{figure}

In our analysis, we call sources with a reflection dominated spectrum Compton thick, of which there are 12. We define a reflection dominated spectrum to be where R$_{pexmon}>8$. The R$_{pexmon}>8$ criterion is arbitrary and chosen to be low enough to include known Compton thick AGN such as NGC 1068. Figure \ref{xrayanalysis_nhew} shows the EW of the Fe K$\alpha$ line against \nh\ for our sample for reflection dominated sources (green), mildly reflection dominated sources (blue, $1<$R$<8$) and sources with weak or no reflection (red, R$<1$). It can be seen from this that reflection dominated sources also have high iron line EWs. We plot only narrow line EWs here, which are most likely to originate in cold distant obscuring matter, unless no narrow line has been measured, in which case we plot the broad line EW, if one has been measured

There are two sources, however, that have high iron line EWs which are constrained to be $>300$ keV, but have a measured \nh$<10^{22}$ \cmsq. Our Monte-Carlo simulations have shown that this is not possible for either a spherical or toroidal matter distribution, and so we investigate them further. These sources are NGC 3690 and IRASF13349+2438. NGC 3690 is the merging system ARP 299, which was shown by {\it Beppo-SAX} to host a Compton thick AGN \citep{dellaceca02}, confirming our assertion. We therefore add it to our sample of Compton thick AGN. \citet{longinotti03} examine the Fe K$\alpha$ features in the X-ray spectrum of IRASF13349+2438, a narrow-line Seyfert 1, finding them to be broad and complex. This line is therefore unlikely to originate from distant Compton thick obscuring matter. This shows that Fe K$\alpha$ EW diagnostics for obscuration are only suitable when we can be sure that the line originates in cold, distant matter such as the torus.

For sources where their Compton thick nature has been inferred by the EW and the reflection fraction, we can only put a lower limit on the \nh\ of $1.5 \times 10^{24}$ \cmsq, but in reality, this could be higher and requires data above 10 keV to constrain. We find a total of 16 Compton thick sources in our sample, the details of which are given in Table \ref{table:xrayanalysis_cthick}. The Compton thick fraction for all 126 galaxies is 16/126 (=13 $\pm$ 3\%), where the errors have been calculated for a binomial distribution ($\sigma_p = \sqrt{p(1-p)/N}$, where p is the proportion and N the total number of sources). The Compton thick fraction for unambiguous AGN ($L_{\rm X}>10^{42}$ \ergs) is 11/60 (=18 $\pm$ 5\%).

\subsubsection{\nh\ distribution and obscured fraction}

\begin{figure}
\begin{center}
\includegraphics[width=90mm]{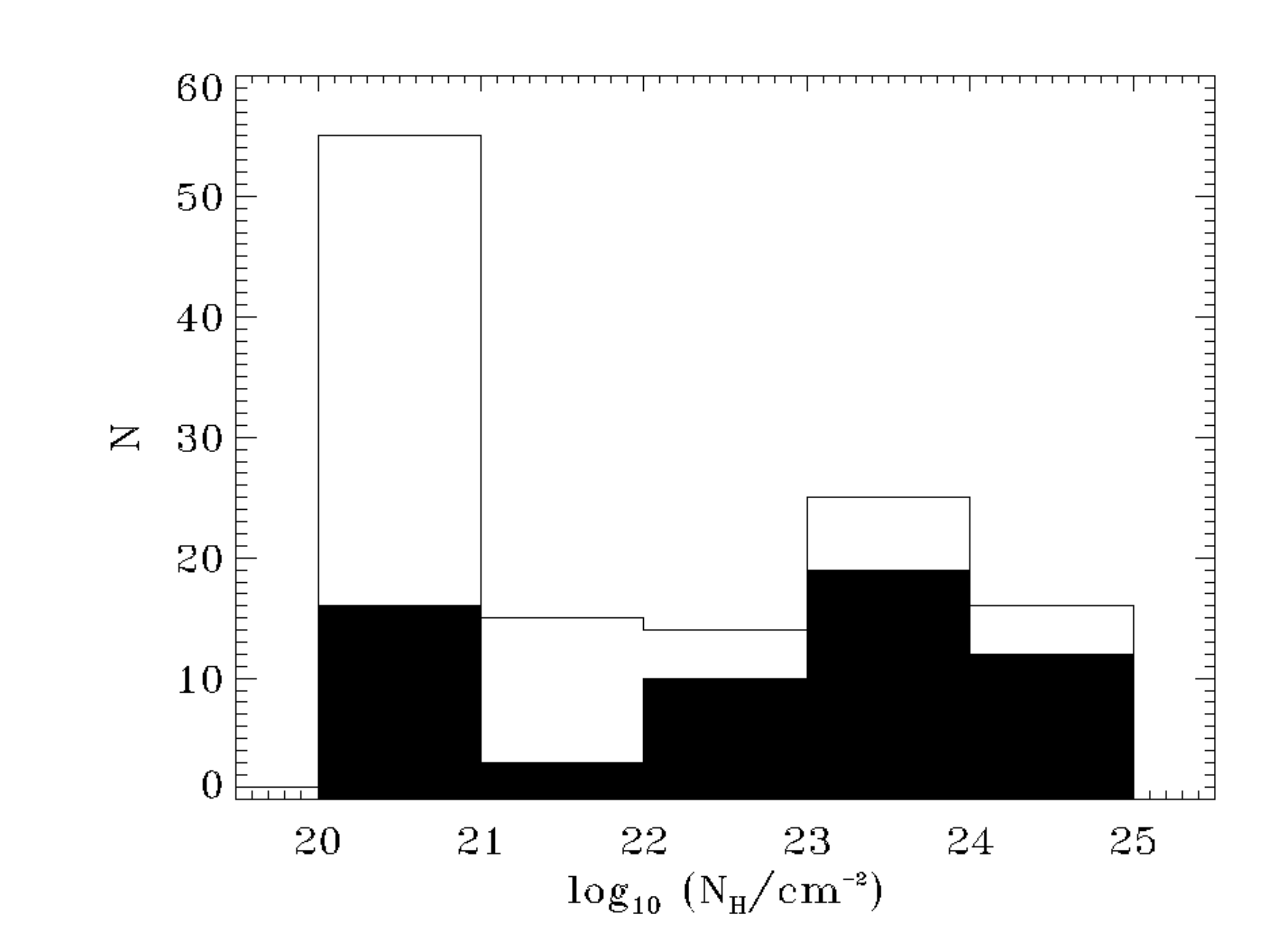}
\caption{\nh\ distribution of all sources (empty histogram), and unambiguous AGN, defined by $L_{\rm X}>10^{42}$ ergs s$^{-1}$ (black histogram). The \nh\ presented is that measured from the primary power-law only.}
\label{xrayanalysis_nhdist}
\end{center}
\end{figure}

In Fig. \ref{xrayanalysis_nhdist}, we present the \nh\ distribution for the whole sample, and over-plotted is the \nh\ distribution of those with $L_{\rm X}>10^{42}$ \ergs. The main difference in the two distributions is the strong peak of unobscured sources at low X-ray luminosities and the greater proportion of heavily absorbed X-ray luminous AGN. The low luminosity, unabsorbed sources may consist of unobscured star forming galaxies, LINERs or lower luminosity AGN, but we will investigate this further in paper II. The \nh\ distribution of unambiguous AGN shows a smaller unobscured peak, but with a second obscured peak at \nh=10$^{23-24}$. 


\begin{figure}
\begin{center}
\includegraphics[width=90mm]{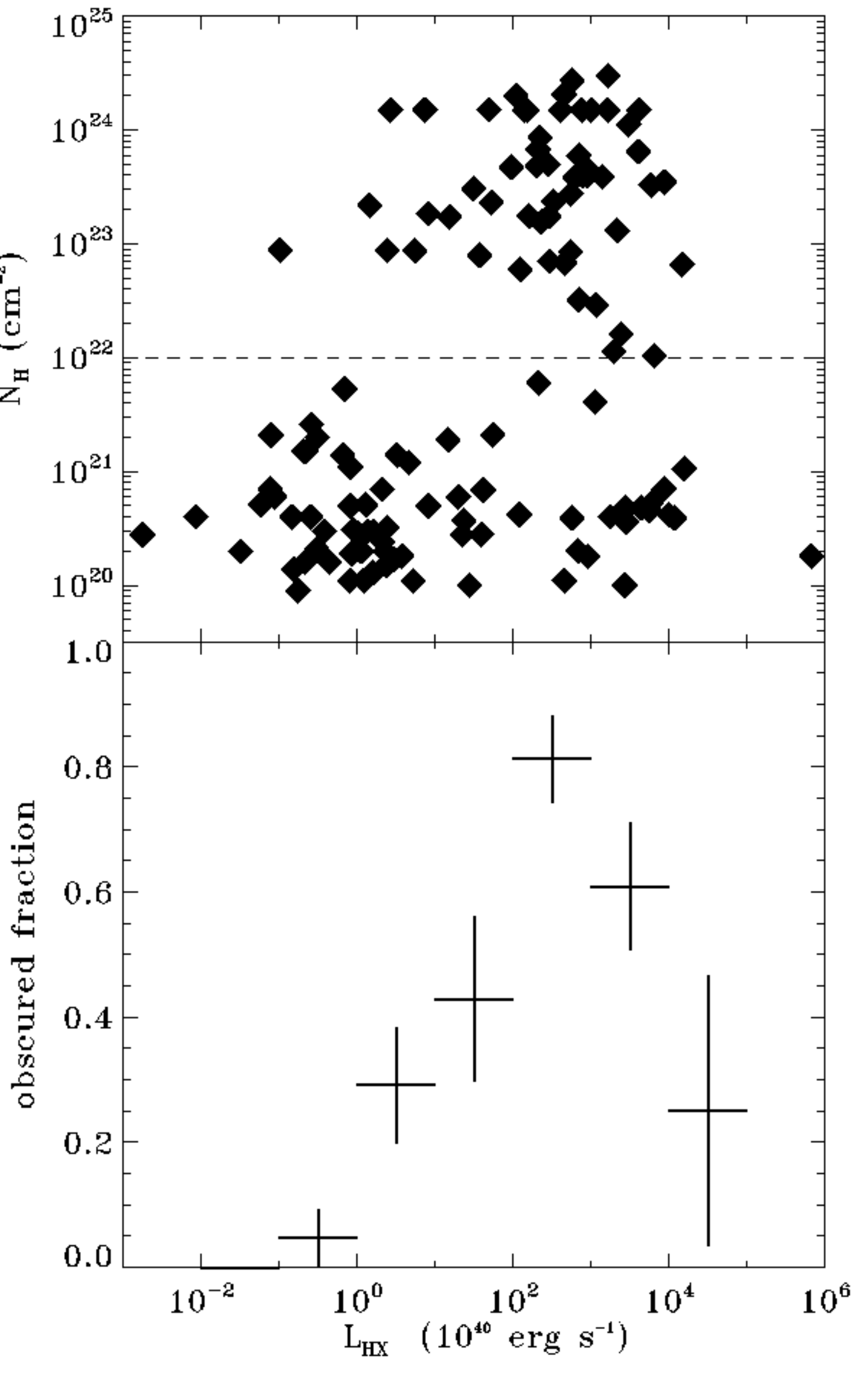}
\caption{\nh\ and obscured fraction versus 2-10 keV luminosity for all sources. Obscured fraction is calculated in equal logarithmic luminosity bins, as the number of sources with \nh\ $> 10^{22}$ \cmsq\ divided by the total number of sources in that bin, where more than one source exists per bin.}
\label{xrayanalysis_lxnh}
\end{center}
\end{figure}

Fig. \ref{xrayanalysis_lxnh} plots  \nh\ vs. $L_{\rm X}$ and also the obscured fraction, defined as the ratio of the number of sources with \nh\ $>10^{22}$ \cmsq\ to the total number of sources in each luminosity bin. There is significant variation in the obscured fraction with X-ray luminosity, particularly above  and below 10$^{42}$ \ergs\ where the ratio declines steeply from a peak.

\subsubsection{Reflection dominated sources}

\begin{figure}
\begin{center}
\includegraphics[width=90mm]{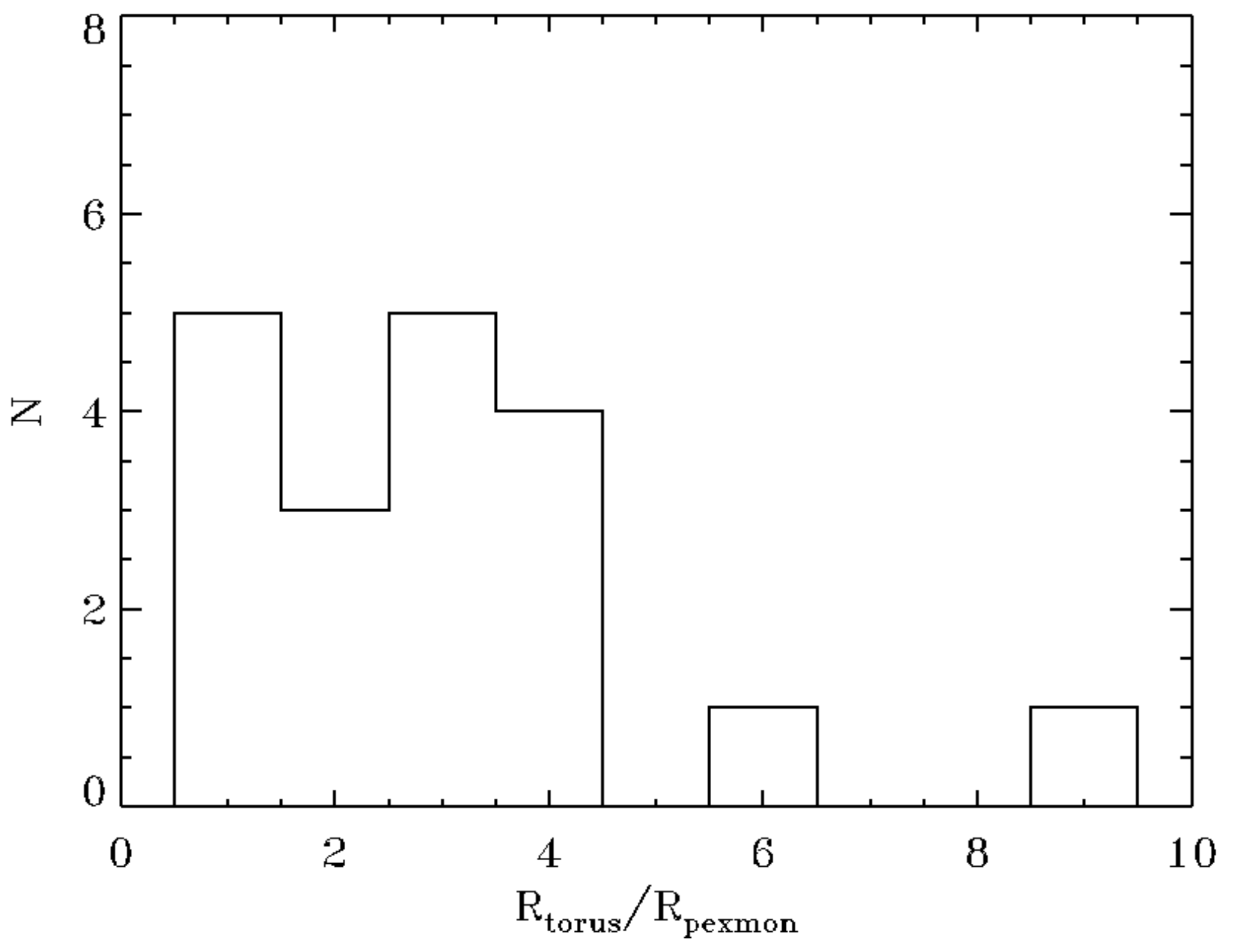}
\includegraphics[width=90mm]{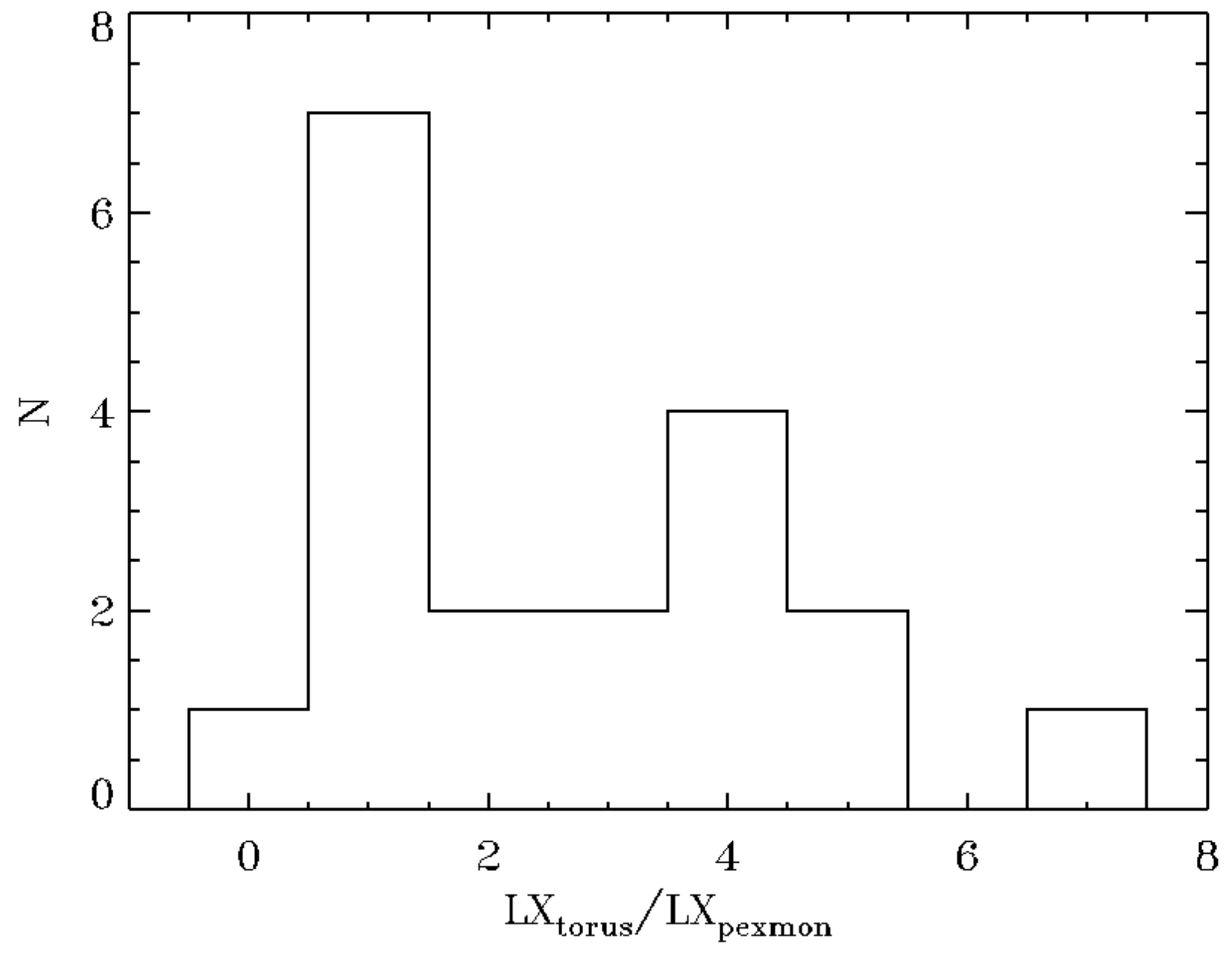}

\caption{Histograms showing the distribution of (top) the reflection fraction and (bottom) intrinsic 2-10 keV luminosities as determined from the {\tt torus} model with respect to the {\tt pexmon} model.}
\label{xrayanalysis_torcomp}
\end{center}
\end{figure}

For sources where the X-ray spectrum includes a strong reflection component, we have also fitted our new torus model, firstly to investigate the reflection fraction and intrinsic source luminosities as determined using this model with respect to the {\tt pexmon} slab model. Secondly, we investigate if it is possible to determine the torus geometric parameters, $\theta_{tor}$ and $\theta_i$ from the 2-10 keV spectra. In Fig. \ref{xrayanalysis_torcomp} we plot the distributions of the ratio of the reflection fraction as determined by the {\tt torus} model to the reflection fraction determined by the {\tt pexmon} model and the ratio of the intrinsic luminosities as determined by each model. We find that for most sources, both the reflection fraction and intrinsic source luminosity as determined using the torus model is greater than that determined by the {\tt pexmon} model. The means and medians of these ratios are 3.08 and 2.70 for $R_{\rm torus}/R_{\rm pexmon}$ and 2.62 and 1.99 for $LX_{\rm torus}/LX_{\rm pexmon}$. We also find that it is not possible to constrain the inclination angle or opening angle of the torus with these data below 10 keV, as neither parameter can be constrained, even with the other parameter fixed.

\section{discussion}\label{discussion}


\subsection{X-ray transmission in heavily obscured sources}


We have made predictions of the equivalent width (EW) of the iron line for different model parameters and for two matter geometries, a sphere and a torus. The predictions we make for the torus are useful for comparing observed EWs and making inferences from these about the matter geometry and parameters in the observed system. 
For example, we find that a maximum EW of 150 eV is achievable when considering a torus geometry for unobscured sight-lines with a typical $\Gamma$ (2.0) and solar abundances. To achieve greater EWs than this, the line of sight must be obscured with \nh$>10^{23}$ \cmsq. Therefore, if an EW higher than this is measured for an unabsorbed spectrum, it may cast doubts on the reliability of the measured line of sight \nh. We showed this for NGC 3690, which shows an Fe K$\alpha$ EW of 913$^{+380}_{-346}$ eV, but a measured \nh\ of only 6.14$^{+5.57}_{-6.14}\times10^{20}$ \cmsq\ which is clearly at odds with our model predictions. Data above 10 keV from {\it Beppo-SAX} shows, however, that this source is indeed heavily obscured with \nh$>1.5\times10^{24}$\cmsq\ \citep{dellaceca02}, in full support of our model predictions. Our Fe K$\alpha$ EW predictions also agree well with predictions of the same nature as made by \citet{ghisellini94}, \citet{ikeda09} and \citet{murphy09}. 

We also present flux suppression factors for both spherical and toroidal geometries as a function of \nh\ for different bands and redshifts. These are particularly relevant for AGN surveys that use the X-ray band as they indicate the effectiveness of each band at detecting heavily obscured systems. It is known that even the hardest X-ray energies are affected by extreme obscuration. Here we quantify the flux suppression in these bands, finding that locally, flux suppression reaches a factor of 10 at \nh=$10^{25}$ \cmsq\ for the $>10$ keV bands. This may explain the fewer than expected local Compton thick AGN detected in hard X-ray surveys conducted by {\it Swift/BAT} and {\it INTEGRAL}. For the $>10$ keV X-ray bands, the flux suppression factor is 10 at $\sim10^{25}$ \cmsq\ for all redshifts. This is because at these energies, Compton scattering is the dominant interaction and source of attenuation in the spectrum. The cross section of this interaction is roughly constant as a function of energy and therefore the effect of redshift is negligible. Flux suppression in the 2-10 keV band reaches 10 at $6\times10^{23}$ \cmsq\, but this improves as redshift increases, becoming $\sim10^{25}$ \cmsq\ at z=3. For the torus geometry, the flux suppression factors behave similarly to the spherical geometry up to $\sim4\times10^{24}$ \cmsq. Here, the flux suppression factors level off due to reflected flux from the inner part of the torus, which remains constant regardless of the line of sight \nh.  In all cases though, an absorbing column of \nh=$10^{25}$ leaves all X-ray bands severely insensitive to detection of X-ray sources.

Using model spectral data from the torus simulations, we also investigated the difference in the reflection fractions and intrinsic luminosities as determined by using these new torus data, to that derived from fits with the slab model, {\tt pexmon}. \citet{murphy09} find that the reflection fraction is underestimated by a factor of $\sim6$ when they compare their torus data to slab data. We also find that the reflection fraction is systematically higher when using the torus model data, by a factor of 3 on average. Though this is not as high as the underestimation presented by \citet{murphy09}, the discrepancy is probably due to differing torus geometries, as they use a cylindrical torus geometry which gives varying \nh\ as a function of inclination angle, whereas we use a spherical torus with constant \nh. We also find though, that the intrinsic luminosity derived from the slab models are underestimated with respect to the torus model by a factor of 2-3. Additionally, using the torus model, we find that we are unable to constrain the torus geometrical parameters, the opening angle or the inclination angle using the \xmm\ spectral data. \citet{ikeda09} also find that it wasn't possible to constrain these parameters with their model, even with spectral data above 10 keV with {\it Suzaku}

\subsection{Properties of the primary X-ray emission from 12 micron selected galaxies}

We have presented the properties of the primary X-ray emission of our sample of 12 micron selected galaxies. As our sample is a heterogeneous mix of galaxies, these results reflect the properties of all X-ray sources. We will present results on individual optical classes in paper II. We do however define a subset of unambiguous AGN using their observed 2-10 keV luminosity though this will of course not include AGN with low X-ray luminosity which may be due to significant absorption or being of intrinsic low luminosity.

We find that the mean power-law index is $<\Gamma>=1.90_{-0.07}^{+0.05}$ for the whole sample, with an intrinsic spread of $\sigma = 0.31_{-0.05}^{+0.05}$. The mean is consistent with previous works on predominantly X-ray selected samples. For example \citet{nandra94} who find $<\Gamma>=1.95 \pm 0.05$ and $\sigma_p = 0.15 \pm 0.04$ for their sample of Seyfert galaxies observed with {\it Ginga}. We do find a larger intrinsic spread in the indices though than \citet{nandra94}. Our result is also consistent with the results from harder X-ray selected samples \citep[eg. 1.93 for all Seyferts in the {\it INTEGRAL} catalogue of][]{beckmann09}. 

Comparing the distribution of power-law indices for obscured sources and unobscured sources is a key test of the unification scheme, which states that the central engine is the same for both, the difference being the orientation of the system to the observer \citep{antonucci93, urry95}. If a systematic difference is found in these distributions, it would suggest that  this is not simply the case. Comparing the average X-ray spectra of Seyfert galaxies with {\it Ginga} and {\it OSSE} data, \citet{zdziarski95} originally found a systematic difference between the $\Gamma$ distributions of Seyfert 1s and Seyfert 2s, however, more recent works have tended to find that the distributions are indeed consistent with each other, generally due to the inclusion of more complex spectral models, such as Compton reflection components \citep[e.g.][]{dadina08}. We investigated these distributions for obscured and unobscured sources, as well as obscured and unobscured unambiguous AGN. Our analysis showed that the means of these distributions in both cases are not significantly different (the means lie within one sigma of each other) and a K-S test supports the same conclusion, implying that the X-ray generation mechanism in both obscured and unobscured AGN is the same, lending support to AGN unification. The fact that the intrinsic dispersions are similar also supports this conclusion. \citet{beckmann09} carried out the same analysis for obscured and unobscured sources on {\it INTEGRAL} data, which should be less affected by absorption effects, and also find no significant differences between their obscured and unobscured sources. Despite all of this, however, we do notice that the $\Gamma$ distribution of obscured AGN is skewed towards low $\Gamma$ values. We discount the possibility of unidentified reflection in these sources, but results presented by \citet{malizia03} suggest an alternative cause. They also find a harder distribution of $\Gamma$ values for Seyfert 2s with respect to Seyfert 1s from their analysis of {\it Beppo-SAX} data, but reconcile this difference by including more complex absorption models, such as partial covering scenarios, which also improve their fits. There have been several instances where absorption in AGN has been shown to be more complex than a single smooth obscurer, for example NGC 1365, which shows rapid variability in the line of sight obscuration \citep{risaliti09} and further examples by \citet{ricci10}. As we do not use partial covering models, this may be the reason for the skewed distribution towards low $\Gamma$ values in our obscured AGN. 

We have also investigated any correlation between the power-law index and the intrinsic X-ray luminosity (Fig. \ref{xrayanalysis_lxgamm}). Our results don't support a correlation between these two quantities however. This was also found by \citet{george00} in their sample of local AGN using {\it ASCA} data. At higher redshifts however, there have been reports of a correlation between $\Gamma$ and L$_X$. For example \citet{dai04} find an anti-correlation between the two for their mini-survey of $1.7 < z < 4$ sources observed with \xmm\ and {\it Chandra} in the luminosity range of 10$^{43-45}$ \ergs. This finding was later supported by \citet{saez08} for the {\it Chandra} deep fields who use a larger range in luminosity (10$^{42-45}$ \ergs) and redshift (0.1-4) with 173 AGN. Our luminosity range is 10$^{38-44}$ \ergs, which is somewhat lower than the higher redshift studies, though the local sample of \citet{george00} ranges from 10$^{42-46}$ \ergs, with no correlation. This may suggest an evolutionary effect, or a selection effect in the X-ray samples. Our sample does not suffer from such effects, as it is MIR selected.

\subsection{Properties of X-ray obscuration in 12 micron selected galaxies}

\subsubsection{Compton thick sources}

We find that 16/126 (=13 $\pm$ 3\%) of the whole sample of sources studied in this sample are Compton thick. For X-ray bright AGN ($L_{\rm X}>10^{42}$ \ergs), there are 11/60 (=18 $\pm$ 5\%) Compton thick sources. This fraction is important for synthesis models of the XRB and the accretion history of the universe. For X-ray samples selected above 10 keV, the Compton thick fraction has been constrained to be less than that predicted by the XRB. \citet{beckmann09} found that only 4 $\pm$ 2\% of the AGN in their sample are Compton thick, whereas the XRB predicts a fraction close to 15\% at the corresponding limiting flux. \citet{treister09} claim that the observed differences between observation and models is due to degeneracies in the parameters of the models, and that only direct observations of Compton thick AGN can determine their relative number.  Howeverm when accounting for the biases present in hard X-ray selection, \citet{burlon10} recover a 20\% Compton thick fraction in the {\it Swift/BAT} survey, which is consistent with the XRB and our results.

The Compton thick fraction that we measure in 12 micron selected sources is almost 3-$\sigma$ greater than that measured in hard X-ray selected sources, suggesting that mid-IR selection is more efficient than hard X-ray selection in finding Compton thick AGN. This is explained by the suppression of X-ray flux in the $>$10 keV bands at high column densities, which we have shown to be a factor $>$10 at \nh\ =$10^{25}$ \cmsq\ using our Monte-Carlo X-ray transmission results. 

The XRB synthesis models of \citet{gilli07} allow the prediction of the number counts expected from an X-ray survey given the set of parameters of that survey\footnote{An online calculator can be found here: http://www.bo.astro.it/~gilli/counts.html}. As this sample is not X-ray flux limited, the comparison is not straightforward. However, the Compton thick fraction expected for a survey with our redshift range ($0<z<0.1$) and 2-10 keV luminosity range ($10^{42}<L_{\rm X}<10^{44}$, a lower limit of $L_{\rm X}>10^{42}$ as we are only considering AGN) lies between $\sim$2\% and $\sim$20\% for X-ray limiting fluxes between $\sim10^{-13}$ and $\sim10^{-14}$ ergs cm$^{-2}$ s$^{-1}$, in between which our `effective' X-ray flux limit lies (see Fig. \ref{fig_obsstats}). Thus our results would seem roughly consistent with the XRB models.

From the four Compton thick AGN which we manage to measure the \nh\ directly with {\tt trans}, in three of these cases, a prominent iron line aids in the determination of the \nh. However in the fourth source, NGC 1667, the spectrum does not present an prominent iron line. Here the model is required to fit a hard excess seen in spectrum, but the \nh\ is not well constrained (\nh\ = 271$_{-251}^{+5730}$). This is a low signal to noise spectrum, and should be considered a marginal Compton thick source. This technique of fitting hard excesses in \citet{brightman08}, where we show that it can be an effective method of identifying heavily obscured nuclei (e.g. the {\it Chandra} observation of NGC 4501). In the absence of high signal to noise data around the Fe K$\alpha$ line, or data above 10 keV, this method is the next best thing for picking out Compton thick candidates from X-ray spectral data alone. This may then motivate longer exposures to gain higher signal to noise data, or observations of broader banded telescopes such as {\it Suzaku}.

We also infer the Compton thick nature of 12 sources from their reflection dominated spectra, which we define as being those with $R_{pexmon}>8$. Though this definition is arbitrary, it is well supported in most sources by high Fe K$\alpha$ EWs or a directly measured \nh$>10^{24}$ \cmsq. Lowering this criteron to, e.g. $R_{pexmon}>1$ would include two further sources, NGC 4631 and NGC 4666, without prior indications of being heavily absorbed. However, these sources don't have well constrained Fe K$\alpha$ EWs and hence don't support a lower $R_{pexmon}$ criterion. If a higher value was chosen,  e.g. $R_{pexmon}>20$, this would exclude two sources, NGC 1068 and 2MASXJ15504152. However, NGC 1068 is well known to be Compton thick, and as such motivated the criterion in the first place. We therefore conclude that the $R_{pexmon}>8$ criterion is best suited to selecting Compton thick AGN.

\subsubsection{\nh\ distribution and obscured fraction}

The \nh\ distribution of AGN has a peak at low column densities, which consists of all unobscured sources. For obscured sources, the distribution peaks in the \nh\ = 10$^{23-24}$ \cmsq\ bin. We find that 29/60 (=48 $\pm$ 6\%) of X-ray luminous AGN are heavily obscured (\nh\ $\geq 10^{23}$ \cmsq). Recent studies into the distribution of absorbing columns, such as that by \citet{tueller08}, reveal that for their sample of {\it Swift/BAT} 14-195 keV selected AGN, 31 $\pm$ 5\% are heavily obscured, which is a greater than 3-$\sigma$ difference. Their sample does however extend to higher X-ray luminosities, which will include a greater fraction of unobscured sources due to the dependence they find of obscuration on luminosity. We also intend to investigate the \nh\ distribution for the AGN in our sample, but using optical line emission to define the AGN in paper II. It is likely here that AGN definition based on X-ray luminosity alone will bias a sample. Those sources which do not meet the luminosity criterion, but have \nh\ $>10^{23}$ \cmsq\ are also likely to be AGN.

Our analysis of the dependence of obscuration on luminosity shows that obscuration depends heavily on the intrinsic power of the X-ray source. The data show that obscuration peaks at $L_{\rm X}=10^{42}$ \ergs, and declines steeply both above and below this. Several previous studies have reported the decline in the obscured fraction above $10^{42}$ \ergs\ in X-ray selected samples \citep[][]{ueda03, lafranca05, akylas06, tueller08, beckmann09}, which we confirm here in our MIR selected sample. These results are in support of the `receding torus' model of \citet{lawrence91} and \citet{simpson05} which describe the decrease in the covering fraction of the torus in AGN with source luminosity, due to sublimation of dust. The decline that we show below $10^{42}$ \ergs\  may be caused in part by the inclusion of non-AGN powered sources which emit at these lower luminosities. However, recently, \citet{burlon10} find compelling evidence for a decrease in the obscured fraction at low X-ray luminosities from analysis of the X-ray luminosity functions of obscured and unobscured AGN from the {\it Swift/BAT} survey.  \citet{akylas09} also evidence for a lower incidence of obscuration at low X-ray luminosities in their optically selected sample. Both sets of authors suggest this may be due to the sustenance of torus by accretion driven winds modelled by \citet{elitzur06}. Our result must be confirmed though with a non-X-ray luminosity dependent method of identifying AGNs, which we intend to do in paper II using optical AGN classifications.

\section{Conclusions}
\label{conclusions}

In summary, we have presented an X-ray spectral analysis of the 126 galaxies with \xmm\ coverage in the 12MGS, which is a relatively unbiased and representative selection of galaxies. We have done so with the help of new X-ray radiative transfer calculations, based on Monte-Carlo simulations which take into account photo-electric absorption, Compton scattering and includes fluorescent iron line emission. We have compiled new X-ray spectral models from these data, {\tt trans}  and {\tt torus} which we have presented here. These table models are available publicly on the web at 

{\tt http://astro.ic.ac.uk/mbrightman/home}

Our main results from these models have been:

\begin{itemize}

 \item unobscured AGN with a presumed toroidal distribution of matter around it can achieve a maximum iron K$\alpha$ equivalent width of $\sim150$ eV, a useful limit to consider when assessing if a source is truly unobscured or not. From our calculations, we show that in order to exceed this EW, the line of sight must be obscured with \nh$>10^{23}$ \cmsq\ (e.g. NGC 3690).

 \item for \nh=$10^{25}$ \cmsq, the flux seen in the 10-40 keV, 20-100 keV ({\it INTEGRAL}), and 14-195 keV ({\it Swift/BAT}) bands is only 10\% of the intrinsic flux, which is important for considering the biases present against hard X-ray selected, heavily obscured AGN. At this \nh, 10\% or less of the intrinsic flux is seen in any X-ray band at all redshifts, which reveals the bias present in all X-ray surveys against such heavily obscured systems.
 
\item using spectral models based on slab geometries such as {\tt pexrav} or {\tt pexmon} will underestimate the reflection fraction and intrinsic luminosities with respect to toroidal geometries by a factor of 2-3, leading to underestimation of the intrinsic luminosity of the source, as also found by \citet{murphy09}.

\end{itemize}

By combining the positive attributes of 12 micron selection with the penetrative power of X-rays, we have been able to determine the intrinsic X-ray properties of galaxies such as source power, power-law index, and in addition we have assessed the absorption occurring in the sources. We have improved upon previous works on this sample with better signal to noise, spectral resolution and larger band pass by using \xmm\ data. We have developed a detailed systematic approach to the fitting of the spectral data, with the aim of determining the intrinsic power, $\Gamma$ and \nh\ as accurately as possible. We have paid particular attention to the uncovering of Compton thick sources using our new model, {\tt trans} and the {\tt pexmon} model. The main conclusions from this study have been:

\begin{itemize}

 \item we find the mean primary power-law index to be $<\Gamma>=1.90_{-0.07}^{+0.05}$ for our sample with an intrinsic spread of $\sigma = 0.31_{-0.05}^{+0.05}$. The index is consistent with previous works, though we find a larger intrinsic dispersion in our results than previously reported. 
 \item we also find that the mean power-law index for obscured and unobscured sources is consistent with the two source populations being the same, supporting unified schemes.
 \item we find no dependence of $\Gamma$ on X-ray luminosity as has been previously reported for local AGN.
 \item we find an 18\% Compton thick fraction for the X-ray AGN defined here, which is higher than the hard X-ray selected samples, but roughly consistent with XRB predictions.
 \item the obscured fraction for our sample is a strong function of X-ray luminosity, peaking at a luminosity of $\sim10^{42-43}$ \ergs. The decline in the obscured fraction at high luminosities, where these sources are unambiguously AGN, is a confirmation of previous works. The decline at lower luminosities has also been suggested but needs further explanation  accounting for optical spectroscopic classifications.
 
 \end{itemize}

\section{Acknowledgements}
We would like to thank the anonymous referee for the careful reading of this manuscript and the constructive criticism provided. MB would like to thank STFC for financial support. This research has made use of data obtained from the High Energy Astrophysics Science Archive Research Center (HEASARC), provided by NASA's Goddard Space Flight Center. This research has made use of the NASA/IPAC Extragalactic Database (NED) which is operated by the Jet Propulsion Laboratory, California Institute of Technology, under contract with the National Aeronautics and Space Administration. We thank those who built and operate \xmm.

\bibliographystyle{mn2e}
\bibliography{bibdesk}

\onecolumn
\begin{landscape}
\begin{center}
\begin{scriptsize}
\begin{longtable}{llrrlrrrrrll}

\caption[]{Observational data for the 126 \xmm\ observations used here. Col. (1) Galaxy name as given by NED; Col. (2) Galaxy name as given by \cite{rush93}; Col. (3) Right ascension in decimal J2000 co-ordinates (RMS); Col. (4) Declination in decimal J2000 co-ordinates (RMS); Col. (5) Redshift as reported in NED. $\ast$ Messier 81, NGC 4214 and Messier 90 have reported negative redshifts, indicating that they have peculiar velocities in our direction. We use the luminosity distance reported in NED when calculating the luminosities for these objects (0.662, 7.54 ad 1.22 Mpc respectively); Col. (6) \xmm\ observation identifier; Col. (7) \xmm\ revolution number; Col. (8) Date of \xmm\ observation; Col. (9) Duration of the \xmm\ observation before background screening (ks); Col. (10) Exposure time after screening and dead-time correction; Col. (11) \xmm\ readout mode. FW = full window, FWe = full window extended, LW= large window, SW = small window; Column
(12) Optical filter in place during \xmm\ observation.}
\label{table:xrayanalysis_obsdat}
\\
\hline\\
Name (NED) 	& Name (RMS) & RA (J2000) & Dec. (J2000) & z & Obs ID	& Rev & Date & Duration & Exposure & Mode & Filter \\
(1) & (2) & (3) & (4) & (5) & (6) & (7) & (8) & (9) & (10) & (11) & (12) \\
\hline\\
MRK 0335 	 & 	MK335	 & 1.576 & 20.201 & 	0.0258	 & 	306870101	 & 	1112	 & 	2006-01-03	 & 	133.3	 & 	82.5	 & 	 SW	 & 	 Thin1	\\
NGC 0017 	 & 	N34=MK938	 & 2.778 & -12.107 & 	0.0196	 & 	150480501	 & 	556	 & 	2002-12-22	 & 	22.2	 & 	10.9	 & 	 FW	 & 	 Medium	\\
NGC 0150 	 & 	N150	 & 8.567 & -27.803 & 	0.0053	 & 	44350101	 & 	452	 & 	2002-05-29	 & 	22.9	 & 	11.9	 & 	 FW	 & 	 Medium	\\
NGC 0214 	 & 	N214	 & 10.367 & 25.499 & 	0.0151	 & 	153030101	 & 	653	 & 	2003-07-04	 & 	15.9	 & 	8.1	 & 	 FW	 & 	 Thin1	\\
NGC 0262 	 & 	N262=MK348	 & 12.201 & 31.957 & 	0.0150	 & 	67540201	 & 	477	 & 	2002-07-18	 & 	49.5	 & 	21	 & 	 FW	 & 	 Medium	\\
UGC 00545 	 & 	IZW1	 & 13.405 & 12.697 & 	0.0611	 & 	300470101	 & 	1027	 & 	2005-07-18	 & 	85.5	 & 	50.8	 & 	 SW	 & 	 Medium	\\
NGC 0424 	 & 	N424=TOL0109	 & 17.862 & -38.085 & 	0.0118	 & 	2942301	 & 	367	 & 	2001-12-10	 & 	8.2	 & 	4.5	 & 	 FW	 & 	 Medium	\\
NGC 0526A 	 & 	N526A	 & 20.978 & -35.065 & 	0.0191	 & 	150940101	 & 	647	 & 	2003-06-21	 & 	47.9	 & 	37.5	 & 	 LW	 & 	 Thin1	\\
NGC 0513 	 & 	N513	 & 21.114 & 33.800 & 	0.0195	 & 	301150401	 & 	1124	 & 	2006-01-28	 & 	16.9	 & 	8.9	 & 	 FW	 & 	 Thin1	\\
NGC 0520 	 & 	N520	 & 21.143 & 3.791 & 	0.0076	 & 	25541601	 & 	378	 & 	2002-01-01	 & 	13.1	 & 	8.9	 & 	 FW	 & 	 Medium	\\
NGC 0660 	 & 	N660	 & 25.759 & 13.646 & 	0.0028	 & 	93641001	 & 	198	 & 	2001-01-07	 & 	12.2	 & 	6.3	 & 	 FWe	 & 	 Medium	\\
2MASX J01500266-0725482 	 & 	F01475-0740	 & 27.515 & -7.428 & 	0.0177	 & 	200431101	 & 	754	 & 	2004-01-21	 & 	11.9	 & 	9	 & 	 FW	 & 	 Thin1	\\
NGC 0695 	 & 	N695	 & 27.808 & 22.582 & 	0.0325	 & 	204410101	 & 	751	 & 	2004-01-14	 & 	11.9	 & 	2.4	 & 	 FW	 & 	 Medium	\\
NGC 1052 	 & 	N1052	 & 40.263 & -8.257 & 	0.0050	 & 	306230101	 & 	1116	 & 	2006-01-12	 & 	54.9	 & 	46.9	 & 	 FW	 & 	 Medium	\\
MESSIER 077 	 & 	N1068	 & 40.672 & -0.013 & 	0.0038	 & 	111200101	 & 	117	 & 	2000-07-29	 & 	42.3	 & 	33.1	 & 	 LW	 & 	 Medium	\\
ARP 118 	 & 	N1143/4	 & 43.803 & -0.183 & 	0.0288	 & 	312190401	 & 	1124	 & 	2006-01-28	 & 	11.9	 & 	8.9	 & 	 FW	 & 	 Thin1	\\
MCG -02-08-039 	 & 	M-2-8-39	 & 45.129 & -11.415 & 	0.0299	 & 	301150201	 & 	1122	 & 	2006-01-23	 & 	23.4	 & 	7	 & 	 FW	 & 	 Thin1	\\
NGC 1194 	 & 	N1194	 & 45.952 & -1.104 & 	0.0136	 & 	307000701	 & 	1135	 & 	2006-02-19	 & 	16.1	 & 	12.4	 & 	 FW	 & 	 Medium	\\
NGC 1291 	 & 	N1291	 & 49.327 & -41.108 & 	0.0028	 & 	201690201	 & 	929	 & 	2005-01-03	 & 	34.9	 & 	18.7	 & 	 FW	 & 	 Medium	\\
NGC 1313 	 & 	N1313	 & 49.570 & -66.492 & 	0.0016	 & 	405090101	 & 	1255	 & 	2006-10-16	 & 	123.2	 & 	91.6	 & 	 FW	 & 	 Medium	\\
NGC 1316 	 & 	N1316	 & 50.672 & -37.209 & 	0.0059	 & 	302780101	 & 	1039	 & 	2005-08-11	 & 	107.6	 & 	76.8	 & 	 FWe	 & 	 Thin1	\\
NGC 1320 	 & 	N1320=MK607	 & 51.200 & -3.043 & 	0.0089	 & 	405240201	 & 	1219	 & 	2006-08-06	 & 	17.1	 & 	12.5	 & 	 FW	 & 	 Thin1	\\
NGC 1365 	 & 	N1365	 & 53.402 & -36.140 & 	0.0055	 & 	205590301	 & 	752	 & 	2004-01-17	 & 	59.7	 & 	51.2	 & 	 FW	 & 	 Medium	\\
NGC 1386 	 & 	N1386	 & 54.190 & -36.000 & 	0.0029	 & 	140950201	 & 	560	 & 	2002-12-29	 & 	17.4	 & 	13.8	 & 	 FW	 & 	 Medium	\\
NGC 1482 	 & 	N1482	 & 58.666 & -20.502 & 	0.0064	 & 	150910101	 & 	583	 & 	2003-02-14	 & 	18.9	 & 	7.4	 & 	 FW	 & 	 Medium	\\
3C 120 	 & 	3C120	 & 68.294 & 5.355 & 	0.0330	 & 	152840101	 & 	680	 & 	2003-08-26	 & 	133.9	 & 	78.5	 & 	 SW	 & 	 Medium	\\
NGC 1614 	 & 	N1614	 & 68.501 & -8.579 & 	0.0159	 & 	150480201	 & 	583	 & 	2003-02-13	 & 	23.9	 & 	15.7	 & 	 FW	 & 	 Medium	\\
MRK 0618 	 & 	MK618	 & 69.097 & -10.375 & 	0.0356	 & 	307001301	 & 	1133	 & 	2006-02-15	 & 	18.1	 & 	6.9	 & 	 FW	 & 	 Medium	\\
NGC 1672 	 & 	N1672	 & 71.428 & -59.247 & 	0.0044	 & 	203880101	 & 	910	 & 	2004-11-27	 & 	49.9	 & 	42.6	 & 	 FW	 & 	 Medium	\\
NGC 1667 	 & 	N1667	 & 72.157 & -6.320 & 	0.0152	 & 	200660401	 & 	876	 & 	2004-09-20	 & 	10.4	 & 	7.4	 & 	 LW	 & 	 Thick	\\
NGC 1808 	 & 	N1808	 & 76.926 & -37.514 & 	0.0033	 & 	110980801	 & 	426	 & 	2002-04-06	 & 	43.6	 & 	31.3	 & 	 FWe	 & 	 Thin1	\\
ESO 362- G 018 	 & 	M-5-13-17	 & 79.903 & -32.658 & 	0.0124	 & 	312190701	 & 	1124	 & 	2006-01-28	 & 	11.9	 & 	8.9	 & 	 FW	 & 	 Thin1	\\
2MASX J05210136-2521450 	 & 	F05189-2524	 & 80.256 & -25.363 & 	0.0426	 & 	85640101	 & 	233	 & 	2001-03-17	 & 	12.3	 & 	7	 & 	 FW	 & 	 Thin1	\\
2MASX J05580206-3820043 	 & 	F05563-3820	 & 89.512 & -38.335 & 	0.0339	 & 	109131001	 & 	508	 & 	2002-09-17	 & 	9.1	 & 	3.3	 & 	 LW	 & 	 Thin1	\\
IC 0450 	 & 	MK6	 & 103.055 & 74.427 & 	0.0188	 & 	61540101	 & 	238	 & 	2001-03-27	 & 	46.5	 & 	17.6	 & 	 FW	 & 	 Medium	\\
UGC 03973 	 & 	MK79	 & 115.633 & 49.810 & 	0.0222	 & 	103862101	 & 	253	 & 	2001-04-26	 & 	6.5	 & 	3.6	 & 	 SW	 & 	 Medium	\\
IRASF07599+6508	 & 	F07599+6508 	 & 121.138 & 64.997 & 	0.1488	 & 	94400301	 & 	344	 & 	2001-10-25	 & 	23.2	 & 	17.3	 & 	 FW	 & 	 Medium	\\
NGC 2639 	 & 	N2639	 & 130.913 & 50.205 & 	0.0111	 & 	301651101	 & 	973	 & 	2005-04-02	 & 	26.1	 & 	2.2	 & 	 FW	 & 	 Thin1	\\
NGC 2655 	 & 	N2655	 & 133.911 & 78.224 & 	0.0047	 & 	301650301	 & 	1051	 & 	2005-09-04	 & 	11.8	 & 	3.6	 & 	 FW	 & 	 Thin1	\\
IC 2431 	 & 	I2431=U4756	 & 136.147 & 14.594 & 	0.0499	 & 	303562101	 & 	1091	 & 	2005-11-24	 & 	6.8	 & 	4.3	 & 	 FW	 & 	 Thin1	\\
MRK 0704 	 & 	MK704	 & 139.607 & 16.307 & 	0.0292	 & 	300240101	 & 	1074	 & 	2005-10-21	 & 	21.7	 & 	12.5	 & 	 SW	 & 	 Medium	\\
NGC 2841 	 & 	N2841	 & 140.516 & 50.972 & 	0.0021	 & 	201440101	 & 	901	 & 	2004-11-09	 & 	41.9	 & 	23.7	 & 	 FW	 & 	 Thin1	\\
UGC 05101 	 & 	U5101	 & 143.953 & 61.356 & 	0.0394	 & 	85640201	 & 	353	 & 	2001-11-12	 & 	35	 & 	26.1	 & 	 FWe	 & 	 Thin1	\\
NGC 2992 	 & 	N2992	 & 146.418 & -14.323 & 	0.0077	 & 	147920301	 & 	630	 & 	2003-05-19	 & 	28.9	 & 	21.8	 & 	 FW	 & 	 Medium	\\
MESSIER 081 	 & 	N3031=M81	 & 148.890 & 69.066 & 	-0.0001$\ast$	 & 	111800101	 & 	251	 & 	2001-04-22	 & 	132.7	 & 	79.7	 & 	 SW	 & 	 Medium	\\
MESSIER 082 	 & 	N3034=M82	 & 148.971 & 69.678 & 	0.0007	 & 	206080101	 & 	800	 & 	2004-04-21	 & 	104.4	 & 	63.3	 & 	 FW	 & 	 Medium	\\
NGC 3079 	 & 	N3079	 & 150.491 & 55.681 & 	0.0037	 & 	110930201	 & 	246	 & 	2001-04-13	 & 	25.3	 & 	13.6	 & 	 FWe	 & 	 Thin1	\\
NGC 3147 	 & 	N3147	 & 154.231 & 73.398 & 	0.0094	 & 	405020601	 & 	1250	 & 	2006-10-06	 & 	17.2	 & 	13.6	 & 	 LW	 & 	 Thin1	\\
NGC 3227 	 & 	N3227	 & 155.881 & 19.864 & 	0.0039	 & 	101040301	 & 	178	 & 	2000-11-28	 & 	40.1	 & 	30.3	 & 	 FW	 & 	 Medium	\\
NGC 3310 	 & 	N3310	 & 159.693 & 53.502 & 	0.0033	 & 	112810301	 & 	260	 & 	2001-05-11	 & 	19.8	 & 	2.7	 & 	 FW	 & 	 Thin1	\\
NGC 3486 	 & 	N3486	 & 165.099 & 28.977 & 	0.0023	 & 	112550101	 & 	259	 & 	2001-05-09	 & 	15.2	 & 	3.7	 & 	 FWe	 & 	 Medium	\\
NGC 3516 	 & 	N3516	 & 166.703 & 72.567 & 	0.0088	 & 	107460701	 & 	352	 & 	2001-11-09	 & 	130	 & 	80.6	 & 	 SW	 & 	 Thin1	\\
MESSIER 066 	 & 	N3627	 & 170.064 & 12.992 & 	0.0024	 & 	93641101	 & 	268	 & 	2001-05-26	 & 	11.2	 & 	5.3	 & 	 FWe	 & 	 Medium	\\
NGC 3690 	 & 	N3690=MK171	 & 172.136 & 58.561 & 	0.0104	 & 	112810101	 & 	258	 & 	2001-05-07	 & 	22.2	 & 	14.6	 & 	 FW	 & 	 Thin1	\\
NGC 3735 	 & 	N3735	 & 173.989 & 70.536 & 	0.0090	 & 	301650501	 & 	1062	 & 	2005-09-27	 & 	16.7	 & 	0.4	 & 	 FW	 & 	 Thin1	\\
NGC 3976 	 & 	N3976	 & 178.991 & 6.748 & 	0.0083	 & 	301651801	 & 	1194	 & 	2006-06-16	 & 	14.4	 & 	4.4	 & 	 FW	 & 	 Thin1	\\
NGC 3982 	 & 	N3982	 & 179.114 & 55.127 & 	0.0037	 & 	204651201	 & 	827	 & 	2004-06-15	 & 	11.9	 & 	9.1	 & 	 LW	 & 	 Thin1	\\
NGC 4013 	 & 	N4013	 & 179.631 & 43.948 & 	0.0028	 & 	306060201	 & 	1086	 & 	2005-11-13	 & 	157.2	 & 	49.7	 & 	 FW	 & 	 Medium	\\
ARP 244 	 & 	N4038/9	 & 180.480 & -18.879 & 	0.0057	 & 	85220201	 & 	382	 & 	2002-01-08	 & 	52.8	 & 	34.9	 & 	 FWe	 & 	 Medium	\\
NGC 4051 	 & 	N4051	 & 180.791 & 44.531 & 	0.0023	 & 	157560101	 & 	541	 & 	2002-11-22	 & 	51.9	 & 	44	 & 	 LW	 & 	 Medium	\\
NGC 4151 	 & 	N4151	 & 182.644 & 39.402 & 	0.0033	 & 	112830201	 & 	190	 & 	2000-12-22	 & 	62.2	 & 	50.9	 & 	 FW	 & 	 Medium	\\
NGC 4214 	 & 	N4214	 & 183.917 & 36.324 & 	-0.0003$\ast$	 & 	35940201	 & 	358	 & 	2001-11-22	 & 	20.3	 & 	12.9	 & 	 FW	 & 	 Medium	\\
NGC 4253 	 & 	MK766	 & 184.608 & 29.814 & 	0.0129	 & 	109141301	 & 	265	 & 	2001-05-20	 & 	129.9	 & 	72.2	 & 	 SW	 & 	 Medium	\\
MESSIER 099 	 & 	N4254	 & 184.713 & 14.414 & 	0.0080	 & 	147610101	 & 	651	 & 	2003-06-29	 & 	51	 & 	12.8	 & 	 FWe	 & 	 Thin1	\\
MESSIER 100 	 & 	N4321	 & 185.724 & 15.823 & 	0.0052	 & 	106860201	 & 	376	 & 	2001-12-28	 & 	36.6	 & 	20.6	 & 	 FWe	 & 	 Medium	\\
NGC 4388 	 & 	N4388	 & 186.441 & 12.664 & 	0.0084	 & 	110930701	 & 	551	 & 	2002-12-12	 & 	12	 & 	7.3	 & 	 FWe	 & 	 Thin1	\\
NGC 4414 	 & 	N4414	 & 186.612 & 31.223 & 	0.0024	 & 	402830101	 & 	1200	 & 	2006-06-29	 & 	42.2	 & 	17.8	 & 	 FW	 & 	 Thin1	\\
NGC 4449 	 & 	N4449	 & 187.034 & 44.099 & 	0.0007	 & 	112521601	 & 	450	 & 	2002-05-25	 & 	24	 & 	1.9	 & 	 FW	 & 	 Thin1	\\
3C 273 	 & 	3C273	 & 187.282 & 2.051 & 	0.1583	 & 	136550101	 & 	277	 & 	2001-06-13	 & 	89.8	 & 	61.4	 & 	 SW	 & 	 Medium	\\
NGC 4490 	 & 	N4490	 & 187.646 & 41.647 & 	0.0019	 & 	112280201	 & 	451	 & 	2002-05-27	 & 	18.1	 & 	11.5	 & 	 FWe	 & 	 Medium	\\
MESSIER 088 	 & 	N4501	 & 187.990 & 14.422 & 	0.0076	 & 	112550801	 & 	364	 & 	2001-12-04	 & 	14	 & 	2.5	 & 	 FWe	 & 	 Medium	\\
NGC 4559 	 & 	N4559	 & 188.987 & 27.960 & 	0.0027	 & 	152170501	 & 	634	 & 	2003-05-27	 & 	42.2	 & 	34.1	 & 	 FW	 & 	 Medium	\\
MESSIER 090 	 & 	N4569=M90	 & 189.212 & 13.162 & 	-0.0008$\ast$	 & 	200650101	 & 	918	 & 	2004-12-13	 & 	66	 & 	44.2	 & 	 FWe	 & 	 Thin1	\\
MESSIER 058 	 & 	N4579	 & 189.428 & 11.820 & 	0.0051	 & 	112840101	 & 	642	 & 	2003-06-12	 & 	23.7	 & 	15.2	 & 	 FW	 & 	 Thin1	\\
NGC 4593 	 & 	N4593	 & 189.917 & -5.346 & 	0.0090	 & 	109970101	 & 	103	 & 	2000-07-02	 & 	28.1	 & 	5	 & 	 SW	 & 	 Thin1	\\
MESSIER 104 	 & 	N4594	 & 190.001 & -11.624 & 	0.0005	 & 	84030101	 & 	376	 & 	2001-12-28	 & 	86.9	 & 	15.6	 & 	 FWe	 & 	 Thin1	\\
NGC 4631 	 & 	N4631	 & 190.529 & 32.543 & 	0.0020	 & 	110900201	 & 	467	 & 	2002-06-28	 & 	54.8	 & 	36.9	 & 	 FWe	 & 	 Thin1	\\
NGC 4666 	 & 	N4666	 & 191.282 & -0.461 & 	0.0051	 & 	110980201	 & 	467	 & 	2002-06-27	 & 	58.2	 & 	48	 & 	 FWe	 & 	 Thin1	\\
NGC 4725 	 & 	N4725	 & 192.574 & 25.489 & 	0.0040	 & 	112550401	 & 	460	 & 	2002-06-14	 & 	18.6	 & 	12.1	 & 	 FWe	 & 	 Medium	\\
UGC 08058 	 & 	MK231=U8058	 & 194.063 & 56.871 & 	0.0422	 & 	81340201	 & 	274	 & 	2001-06-07	 & 	22.3	 & 	17.3	 & 	 FW	 & 	 Medium	\\
NGC 4968 	 & 	N4968	 & 196.774 & -23.677 & 	0.0099	 & 	200660201	 & 	837	 & 	2004-07-05	 & 	17.7	 & 	4.8	 & 	 LW	 & 	 Thick	\\
NGC 5005 	 & 	N5005	 & 197.732 & 37.059 & 	0.0032	 & 	110930501	 & 	551	 & 	2002-12-13	 & 	13.6	 & 	8.6	 & 	 FWe	 & 	 Thin1	\\
MESSIER 063 	 & 	N5055	 & 198.956 & 42.028 & 	0.0017	 & 	405080301	 & 	1367	 & 	2007-05-28	 & 	51.8	 & 	7.1	 & 	 FW	 & 	 Thin1	\\
MCG -03-34-064 	 & 	M-3-34-63	 & 200.598 & -16.726 & 	0.0165	 & 	206580101	 & 	939	 & 	2005-01-24	 & 	44.9	 & 	35	 & 	 FW	 & 	 Medium	\\
NGC 5170 	 & 	N5170	 & 202.460 & -17.972 & 	0.0050	 & 	206300101	 & 	757	 & 	2004-01-27	 & 	31.9	 & 	17.6	 & 	 FW	 & 	 Thin1	\\
NGC 5194	 & 	N5194=M51	 & 202.473 & 47.195 & 	0.0015	 & 	112840201	 & 	568	 & 	2003-01-15	 & 	20.9	 & 	17.2	 & 	 FW	 & 	 Thin1	\\
ESO 383- G 035 	 & 	M-6-30-15	 & 203.977 & -34.297 & 	0.0077	 & 	29740801	 & 	303	 & 	2001-08-04	 & 	130.5	 & 	83.4	 & 	 SW	 & 	 Medium	\\
MESSIER 083 	 & 	N5236=M83	 & 204.245 & -29.863 & 	0.0017	 & 	110910201	 & 	574	 & 	2003-01-27	 & 	30.7	 & 	22.4	 & 	 FWe	 & 	 Thin1	\\
IRASF13349+2438	 & 	F13349+2438	 & 204.325 & 24.386 & 	0.1076	 & 	96010101	 & 	97	 & 	2000-06-20	 & 	130.4	 & 	28.5	 & 	 SW	 & 	 Thin1	\\
NGC 5256 (S)	 & 	N5256=MK266	 & 204.576 & 48.275 & 	0.0279	 & 	55990501	 & 	445	 & 	2002-05-15	 & 	23.1	 & 	14.2	 & 	 FWe	 & 	 Thin1	\\
NGC 5253 	 & 	N5253	 & 204.980 & -31.639 & 	0.0014	 & 	35940301	 & 	305	 & 	2001-08-08	 & 	47	 & 	27.5	 & 	 FWe	 & 	 Thin1	\\
MRK 0273 	 & 	MK273=U8696	 & 206.174 & 55.887 & 	0.0378	 & 	101640401	 & 	441	 & 	2002-05-07	 & 	22.8	 & 	17.9	 & 	 FW	 & 	 Thick	\\
IC 4329A 	 & 	I4329A	 & 207.331 & -30.310 & 	0.0161	 & 	147440101	 & 	670	 & 	2003-08-06	 & 	136	 & 	68	 & 	 SW	 & 	 Thin1	\\
UGC 08850 (E)	 & 	MK463	 & 209.011 & 18.372 & 	0.0504	 & 	94401201	 & 	373	 & 	2001-12-22	 & 	26.8	 & 	21	 & 	 FW	 & 	 Medium	\\
NGC 5506 	 & 	N5506	 & 213.309 & -3.207 & 	0.0062	 & 	201830201	 & 	840	 & 	2004-07-11	 & 	21.6	 & 	14.6	 & 	 SW	 & 	 Medium	\\
NGC 5548 	 & 	N5548	 & 214.499 & 25.137 & 	0.0172	 & 	89960301	 & 	290	 & 	2001-07-09	 & 	95.8	 & 	46.7	 & 	 SW	 & 	 Thin1	\\
NGC 5775 	 & 	N5775	 & 223.492 & 3.542 & 	0.0056	 & 	150350101	 & 	665	 & 	2003-07-28	 & 	47.2	 & 	22.4	 & 	 FW	 & 	 Thin1	\\
2MASX J15115979-2119015 	 & 	F15091-2107	 & 227.998 & -21.317 & 	0.0446	 & 	300240201	 & 	1031	 & 	2005-07-26	 & 	23	 & 	7.6	 & 	 LW	 & 	 Medium	\\
VV 705 	 & 	IZW107	 & 229.528 & 42.745 & 	0.0402	 & 	203390601	 & 	864	 & 	2004-08-28	 & 	25.2	 & 	9.1	 & 	 FW	 & 	 Medium	\\
UGC 09944 	 & 	U9944	 & 233.950 & 73.451 & 	0.0245	 & 	307002401	 & 	1014	 & 	2005-06-23	 & 	16.1	 & 	10.8	 & 	 FW	 & 	 Medium	\\
2MASX J15504152-0353175 	 & 	F15480-0344	 & 237.671 & -3.889 & 	0.0303	 & 	138951501	 & 	667	 & 	2003-07-31	 & 	12.4	 & 	1.5	 & 	 FW	 & 	 Medium	\\
NGC 6240 	 & 	N6240	 & 253.244 & 2.401 & 	0.0245	 & 	101640101	 & 	144	 & 	2000-09-22	 & 	30.1	 & 	12.5	 & 	 FW	 & 	 Medium	\\
NGC 6286 	 & 	N6286	 & 254.629 & 58.939 & 	0.0183	 & 	203390701	 & 	765	 & 	2004-02-12	 & 	41.6	 & 	1.8	 & 	 FW	 & 	 Medium	\\
NGC 6552 	 & 	N6552	 & 270.030 & 66.615 & 	0.0265	 & 	112310801	 & 	523	 & 	2002-10-18	 & 	7.8	 & 	4.3	 & 	 FW	 & 	 Medium	\\
AM 1925-724 	 & 	F19254-7245	 & 292.844 & -72.656 & 	0.0617	 & 	81341001	 & 	239	 & 	2001-03-30	 & 	23.2	 & 	13	 & 	 FW	 & 	 Thin1	\\
NGC 6810 	 & 	N6810	 & 295.890 & -58.656 & 	0.0068	 & 	205220101	 & 	802	 & 	2004-04-25	 & 	48.7	 & 	31	 & 	 FW	 & 	 Medium	\\
NGC 6890 	 & 	N6890	 & 304.577 & -44.806 & 	0.0081	 & 	301151001	 & 	1063	 & 	2005-09-29	 & 	12.8	 & 	1.2	 & 	 FW	 & 	 Thin1	\\
MRK 0509 	 & 	MK509	 & 311.037 & -10.724 & 	0.0344	 & 	306090201	 & 	1073	 & 	2005-10-18	 & 	86	 & 	58.1	 & 	 SW	 & 	 Thin1	\\
ESO 286-IG 019 	 & 	E286-IG19	 & 314.614 & -42.649 & 	0.0430	 & 	81340401	 & 	250	 & 	2001-04-21	 & 	21.7	 & 	10	 & 	 FW	 & 	 Thin1	\\
NGC 7090 	 & 	N7090	 & 324.118 & -54.558 & 	0.0028	 & 	200230101	 & 	793	 & 	2004-04-08	 & 	28	 & 	3.9	 & 	 FWe	 & 	 Thin1	\\
NGC 7172 	 & 	N7172	 & 330.509 & -31.869 & 	0.0087	 & 	202860101	 & 	902	 & 	2004-11-11	 & 	58.9	 & 	40	 & 	 LW	 & 	 Thin1	\\
NGC 7213 	 & 	N7213	 & 332.319 & -47.166 & 	0.0058	 & 	111810101	 & 	269	 & 	2001-05-29	 & 	49.6	 & 	29.3	 & 	 SW	 & 	 Medium	\\
IC 5169 	 & 	I5169	 & 332.541 & -36.089 & 	0.0104	 & 	307002001	 & 	1167	 & 	2006-04-23	 & 	18.6	 & 	0.9	 & 	 FW	 & 	 Medium	\\
NGC 7252 	 & 	N7252	 & 335.189 & -24.677 & 	0.0160	 & 	49340201	 & 	354	 & 	2001-11-13	 & 	28.2	 & 	21.2	 & 	 FW	 & 	 Medium	\\
3C 445 	 & 	3C445	 & 335.955 & -2.104 & 	0.0562	 & 	90050601	 & 	365	 & 	2001-12-06	 & 	23.9	 & 	15.1	 & 	 SW	 & 	 Thin1	\\
NGC 7314 	 & 	N7314	 & 338.941 & -26.050 & 	0.0048	 & 	111790101	 & 	256	 & 	2001-05-02	 & 	44.7	 & 	26.7	 & 	 SW	 & 	 Medium	\\
MCG -03-58-007 	 & 	M-3-58-7	 & 342.410 & -19.272 & 	0.0315	 & 	301150301	 & 	992	 & 	2005-05-09	 & 	15.2	 & 	2.7	 & 	 FW	 & 	 Thin1	\\
NGC 7469 	 & 	N7469	 & 345.815 & 8.874 & 	0.0163	 & 	207090101	 & 	912	 & 	2004-11-30	 & 	85	 & 	57.8	 & 	 SW	 & 	 Medium	\\
NGC 7479 	 & 	N7479	 & 346.240 & 12.325 & 	0.0079	 & 	25541001	 & 	280	 & 	2001-06-19	 & 	13.1	 & 	6.1	 & 	 FW	 & 	 Medium	\\
ESO 148-IG 002 	 & 	E148-IG02	 & 348.944 & -59.054 & 	0.0446	 & 	81340301	 & 	539	 & 	2002-11-19	 & 	23.4	 & 	8.8	 & 	 FW	 & 	 Medium	\\
NGC 7552 	 & 	N7552	 & 349.040 & -42.586 & 	0.0054	 & 	93640701	 & 	270	 & 	2001-05-30	 & 	20.6	 & 	1.5	 & 	 FWe	 & 	 Medium	\\
NGC 7582 	 & 	N7582	 & 349.593 & -42.372 & 	0.0053	 & 	204610101	 & 	987	 & 	2005-04-29	 & 	101.9	 & 	65.6	 & 	 FW	 & 	 Medium	\\
NGC 7674 	 & 	N7674	 & 351.987 & 8.779 & 	0.0289	 & 	200660101	 & 	821	 & 	2004-06-02	 & 	10.4	 & 	7.6	 & 	 LW	 & 	 Thick	\\
NGC 7714 	 & 	N7714	 & 354.059 & 2.155 & 	0.0093	 & 	112521301	 & 	457	 & 	2002-06-07	 & 	22.2	 & 	13.8	 & 	 FW	 & 	 Medium	\\
NGC 7771 	 & 	N7771	 & 357.853 & 20.111 & 	0.0143	 & 	93190301	 & 	464	 & 	2002-06-21	 & 	32.1	 & 	26.8	 & 	 FW	 & 	 Medium	\\
MRK 0331 	 & 	MK331	 & 357.859 & 20.586 & 	0.0185	 & 	203390801	 & 	827	 & 	2004-06-15	 & 	18.4	 & 	12.2	 & 	 FW	 & 	 Medium	\\

\hline\\
\end{longtable}
\end{scriptsize}

\end{center}
\end{landscape}
 
\begin{landscape}
\centering
\begin{longtable}{l r r r l r l l r}\\
\caption{Parameters of the basic model consisting of the primary power-law and the thermal plasma component, where present. More complex model parameters are presented in Tables \ref{table:xrayanalysis_specfit_pl2} to \ref{table:xrayanalysis_specfit_feka}. Column (1) Galaxy name; Column (2) \chisq/DOF of the fit. $^{(F)}$ indicates that the full band spectral fit method was used, $^{(D)}$ indicates that although the \rchisq\ of this fit is bad (\rchisq$>2$), the data/model ratio never exceded 5\%, which is the level of uncertainties in the \xmm\ calibration, $^{(H)}$ indicates that only the 2.5-10 keV fit parameters have been presented here due to spectral complexities below 2 keV. For the 2.5-10 keV fits, only NGC 1068, NGC 1365 and M82 have bad fits. These sources have notoriously complex spectra, which are high signal to noise in this case, with several emission and absorption features. We do not attempt to fit these additional complexities, however, and present only the basic hard band parameters for these fits. No errors are quoted because the standard formalism does not apply unless the model fits the data \citep{lampton76}. ; Column (3) Galactic column density (10$^{20}$ \cmsq); Column (4) the neutral column density measured in the primary power-law (10$^{22}$ \cmsq). $\ast$ indicates that an ionised absorber has been detected in this spectral fit, details of which are in Table \ref{table:xrayanalysis_specfit_cwa}; Column (5) The power-law index of the primary power-law. $^{\dag}$ indicates that this parameter was fixed in the fit;  Column (6) The normalisation of the primary power-law ($\times 10^{-5}$ Photons cm$^{-2}$s$^{-1}$keV$^{-1}$ at 1 keV) Column (7)  the temperature, in keV, of the thermal plasma component; Column (8) The ratio of the normalisation of the thermal plasma component to the normalisation of the primary power-law. (9) Observed flux in the 2-10 keV band in units of 10$^{-14}$ ergs s$^{-1}$ cm$^{-2}$ and the logarithm of the intrinsic luminosity in this band in units of \ergs.
}
\label{table:xrayanalysis_specfit}
\\
\hline
\\
Name & $\chi^2$/DOF & $N_{\rm{H}}$(Gal) & $N_{\rm{H}}$ & $\Gamma_1$ &$A_{\rm pl}$ & kT$_{therm}$ & R$_{therm}$ & F$_{HX}$/L$_{HX}$ \\
(1) & (2) & (3) & (4) & (5)  & (6)  & (7) & (8) & (9) \\
\hline
MRK0335 &1758.79/143$^{(D)}$&3.6& - $\ast$& 2.03 $_{-0.02}^{+0.02}$ & 704 $_{-19}^{+20}$ &0.27&0.22&1869/43.5\\
NGC0017 &35.15/39$^{}$&2.1& 47.0 $_{-21.2}^{+30.4}$ &1.9$^{\dag}$& 36.6 $_{-15.9}^{+33.1}$ & 0.70 $_{-0.10}^{+0.11}$ & 0.0271 $_{-0.0093}^{+0.0285}$ &28/42.0\\
NGC0150 &76.04/61$^{}$&1.5& - & 1.73 $_{-0.3}^{+0.3}$ & 9.11 $_{-3.40}^{+5.25}$ & 0.35 $_{-0.05}^{+0.06}$ & 0.274 $_{-0.052}^{+0.279}$ &38/40.4\\
NGC0214 &6.02/6$^{}$&3.8&17.3 $_{-8.0}^{+13.1}$&1.9$^{\dag}$&9.84 $_{-3.91}^{+5.73}$&0.35 $_{-0.10}^{+0.35}$&0.0507 $_{-0.0374}^{+0.0424}$&15/41.2\\
NGC0262 &210.06/140$^{}$&5.8& 13.1 $_{-1.3}^{+0.7}$ $\ast$& 1.68 $_{-0.1}^{+0.07}$ & 1020 $_{-329}^{+179}$ & 0.20 $_{-0.05}^{+0.05}$ & 0.0016 $_{-0.0007}^{+0.0082}$ &2809/43.3\\
UGC00545 &79.64/97$^{(H)}$&4.8&-& 2.16 $_{-0.04}^{+0.05}$ & 239 $_{-12}^{+15}$ &-&-&494/43.6\\
NGC0424 &63.56/46$^{}$&1.5& 23.6 $_{-15.8}^{+44.7}$ &1.9$^{\dag}$& 36.7 $_{-27.3}^{+822}$ & 0.74 $_{-0.11}^{+0.18}$ & 0.0514 $_{-0.0279}^{+0.0593}$ &135/42.5\\
NGC0526A &179.09/149$^{}$&2.3& 1.14 $_{-0.26}^{+0.26}$ $\ast$& 1.51 $_{-0.05}^{+0.05}$ & 432 $_{-35}^{+38}$ & 1.2 $_{-0.2}^{+0.0}$ & 0.0082 $_{-0.0036}^{+0.0087}$ &2255/43.3\\
NGC0513 &86.71/123$^{}$&5.2& 6.82 $_{-1.55}^{+1.60}$ & 1.69 $_{-0.2}^{+0.2}$ & 128. $_{-45.2}^{+70.2}$ & 0.30 $_{-0.07}^{+0.09}$ & 0.0239 $_{-0.0107}^{+0.0345}$ &406/42.7\\
NGC0520 &23.05/24$^{}$&3.1& - $\ast$&1.9$^{\dag}$& 2.25 $_{-0.87}^{+0.87}$ & 0.71 $_{-0.21}^{+0.13}$ & 0.347 $_{-0.116}^{+0.365}$ &6/40.0\\
NGC0660 &20.3/22$^{}$&4.7& 0.261 $_{-0.078}^{+0.146}$ &1.9$^{\dag}$& 4.95 $_{-1.27}^{+1.27}$ & 0.62 $_{-0.12}^{+0.13}$ & 0.303 $_{-0.098}^{+0.319}$ &15/39.4\\
2MASXJ01500266 &81.95/75$^{}$&2& 0.211 $_{-0.006}^{+0.055}$ $\ast$& 2.19 $_{-0.2}^{+0.3}$ & 40.1 $_{-11.5}^{+17.2}$ & - & - &78/41.7\\
NGC0695 &3.89/8$^{}$&6.9& - $\ast$&1.9$^{\dag}$& 5.67 $_{-2.29}^{+2.29}$ & 0.68 $_{-0.21}^{+0.12}$ & 0.357 $_{-0.164}^{+0.395}$ &16/41.6\\
NGC1052 &169.33/139$^{}$&2.8& 30.6 $_{-26.3}^{+11.4}$ $\ast$& 1.72 $_{-0.2}^{+0.1}$ & 142 $_{-67}^{+64}$ &0.61 $_{-0.04}^{+0.04}$&0.0183 $_{-0.0021}^{+0.0184}$&465/41.5\\
NGC1068 &235.38/97$^{}$&3& - &1.84&74& - & - &497/42.2\\
ARP118 &80.27/85$^{}$&5.9& 64.7 $_{-6.9}^{+7.8}$ &1.9$^{\dag}$& 545 $_{-67}^{+79}$ & 0.60 $_{-0.49}^{+0.17}$ & 0.0033 $_{-0.0022}^{+0.073}$ &331/43.6\\
MCG-02-08-039 &64.24/44$^{}$&5.1& 46.9 $_{-11.9}^{+13.1}$ &1.9$^{\dag}$& 138 $_{-36}^{+44}$ & 0.19 $_{-0.02}^{+0.03}$ & 1.49 $_{-0.30}^{+1.53}$ &83/42.9\\
NGC1194 &55.18/57$^{}$&6.5& 67.8 $_{-16.4}^{+18.3}$ $\ast$&1.9$^{\dag}$& 168 $_{-79}^{+84}$ & 0.76 $_{-0.18}^{+0.19}$ & 0.0671 $_{-0.0559}^{+0.178}$ &101/42.3\\
NGC1291 &93.19/66$^{}$&1.6& - $\ast$& 2.21 $_{-0.3}^{+0.3}$ & 13.6 $_{-4.8}^{+7.1}$ & 0.38 $_{-0.03}^{+0.03}$ & 0.466 $_{-0.190}^{+0.681}$ &25/39.7\\
NGC1313 &154.82/133$^{}$&4& 0.151 $_{-0.012}^{+0.013}$ &1.9$^{\dag}$& 11.0 $_{-0.3}^{+0.3}$ & 0.26 $_{-0.02}^{+0.04}$ & 0.262 $_{-0.081}^{+0.089}$ &37/39.3\\
NGC1316 &83.29/77$^{(H)}$&2.4&-& 2.12 $_{-0.2}^{+0.2}$ & 12.2 $_{-2.5}^{+3.0}$ &-&-&28/40.3\\
NGC1320 &84.61/53$^{}$&3.7& 204 $_{-99}^{+91}$ &1.9$^{\dag}$& 854 $_{-708}^{+1490}$ & 0.82 $_{-0.08}^{+0.08}$ & 0.0029 $_{-0.0008}^{+0.0030}$ &48/42.7\\
NGC1365 &830.55/95$^{}$&1.3&17.51&2.81&5271& - & - &1296/42.5\\
NGC1386 &96.32/50$^{}$&1.3& 0.0459 $_{-0.0137}^{+0.0161}$ & 3.35 $_{-0.2}^{+0.2}$ & 7.03 $_{-0.77}^{+0.84}$ & 0.73 $_{-0.05}^{+0.04}$ & 0.523 $_{-0.0954}^{+0.124}$ &32/40.9\\
NGC1482 &56.64/50$^{(F)}$&3.4& 8.68 $_{-3.86}^{+7.19}$ & 1.93 $_{-0.5}^{+0.5}$ & 21.5 $_{-16.1}^{+42.0}$ & 0.66 $_{-0.05}^{+0.04}$ & 0.183 $_{-0.140}^{+0.359}$ &59/40.7\\
3C120 &416.19/145$^{(D)}$&10.6& - & 1.75 $_{-0.02}^{+0.02}$ & 1170 $_{-22}^{+21}$ &0.09&2.37&4569/44.2\\

Name & $\chi^2$/DOF & $N_{\rm{Gal}}$ & $N_{\rm{H,1}}$ & $\Gamma_1$ &norm & kT$_{therm}$ & R$_{therm}$ & F$_{HX}$/L$_{HX}$ \\
(1) & (2) & (3) & (4) & (5)  & (6)  & (7) & (8) & (9) \\
\hline

NGC1614 &67.43/67$^{}$&6.4& 0.194 $_{-0.027}^{+0.036}$ & 2.06 $_{-0.4}^{+0.4}$ & 11.1 $_{-4.4}^{+7.8}$ & 0.75 $_{-0.06}^{+0.06}$ & 0.238 $_{-0.044}^{+0.242}$ &26/41.2\\
MRK0618 &142.92/127$^{}$&4.8& - $\ast$& 2.08 $_{-0.08}^{+0.1}$ & 390 $_{-41}^{+51}$ & 0.070 $_{-0.070}^{+0.13}$ & 1.45 $_{-1.45}^{+9.59}$ &940/43.4\\
NGC1672 &117.82/63$^{}$&1.7& 0.0304 $_{-0.007}^{+0.009}$ & 1.82 $_{-0.3}^{+0.2}$ & 2.36 $_{-0.82}^{+1.19}$ & 0.62 $_{-0.02}^{+0.02}$ & 16.4 $_{-8.2}^{+9.7}$ &9/39.6\\
NGC1667 &17.67/15$^{(F)}$&5.4& 271 $_{-251}^{+5730}$ & 1.9$^\dag$ & 370 $_{-370}^{+62100}$ & 0.30 $_{-0.05}^{+0.05}$ & 0.0062 $_{-0.006}^{+1.04}$ &10/42.8\\
NGC1808 &229.82/123$^{}$&2.5&8.73 $_{-7.56}^{+6.82}$&1.39 $_{-0.3}^{+0.4}$&10.0 $_{-6.9}^{+16.0}$& 0.66 $_{-0.02}^{+0.02}$ & 15.0 $_{-2.1}^{+15.2}$ &102/40.4\\
ESO362-G018 &165.27/124$^{}$&1.8& 15.6 $_{-9.7}^{+8.1}$ $\ast$&1.53 $_{-0.2}^{+0.4}$& 67.5 $_{-26.0}^{+114}$ & 0.10 $_{-0.01}^{+0.01}$ & 0.614 $_{-0.153}^{+0.633}$ &335/42.4\\
2MASXJ05210136 &96.92/101$^{(F)}$&1.7&6.58 $_{-0.76}^{+0.88}$&2.08 $_{-0.06}^{+0.2}$&199 $_{-61}^{+74}$&-&-&315/44.2\\
2MASXJ05580206 &93.48/100$^{(H)}$&3.9&-& 1.67 $_{-0.05}^{+0.04}$ & 1040 $_{-69}^{+69}$ &-&-&4226/44.1\\
IC0450 &186.44/137$^{}$&5.5& 2.91 $_{-0.22}^{+0.79}$ $\ast$& 1.50 $_{-0.09}^{+0.09}$ & 219 $_{-34}^{+16}$ & - & - &1258/43.1\\
UGC03973 &134.41/127$^{}$&5.3& - $\ast$& 1.57 $_{-0.08}^{+0.09}$ & 233 $_{-26}^{+32}$ & 0.20 $_{-0.03}^{+0.02}$ & 0.646 $_{-0.303}^{+0.647}$ &1212/43.8\\
IRASF07599+6508 &38.26/28$^{}$&4.2& - &1.9$^{\dag}$& 0.726 $_{-0.605}^{+0.605}$ & - & - &2/42.1\\
NGC2639 &65.5/51$^{}$&3& - &1.9$^{\dag}$& 1.56 $_{-1.56}^{+6.55}$ & 0.59 $_{-0.16}^{+0.13}$ & 2.74 $_{-0.60}^{+2.80}$ &5/40.1\\
NGC2655 &40.84/35$^{}$&2.1& 23.0 $_{-5.3}^{+9.5}$ &1.9$^{\dag}$& 100 $_{-21}^{+27}$ & 0.61 $_{-0.06}^{+0.06}$ & 0.0647 $_{-0.011}^{+0.067}$ &105/41.7\\
IC2431 &11.69/16$^{(F)}$&3.7&$<0.04$& 1.9$^\dag$ & 1.32 $_{-0.43}^{+0.51}$ & 0.81 $_{-0.11}^{+0.06}$ & 1.94 $_{-0.88}^{+0.97}$ &4/41.4\\
MRK0704 &251.24/141$^{}$&3& 1.62 $_{-0.09}^{+0.83}$ $\ast$& 1.73 $_{-0.1}^{+0.2}$ & 311 $_{-61}^{+81}$ & - & - &1025/43.4\\
NGC2841 &104.63/79$^{}$&1.4&-&1.9$^{\dag}$& 5.45 $_{-0.74}^{+0.74}$ & 0.48 $_{-0.16}^{+0.11}$ & 0.230 $_{-0.051}^{+0.253}$ &16/39.2\\
UGC05101 &43.8/51$^{}$&3& 49.6 $_{-18.2}^{+25.4}$ &1.9$^{\dag}$& 25.8 $_{-10.3}^{+17.4}$ & 0.78 $_{-0.21}^{+0.28}$ & 0.0196 $_{-0.0108}^{+0.0261}$ &18/42.5\\
NGC2992 &110.95/92$^{(H)}$&4.8& 0.407 $_{-0.165}^{+0.165}$ & 1.59 $_{-0.03}^{+0.03}$ & 1770 $_{-96}^{+102}$ &-&-&8422/43.1\\
MESSIER081 &219.32/145$^{}$&5.1&-& 1.83 $_{-0.02}^{+0.03}$ & 319 $_{-84}^{+10}$ & 0.94 $_{-0.15}^{+0.09}$ & 0.0244 $_{-0.0079}^{+0.0252}$ &1115/38.8\\
MESSIER082 &286.58/92$^{}$&5.1& - &1.74&31& - & - &1205/40.1\\
NGC3079 &89.37/69$^{}$&8.6& 0.0532 $_{-0.041}^{+0.049}$ & 2.42 $_{-0.3}^{+0.3}$ & 9.56 $_{-1.93}^{+2.58}$ & 0.73 $_{-0.10}^{+0.05}$ & 0.482 $_{-0.126}^{+0.126}$ &34/40.9\\
NGC3147 &118.87/115$^{}$&2.9& 0.0133 $_{-0.004}^{+0.004}$ & 1.52 $_{-0.1}^{+0.2}$ & 25.3 $_{-4.9}^{+6.6}$ & 1.2 $_{-0.5}^{+0.0}$ & 0.0915 $_{-0.050}^{+0.097}$ &141/41.4\\
NGC3227 &193.85/137$^{}$&2.2& 7.93 $_{-2.14}^{+2.08}$ $\ast$& 1.54 $_{-0.00}^{+0.03}$ & 193 $_{-6}^{+11}$ & 0.26 $_{-0.01}^{+0.01}$ & 0.154 $_{-0.027}^{+0.0428}$ &883/41.6\\
NGC3310 &85.72/45$^{(F)}$&1.3& 0.0700 $_{-0.0163}^{+0.0202}$ & 1.93 $_{-0.1}^{+0.1}$ & 30.4 $_{-3.5}^{+4.0}$ & 0.67 $_{-0.07}^{+0.07}$ & 0.169 $_{-0.050}^{+0.072}$ &87/40.3\\
NGC3486 &33.92/22$^{}$&1.9& - &1.9$^{\dag}$& 3.32 $_{-2.01}^{+2.01}$ & - & - &10/39.9\\
NGC3516 &92.03/96$^{(H)}$&3.4& 32.7 $_{-3.9}^{+3.9}$ & 2.03 $_{-0.07}^{+0.09}$ & 794 $_{-171}^{+243}$ &-&-&1563/43.8\\
MESSIER066 &32.2/29$^{}$&2.1& - $\ast$&1.9$^{\dag}$& 8.05 $_{-1.44}^{+1.44}$ & 0.63 $_{-0.08}^{+0.10}$ & 0.951 $_{-0.182}^{+1.31}$ &23/39.5\\
NGC3690 &128.81/93$^{}$&0.9& 0.0614 $_{-0.0614}^{+0.0557}$ & 1.88 $_{-0.2}^{+0.2}$ & 24.3 $_{-5.8}^{+7.6}$ & 0.72 $_{-0.03}^{+0.02}$ & 0.655 $_{-0.066}^{+0.663}$ &83/41.3\\
NGC3735 &34.98/43$^{(F)}$&1.2& 0.0270 $_{-0.0270}^{+0.0641}$ & 1.9$^\dag$ & 3.11 $_{-1.39}^{+1.57}$ & - & - &9/40.2\\
NGC3976 &27.96/33$^{}$&1.1& - $\ast$&1.9$^{\dag}$& 2.58 $_{-1.20}^{+1.20}$ & 0.52 $_{-0.13}^{+0.10}$ & 0.354 $_{-0.151}^{+0.390}$ &7/40.1\\
NGC3982 &60.97/29$^{}$&1&21.7&1.83&13&0.31&0.24&19/40.2\\
NGC4013 &47.23/41$^{}$&1.3& 8.86 $_{-4.44}^{+6.18}$ $\ast$&1.9$^{\dag}$& 1.97 $_{-0.69}^{+0.87}$ & 0.22 $_{-0.04}^{+0.05}$ & 0.164 $_{-0.071}^{+0.226}$ &4/39.0\\
ARP244 &154.83/94$^{}$&3.2&-$\ast$& 1.85 $_{-0.2}^{+0.2}$ & 10.4 $_{-2.6}^{+3.0}$ & 0.62 $_{-0.03}^{+0.02}$ & 0.490 $_{-0.075}^{+0.490}$ &33/40.4\\
NGC4051 &118.05/99$^{(H)}$&1.2& 18.5 $_{-2.6}^{+2.7}$ & 2.14 $_{-0.2}^{+0.1}$ & 337 $_{-112}^{+119}$ &-&-&531/40.9\\
NGC4151 &131.63/90$^{(H)}$&2.1& 5.96 $_{-0.46}^{+0.29}$ & 1.57 $_{-0.06}^{+0.05}$ & 1030 $_{-126}^{+112}$ &-&-&4678/42.1\\
NGC4214 &69.47/49$^{}$&1.9& 0.0640 $_{-0.02}^{+0.02}$ &1.9$^{\dag}$& 4.36 $_{-0.93}^{+0.93}$ & 0.30 $_{-0.04}^{+0.07}$ & 0.566 $_{-0.192}^{+0.610}$ &13/38.9\\
NGC4253 &2280.28/143$^{(D)}$&1.8&-$\ast$& 2.08 $_{-0.02}^{+0.02}$ & 1030 $_{-22}^{+23}$ &0.21&0.19&2461/43.0\\
MESSIER099 &66.3/37$^{}$&2.9& - &1.9$^{\dag}$& 2.19 $_{-0.71}^{+0.71}$ & 0.32 $_{-0.02}^{+0.02}$ & 2.36 $_{-0.17}^{+2.37}$ &7/40.0\\
MESSIER100 &114/70$^{}$&2& - &1.9$^{\dag}$& 6.01 $_{-0.79}^{+0.79}$ & 0.63 $_{-0.03}^{+0.03}$ & 2.85 $_{-0.55}^{+2.90}$ &19/40.1\\
NGC4388 &163.61/127$^{}$&2.6& 39.7 $_{-13.1}^{+6.6}$ & 1.60 $_{-0.2}^{+0.1}$ & 1040 $_{-302}^{+450}$ & 0.77 $_{-0.11}^{+0.09}$ & 0.0033 $_{-0.001}^{+0.009}$ &2109/42.9\\
NGC4414 &89.07/55$^{}$&1.6& - $\ast$&1.9$^{\dag}$& 5.71 $_{-0.71}^{+0.71}$ & 0.46 $_{-0.07}^{+0.05}$ & 0.562 $_{-0.073}^{+0.563}$ &17/39.3\\
NGC4449 &62/67$^{}$&1.4& 0.0166 $_{-0.0128}^{+0.0180}$ &1.9$^{\dag}$& 9.83 $_{-9.26}^{+9.26}$ & - & - &29/38.5\\

Name & $\chi^2$/DOF & $N_{\rm{Gal}}$ & $N_{\rm{H,1}}$ & $\Gamma_1$ &norm & kT$_{therm}$ & R$_{therm}$ & F$_{HX}$/L$_{HX}$ \\
(1) & (2) & (3) & (4) & (5)  & (6)  & (7) & (8) & (9) \\
\hline

3C273 &1161.87/145$^{(D)}$&1.8&-& 1.58 $_{-0.01}^{+0.01}$ & 1950 $_{-20}^{+21}$ &0.99&0.04&9724/45.8\\
NGC4490 &50.39/60$^{}$&1.8& 0.204 $_{-0.053}^{+0.023}$ $\ast$& 2.06 $_{-0.3}^{+0.3}$ & 16.5 $_{-5.7}^{+8.6}$ & 0.60 $_{-0.23}^{+0.24}$ & 0.112 $_{-0.082}^{+0.134}$ &38/39.5\\
MESSIER088 &6.77/9$^{(F)}$&2.6& 0.185 $_{-0.053}^{+0.274}$ & 4.05 $_{-1}^{+2}$ & 7.28 $_{-3.57}^{+9.27}$ & 0.77 $_{-0.52}^{+0.26}$ & 0.190 $_{-0.171}^{+0.318}$ &12/41.7\\
NGC4559 &117.77/100$^{}$&1.5& 0.110 $_{-0.01}^{+0.02}$ $\ast$& 2.13 $_{-0.1}^{+0.2}$ & 23.9 $_{-4.3}^{+5.7}$ & - & - &50/39.9\\
MESSIER090 &134.05/77$^{}$&2.8& - &1.9$^{\dag}$& 3.36 $_{-0.40}^{+0.40}$ & 0.63 $_{-0.02}^{+0.02}$ & 0.966 $_{-0.046}^{+0.967}$ &10/37.3\\
MESSIER058 &185.3/134$^{}$&2.8& - & 1.78 $_{-0.09}^{+0.1}$ & 98.2 $_{-11.6}^{+13.6}$ & 0.71 $_{-0.05}^{+0.03}$ & 0.911 $_{-0.109}^{+0.918}$ &391/41.4\\
NGC4593 &168.95/142$^{}$&1.9& $<$0.0156 $\ast$& 1.83 $_{-0.05}^{+0.05}$ & 1060 $_{-68}^{+73}$ & 0.21 $_{-0.01}^{+0.01}$ & 0.0498 $_{-0.0502}^{+0.0918}$ &3795/42.8\\
MESSIER104 &176.27/110$^{}$&3.6& 0.0673 $_{-0.018}^{+0.022}$ $\ast$& 1.95 $_{-0.1}^{+0.1}$ & 50.6 $_{-8.5}^{+9.2}$ & - & - &139/38.9\\
NGC4631 &60.79/54$^{}$&1.3& 0.214 $_{-0.079}^{+0.043}$ $\ast$& 1.9 $^\dag$ & 2.95 $_{-0.36}^{+0.39}$ & 0.34 $_{-0.04}^{+0.08}$ & 1.16 $_{-0.50}^{+1.44}$ &9/38.9\\
NGC4666 &167.4/100$^{}$&1.7& - $\ast$&1.9$^{\dag}$& 8.60 $_{-0.71}^{+0.80}$ & 0.47 $_{-0.06}^{+0.04}$ & 0.292 $_{-0.025}^{+0.012}$ &30/40.5\\
NGC4725 &32.38/26$^{}$&0.9& - &1.9$^{\dag}$& 1.65 $_{-0.73}^{+0.73}$ & 0.59 $_{-0.22}^{+0.16}$ & 0.399 $_{-0.182}^{+0.126}$ &5/39.2\\
UGC08058 &123.59/73$^{(F)}$&1& 7.07 $_{-1.36}^{+1.82}$ & 1.66 $_{-0.08}^{+0.1}$ & 16.3 $_{-2.4}^{+6.9}$ & 0.79 $_{-0.07}^{+0.05}$ & 0.0901 $_{-0.0307}^{+0.0468}$ &60/42.5\\
NGC4968 &22.37/14$^{}$&8.4& 300 $_{-123}^{+2230}$ &1.9$^{\dag}$& 2540 $_{-2040}^{+10000}$ & - & - &26/43.2\\
NGC5005 &88.07/51$^{}$&1.1& - & 1.57 $_{-0.3}^{+0.3}$ & 7.14 $_{-2.84}^{+4.32}$ & 0.72 $_{-0.08}^{+0.03}$ & 0.997 $_{-0.072}^{+0.999}$ &36/39.9\\
MESSIER063 &59.41/40$^{}$&1.3& 0.0405 $_{-0.015}^{+0.026}$ $\ast$&1.9$^{\dag}$& 7.55 $_{-1.38}^{+1.38}$ & 0.30 $_{-0.02}^{+0.02}$ & 1.14 $_{-0.19}^{+1.15}$ &21/39.2\\
MCG-03-34-064 &271.88/139$^{}$&5.2& 40.0 $_{-1.8}^{+5.2}$ & 2.50 $_{-0.2}^{+0.3}$ & 1160 $_{-375}^{+1090}$ & 0.80 $_{-0.02}^{+0.02}$ & 0.007 $_{-0.000}^{+0.007}$ &258/42.9\\
NGC5170 &30.92/28$^{}$&7& 0.145 $_{-0.046}^{+0.063}$ &1.9$^{\dag}$& 1.21 $_{-0.55}^{+0.55}$ & - & - &4/39.3\\
NGC5194 &101.54/67$^{}$&1.8& 0.0377 $_{-0.006}^{+0.007}$ $\ast$& 2.61 $_{-0.3}^{+0.2}$ & 20.2 $_{-5.9}^{+15.1}$ & 0.64 $_{-0.01}^{+0.01}$ & 0.964 $_{-0.282}^{+0.721}$ &42/40.4\\
ESO383-G035 &176.91/96$^{(H)}$&3.9& - & 2.09 $_{-0.03}^{+0.03}$ & 1910 $_{-83}^{+87}$ & - & - &4033/42.8\\
MESSIER083 &162.35/88$^{}$&4&-$\ast$& 2.00 $_{-0.2}^{+0.2}$ & 15.5 $_{-4.3}^{+6.1}$ & 0.71 $_{-0.01}^{+0.01}$ & 1.44 $_{-0.06}^{+1.44}$ &40/39.4\\
IRASF13349+2438 &227.52/125$^{}$&1& 1.04 $_{-0.19}^{+0.24}$ $\ast$& 2.31 $_{-0.06}^{+0.07}$ & 229 $_{-71}^{+107}$ & 0.22 $_{-0.01}^{+0.00}$ & 1.03 $_{-0.19}^{+0.20}$ &217/43.8\\
NGC5256 &82.96/49$^{(F)}$&1.5& 17.5 $_{-3.8}^{+5.6}$ & 2.74 $_{-0.2}^{+0.3}$ & 99.8 $_{-31.2}^{+55.3}$ & 0.80 $_{-0.04}^{+0.03}$ & 0.0413 $_{-0.0145}^{+0.0238}$ &30/42.2\\
NGC5253 &8.3/6$^{(H)}$&4&-&1.9$^{\dag}$& 0.490 $_{-0.283}^{+0.283}$ &-&-&2/37.9\\
MRK0273 &54.16/54$^{}$&0.9& 59.7 $_{-12.8}^{+17.1}$ &1.9$^{\dag}$& 70.1 $_{-19.0}^{+31.9}$ & 0.72 $_{-0.07}^{+0.06}$ & 0.0277 $_{-0.0046}^{+0.0129}$ &37/42.8\\
IC4329A &112.97/97$^{(H)}$&4.6&-& 1.67 $_{-0.01}^{+0.01}$ & 2160 $_{-27}^{+28}$ &-&-&9544/43.7\\
UGC08850 &101.22/79$^{}$&2& 38.7 $_{-5.7}^{+9.2}$ &1.9$^{\dag}$& 78.2 $_{-6.4}^{+20.9}$ & 0.71 $_{-0.09}^{+0.04}$ & 0.0323 $_{-0.0039}^{+0.0235}$ &63/43.1\\
NGC5506 &273.02/149$^{}$&4& 3.24 $_{-0.24}^{+0.24}$ & 1.82 $_{-0.04}^{+0.05}$ & 2440 $_{-187}^{+205}$ & - & - &6816/42.8\\
NGC5548 &675.17/142$^{(D)}$&1.6&0.01$\ast$& 1.64 $_{-0.02}^{+0.01}$ & 874 $_{-21}^{+20}$ &0.2&0.14&4061/43.4\\
NGC5775 &45.72/43$^{}$&3.6& 0.525 $_{-0.455}^{+0.300}$ $\ast$&1.9$^{\dag}$& 2.38 $_{-0.53}^{+0.53}$ & 0.29 $_{-0.04}^{+0.03}$ & 5.64 $_{-3.19}^{+8.80}$ &6/39.8\\
2MASXJ15115979 &252.79/145$^{}$&8.3& 0.0657 $_{-0.0034}^{+0.0035}$ & 1.67 $_{-0.07}^{+0.09}$ & 198 $_{-19}^{+25}$ & - & - &858/43.9\\
VV705 &26.5/18$^{(F)}$&1.9& 85.9 $_{-62.2}^{+2030}$ & 1.9$^\dag$ & 19.8 $_{-17.3}^{+1680}$ & 0.32 $_{-0.07}^{+0.26}$ & 0.0248 $_{-0.0280}^{+2.11}$ &12/42.3\\
UGC09944 &33.35/31$^{}$&2.8& - &1.9$^{\dag}$& 1.55 $_{-0.70}^{+0.70}$ & 0.81 $_{-0.06}^{+0.07}$ & 1.85 $_{-0.34}^{+0.53}$ &5/41.6\\
2MASXJ15504152 &5.54/4$^{(F)}$&9& 0.362 $_{-0.277}^{+0.543}$ & 5.66 $_{-2}^{+4}$ & 11.7 $_{-6.7}^{+32.5}$ & - & - &45/43.0\\
NGC6240 &177.15/111$^{}$&5& 112 $_{-10}^{+20}$ &1.9$^{\dag}$& 747 $_{-151}^{+284}$ & 0.79 $_{-0.04}^{+0.04}$ & 0.0839 $_{-0.0500}^{+0.099}$ &191/43.5\\
NGC6286 &8.34/5$^{(F)}$&1.8&$<0.04$& 1.9$^\dag$ & 1.57 $_{-0.71}^{+0.94}$ & 0.70 $_{-0.13}^{+0.10}$ & 1.72 $_{-0.92}^{+1.27}$ &5/40.6\\
NGC6552 &19.13/18$^{(F)}$&3.9&-& 1.9$^\dag$ & 1.41 $_{-0.54}^{+0.51}$ & 0.23 $_{-0.04}^{+0.04}$ & 1.53 $_{-0.73}^{+0.78}$ &49/42.9\\
AM1925-724 &43.51/54$^{}$&5.2& 38.1 $_{-21.7}^{+39.2}$ &1.9$^{\dag}$& 22.6 $_{-10.3}^{+25.8}$ & 0.58 $_{-0.26}^{+0.20}$ & 0.0563 $_{-0.0295}^{+0.180}$ &25/42.8\\
NGC6810 &85.94/61$^{}$&5& - &1.9$^{\dag}$& 2.58 $_{-0.469}^{+0.469}$ & 0.61 $_{-0.022}^{+0.023}$ & 4.33 $_{-1.00}^{+4.44}$ &8/39.9\\
NGC6890 &2.06/4$^{(F)}$&2.9&0.106 $_{-0.086}^{+0.201}$&3.87 $_{-1}^{+2}$&6.58 $_{-2.03}^{+4.66}$& - & - &28/42.2\\
MRK0509 &550.93/138$^{(D)}$&4.2& - $\ast$& 1.73 $_{-0.02}^{+0.02}$ & 901 $_{-22}^{+21}$ &0.03&248&3660/44.0\\
ESO286-IG019 &41.66/33$^{}$&3.3& 48.7 $_{-35.2}^{+258}$ &1.9$^{\dag}$& 15.7 $_{-10.9}^{+10000}$ & 0.61 $_{-0.34}^{+0.12}$ & 0.103 $_{-0.026}^{+0.53}$ &11/42.3\\
NGC7090 &125.36/134$^{(F)}$&2.4& 0.136 $_{-0.136}^{+53.9}$ & 1.9$^\dag$ & 1.11 $_{-1.11}^{+6.36}$ & - & - &38/39.8\\

Name & $\chi^2$/DOF & $N_{\rm{Gal}}$ & $N_{\rm{H,1}}$ & $\Gamma_1$ &norm & kT$_{therm}$ & R$_{therm}$ & F$_{HX}$/L$_{HX}$ \\
(1) & (2) & (3) & (4) & (5)  & (6)  & (7) & (8) & (9) \\
\hline
NGC7172 &254.94/153$^{}$&1.9& 8.50 $_{-0.39}^{+0.27}$ & 1.54 $_{-0.07}^{+0.06}$ & 611 $_{-52}^{+43}$ & - & - &2108/42.7\\
NGC7213 &186.15/146$^{}$&1.1& - & 1.69 $_{-0.02}^{+0.02}$ & 520 $_{-17}^{+17}$ & 0.068 $_{-0.009}^{+0.003}$ & 0.743 $_{-0.131}^{+0.776}$ &2194/42.7\\
IC5169 &6.01/5$^{}$&1.1& - $\ast$&1.9$^{\dag}$& 7.30 $_{-5.39}^{+5.39}$ & - & - &21/40.7\\
NGC7252 &50.76/34$^{}$&2& - &1.9$^{\dag}$& 1.24 $_{-0.45}^{+0.45}$ & 0.47 $_{-0.07}^{+0.06}$ & 1.38 $_{-0.16}^{+1.39}$ &4/40.4\\
3C445 &157.17/139$^{(F)}$&4.5& 35.0 $_{-7.7}^{+10.8}$ & 1.59 $_{-0.2}^{+0.2}$ & 238 $_{-78}^{+172}$ & 0.14 $_{-0.02}^{+0.03}$ & 0.0170 $_{-0.0066}^{+0.020}$ &687/43.9\\
NGC7314 &233.2/140$^{}$&0.1& 0.600 $_{-0.027}^{+0.013}$ $\ast$& 1.95 $_{-0.01}^{+0.02}$ & 1500 $_{-14}^{+24}$ & 0.22 $_{-0.02}^{+0.02}$ & 0.00961 $_{-0.001}^{+0.008}$ &4008/42.3\\
MCG-03-58-007 &14.01/20$^{}$&2&27.5 $_{-6.2}^{+7.9}$&1.9$^{\dag}$& 89.2 $_{-20.5}^{+26.8}$ & 0.29 $_{-0.09}^{+0.10}$ & 0.0313 $_{-0.0142}^{+0.0347}$ &91/42.7\\
NGC7469 &417.02/143$^{(D)}$&4.4&0.04$\ast$& 1.86 $_{-0.02}^{+0.02}$ & 862 $_{-19}^{+20}$ &0.22& 0.121 $_{-0.00}^{+0.171}$ &2955/43.2\\
NGC7479 &15.82/17$^{}$&5.4& 201 $_{-122}^{+493}$ &1.9$^{\dag}$& 263 $_{-235}^{+3330}$ & 0.39 $_{-0.13}^{+0.16}$ & 0.0104 $_{-0.0046}^{+0.022}$ &33/42.0\\
ESO148-IG002 &17.55/24$^{}$&1.6&-$\ast$&1.9$^{\dag}$& 2.72 $_{-0.59}^{+1.08}$ & 0.65 $_{-0.10}^{+0.11}$ & 0.399 $_{-0.175}^{+0.188}$ &17/43.2\\
NGC7552 &31.36/19$^{(F)}$&1.3&-& 1.9$^\dag$ &8.09 $_{-1.39}^{+1.41}$&0.76 $_{-0.10}^{+0.06}$& 1.16 $_{-0.53}^{+0.60}$ &25/40.2\\
NGC7582 &353.48/130$^{(F)}$&1.3&0.04&2.35&11&-&-&223/42.6\\
NGC7674 &64.74/36$^{(F)}$&4.2&-& 3.28 $_{-0.3}^{+2}$ & 5.11 $_{-4.63}^{+1.78}$ & 0.79 $_{-0.12}^{+0.08}$ & 0.0918 $_{-0.0983}^{+10.3}$ &60/43.6\\
NGC7714 &82.16/51$^{}$&5& 0.137 $_{-0.031}^{+0.048}$ &1.9$^{\dag}$& 5.34 $_{-0.81}^{+0.81}$ & 0.59 $_{-0.10}^{+0.06}$ & 11.9 $_{-4.3}^{+16.0}$ &17/40.5\\
NGC7771 &53/59$^{}$&4& 0.05 $_{-0.05}^{+0.06}$ $\ast$& 1.68 $_{-0.3}^{+0.3}$ & 4.26 $_{-1.48}^{+0.50}$ & 0.59 $_{-0.04}^{+0.04}$ & 0.525 $_{-0.064}^{+0.530}$ &18/40.9\\
MRK0331 &32.32/30$^{}$&4& 0.123 $_{-0.055}^{+0.120}$ &1.9$^{\dag}$& 1.79 $_{-0.64}^{+0.64}$ & 0.73 $_{-0.05}^{+0.05}$ & 1.45 $_{-0.19}^{+1.46}$ &6/40.7\\
\hline
\end{longtable}
\end{landscape}

\begin{table*}
\centering
\caption{Parameters of the second power-law component, if it is present. Column (1) Galaxy name; Column (2) the neutral column density measured in the secondary power-law (10$^{22}$ \cmsq); Column (3) the power-law index of the secondary power-law; Column (4) the ratio of the normalisation of the second power-law to the primary power-law.}
\label{table:xrayanalysis_specfit_pl2}
\begin{center}
\begin{tabular}{l r r r}
\hline
Name & $N_{\rm H,2}$ & $\Gamma_2$ & F$_{pl}$  \\
(1) & (2) & (3) & (4) \\

\hline
NGC0017 & 0.0212 $_{-0.0145}^{+0.0180}$ &$=\Gamma_1$& 0.0740 $_{-0.0112}^{+0.0112}$ \\
NGC0214 & - &$=\Gamma_1$&0.0623 $_{-0.0248}^{+0.0363}$\\
NGC0262 &-& 1.80 $_{-0.1}^{+0.2}$ & 0.00530 $_{-0.00041}^{+0.00039}$ \\
NGC0424 & 0.0334 $_{-0.0179}^{+0.0220}$ & 3.05 $_{-0.3}^{+0.3}$ & 0.292 $_{-0.041}^{+0.046}$ \\
NGC0526A & 0.0946 $_{-0.0171}^{+0.0174}$ & 2.04 $_{-0.2}^{+0.2}$ & 0.0303 $_{-0.0042}^{+0.0045}$ \\
NGC0513 & 0.0816 $_{-0.0596}^{+0.0926}$ & 2.16 $_{-0.5}^{+0.7}$ & 0.0331 $_{-0.0095}^{+0.0154}$ \\
NGC1052 & 6.29 $_{-1.94}^{+1.30}$ &$=\Gamma_1$& 0.766 $_{-0.351}^{+0.361}$ \\
ARP118 & 0.176 $_{-0.116}^{+0.182}$ & 4.11 $_{-0.9}^{+1}$ & 0.00842 $_{-0.00370}^{+0.00857}$ \\
MCG-02-08-039 & - &$=\Gamma_1$& 0.0196 $_{-0.0033}^{+0.0033}$ \\
NGC1194 & - &$=\Gamma_1$& 0.0220 $_{-0.0110}^{+-0.0112}$ \\
NGC1320 & 0.0341 $_{-0.0133}^{+0.0155}$ & 3.36 $_{-0.2}^{+0.2}$ & 0.0110 $_{-0.0011}^{+0.0012}$ \\
NGC1365 & - &$=\Gamma_1$&0.03\\
NGC1482 & 0.101 $_{-0.057}^{+0.066}$ &$=\Gamma_1$& 0.412 $_{-0.308}^{+0.804}$ \\
NGC1667 &-&$=\Gamma_1$& 0.00585 $_{-0.00583}^{+0.981}$ \\
NGC1808 & 0.164 $_{-0.013}^{+0.013}$ &$=\Gamma_1$& 0.868 $_{-0.046}^{+0.120}$ \\
ESO362-G018 &-& 1.78 $_{-0.06}^{+0.07}$ & 0.433 $_{-0.015}^{+0.020}$ \\
2MASXJ05210136 &0.0187 $_{-0.0151}^{+0.0171}$&3.42 $_{-0.3}^{+0.3}$&0.02 $_{-0.12}^{+0.07}$ \\
IC0450 &-& 2.59 $_{-1}^{+0.3}$ & 0.292 $_{-0.116}^{+0.025}$ \\
NGC2655 & 0.0325 $_{-0.0325}^{+0.0777}$ & 2.60 $_{-0.6}^{+0.8}$ & 0.0422 $_{-0.0136}^{+0.0234}$ \\
MRK0704 &-& 3.04 $_{-0.1}^{+0.1}$ & 0.190 $_{-0.043}^{+0.067}$ \\
UGC05101 & 0.199 $_{-0.062}^{+0.111}$ &$=\Gamma_1$& 0.0498 $_{-0.00}^{+0.0498}$ \\
NGC3227 &-&$=\Gamma_1$& 0.123 $_{-0.005}^{+0.006}$ \\
NGC3516 & 2.98 $_{-0.39}^{+0.42}$ &$=\Gamma_1$& 0.669 $_{-0.144}^{+0.204}$ \\
NGC3982 &0.06&2.87&0.24\\
NGC4013 & 0.0350 $_{-0.0254}^{+0.0444}$ &$=\Gamma_1$& 0.238 $_{-0.047}^{+0.045}$ \\
NGC4051 &-&$=\Gamma_1$& 0.498 $_{-0.166}^{+0.177}$ \\
NGC4388 & 18.7 $_{-9.4}^{+2.1}$ &$=\Gamma_1$& 0.409 $_{-0.230}^{+-0.240}$ \\
UGC08058 &-&$=\Gamma_1$& 0.184 $_{-0.027}^{+0.079}$ \\
NGC4968 & - &$=\Gamma_1$& 0.00101 $_{-0.00016}^{+0.00016}$ \\
MCG-03-34-064 & 0.0161 $_{-0.0081}^{+0.0090}$ & 2.73 $_{-0.08}^{+0.09}$ & 0.0147 $_{-0.0008}^{+0.0008}$ \\
IRASF13349+2438 &-& 4.17 $_{-0.1}^{+0.1}$ & 0.743 $_{-0.0644}^{+0.0791}$ \\
NGC5256 & 0.0269 $_{-0.0128}^{+0.0171}$ &$=\Gamma_1$& 0.0535 $_{-0.0167}^{+0.0296}$ \\
MRK0273 & - &$=\Gamma_1$& 0.0341 $_{-0.0037}^{+0.0037}$ \\
UGC08850 & - &$=\Gamma_1$& 0.0364 $_{-0.0031}^{+0.0031}$ \\
NGC5506 & 0.0910 $_{-0.0116}^{+0.0124}$ &$=\Gamma_1$& 0.0183 $_{-0.0011}^{+0.0012}$ \\
VV705 & 0.212 $_{-0.107}^{+0.229}$ &$=\Gamma_1$& 0.121 $_{-0.106}^{+10.3}$ \\
NGC6240 & 0.168 $_{-0.018}^{+0.019}$ &$=\Gamma_1$& 0.0376 $_{-0.0026}^{+0.0026}$ \\
AM1925-724 & 0.172 $_{-0.058}^{+0.071}$ &$=\Gamma_1$& 0.144 $_{-0.022}^{+0.023}$ \\
ESO286-IG019 & 0.0347 $_{-0.0347}^{+0.0554}$ &$=\Gamma_1$& 0.0666 $_{-0.0255}^{+0.0268}$ \\
NGC7172 & 0.133 $_{-0.031}^{+0.035}$ & 3.00 $_{-0.3}^{+0.3}$ & 0.00593 $_{-0.00067}^{+0.00079}$ \\
3C445 & 0.00489 $_{-0.00489}^{+0.0218}$ &$=\Gamma_1$& 0.0352 $_{-0.0116}^{+0.0255}$ \\
NGC7314 & 0.150 $_{-0.038}^{+0.012}$ & 4.15 $_{-0.8}^{+0.3}$ & 0.00672 $_{-0.00059}^{+0.00262}$ \\
MCG-03-58-007 &-& 3.37 $_{-0.4}^{+0.4}$ & 0.0251 $_{-0.0097}^{+1000}$ \\
NGC7479 & 0.0888 $_{-0.0888}^{+0.201}$ & 3.05 $_{-1}^{+2}$ & 0.00643 $_{-0.00337}^{+0.00925}$ \\

\hline
\end{tabular}
\end{center}
\end{table*}

\begin{table*}
\centering
\caption{Parameters of the soft excess component, if it is present. Column (1) Galaxy name; Column (2) the ratio of the data at 0.4 keV to the extrapolated 2.5-10 keV power-law. $\ast$ 3C120 has a soft excess which does not fit our criteria, as it does not have a data/model ratio greater than 20\% at 0.4 keV. However, this source has a clear soft excess at 0.7 keV not well modelled by thermal plasma emission. We find that a ratio of 3.2 at 0.4 keV is recovered when an additional intrinsic absorption component is used; Column (3) the model used; Column (4) the input soft photon (Wien) temperature ({\tt compTT} model only, keV); Column (5) the temperature (keV); Column (6) the plasma optical depth ({\tt compTT} model only); Column (7) the ratio of the normalisation of this component to the normalisation of the primary power-law.}
\label{table:xrayanalysis_specfit_sx}
\begin{center}
\begin{tabular}{l l l l l l l}
\hline
Name & d/m (0.4 keV)  & model  & T$_0$ & kT & $\tau_p$ & F$_{SX}$   \\
(1) & (2) & (3) & (4) & (5) & (6) & (7) \\

\hline
MRK0335 &2.67&{\tt compTT}&0.05&90.7&0.01&0.22\\
MRK0618 &1.59&{\tt compTT}& 0.0941 $_{-0.0142}^{+0.0117}$ & 2.00 $_{-0.00}^{+4.47}$ & 2.73 $_{-1.63}^{+0.24}$ & 2.29 $_{-1.89}^{+2.45}$ \\
3C120&3.20$\ast$&{\tt compTT}&0.09&69.0&0.01&0.11\\
UGC03973 &2.37&{\tt compTT}& 0.0510 $_{-0.0281}^{+0.0548}$ & 2.01 $_{-0.01}^{+3.05}$ & 3.38 $_{-1.25}^{+0.14}$ & 14.8 $_{-12.6}^{+22.2}$ \\
IRASF07599+6508 &2.46&{\tt bbody}&-& 0.13 $_{-0.02}^{+0.02}$ &-& 0.0524 $_{-0.0116}^{+0.0532}$ \\
MESSIER081 &1.29&{\tt compTT}& 0.137 $_{-0.009}^{+0.009}$ & 2.00 $_{-0.00}^{+9.70}$ & 2.75 $_{-2.39}^{+0.16}$ & 0.487 $_{-0.473}^{+0.489}$ \\
NGC4253 &1.62&{\tt compTT}&0.08&34.1&0.1&0.17\\
3C273 &2.1&{\tt compTT}&0.06&66.2&0.03&0.14\\
NGC4593 &1.64&{\tt compTT}& 0.00743 $_{-0.00486}^{+0.0257}$ & 2.06 $_{-0.06}^{+0.85}$ & 3.17 $_{-0.55}^{+0.27}$ & 590 $_{-558}^{+615}$ \\
NGC4725 &3.83&{\tt bbody}&-& 0.11 $_{-0.02}^{+0.02}$ &-& 0.0676 $_{-0.0087}^{+0.0679}$ \\
NGC5548 &1.61&{\tt compTT}&0.02&26.5&0.27&4.44\\
UGC09944 &7.8&{\tt bbody}&-& 0.10 $_{-0.01}^{+0.01}$ &-& 0.136 $_{-0.014}^{+0.138}$ \\
MRK0509 &2.87&{\tt compTT}&0.07&44.6&0.11&0.25\\
NGC7213 &1.3&{\tt compTT}& 0.146 $_{-0.009}^{+0.011}$ & 44.9 $_{-42.9}^{+12.1}$ & 0.0100 $_{-0.00}^{+0.582}$ & 0.0141 $_{-0.0025}^{+0.124}$ \\
NGC7469 &2.15&{\tt compTT}&0.02&40.8&0.07&8.78\\

\hline
\end{tabular}
\end{center}
\end{table*}

\begin{table*}
\centering
\caption{Parameters of the reflection component, if it is present. Column (1) Galaxy name; Column (2) power-law index of the reflection component; Column (3) the ratio of the normalisation of the {\tt pexmon} component to the normalisation of the primary power-law; Column (4) the ratio of the normalisation of the {\tt torus} component to the normalisation of the primary power-law. Presented only where R$_{pexmon}>1$; Column (5) The intrinsic 2-10 keV luminosity as determined from the {\tt pexmon} component; Column (6) The intrinsic 2-10 keV luminosity as determined from the {\tt torus} component. }
\label{table:xrayanalysis_specfit_pex}
\begin{center}
\begin{tabular}{l l l l l l}
\hline
Name & $\Gamma_2$ & R$_{pexmon}$ &  R$_{torus}$ & L$_{HX}$ ({\tt pexmon}) & L$_{HX}$ ({\tt torus})\\
(1) & (2) & (3) & (4) & (5) & (6) \\

\hline
MRK0335 &2.03& 0.308 $_{-0.116}^{+0.115}$ & - & - & - \\
NGC0424 &$=\Gamma_1$& 9.72 $_{-9.02}^{+218.}$ & 24.9    &  42.5    &  43.1\\
NGC0526A &$=\Gamma_1$& 0.193 $_{-0.100}^{+0.105}$  & - & - & - \\
NGC0513 &$=\Gamma_1$& 0.955 $_{-1.01}^{+1.09}$ & - & - & -  \\
NGC1068 &$=\Gamma_1$&10.3 &31.0   &   42.2    &  42.4\\
ARP118 &$=\Gamma_1$& 0.501 $_{-0.213}^{+0.189}$  & - & - & - \\
NGC1194 &$=\Gamma_1$& 49.8 $_{-45.9}^{+45.8}$  &  167.0  &    42.3   &   43.0\\
NGC1320 &$=\Gamma_1$& 40.4 $_{-39.0}^{+44.9}$ &109.0    &  42.7   &   42.3\\
NGC1386 & 1.9$^\dag$ & 145 $_{-147}^{+160}$ &615.0   &   40.9      &41.7\\
3C120 &$=\Gamma_1$& 0.156 $_{-0.143}^{+0.116}$  & - & - & - \\
MRK0618 &$=\Gamma_1$& 0.379 $_{-0.381}^{+0.507}$ & - & - & -  \\
ESO362-G018 &$=\Gamma_1$& 5.47 $_{-6.11}^{+5.24}$& 11.4   &   42.4    &  42.4\\
IC0450 &$=\Gamma_1$& 0.441 $_{-0.168}^{+0.209}$  & - & - & - \\
MRK0704 &$=\Gamma_1$& 0.285 $_{-0.290}^{+0.708}$  & - & - & - \\
NGC2992 &$=\Gamma_1$& 0.220 $_{-0.109}^{+0.089}$  & - & - & - \\
MESSIER081 &$=\Gamma_1$& 0.302 $_{-0.120}^{+0.105}$  & - & - & - \\
NGC3079 & 1.9$^\dag$ & 16.3 $_{-9.7}^{+7.8}$ & 30.9  &    40.9    &  41.2\\
NGC3147 &$=\Gamma_1$& 0.139 $_{-0.141}^{+1.22}$  & - & - & - \\
NGC3227 &$=\Gamma_1$& 5.21 $_{-3.54}^{+3.02}$& 12.4    &  41.6 &     42.0\\
NGC4051 &$=\Gamma_1$& 1.53 $_{-1.55}^{+1.04}$& 9.0      &40.9  &    41.5\\
NGC4151 &$=\Gamma_1$& 0.785 $_{-0.306}^{+0.110}$  & - & - & - \\
NGC4253 &2.08& 0.413 $_{-0.077}^{+0.082}$  & - & - & - \\
MESSIER088 & 1.9$^\dag$ & 17.6 $_{-17.6}^{+24.2}$ &17.5   &   41.7   &   41.7\\
MESSIER058 &$=\Gamma_1$& 0.872 $_{-0.571}^{+0.703}$  & - & - & - \\
NGC4593 &$=\Gamma_1$& 0.888 $_{-0.256}^{+0.299}$ & - & - & -  \\
NGC4631 &$=\Gamma_1$& 1.16 $_{-1.16}^{+2.01}$ &0.8  &    38.9    &  38.8\\
NGC4666 &$=\Gamma_1$& 2.02 $_{-1.71}^{+1.58}$ &8.9   &   40.5  &    41.1\\
NGC5194 &$=\Gamma_1$& 51.3 $_{-41.1}^{+31.6}$ &177.0   &   40.4    &  40.9\\
ESO383-G035 &$=\Gamma_1$& 0.212 $_{-0.068}^{+0.066}$ & - & - & -  \\
IC4329A &$=\Gamma_1$& 0.274 $_{-0.069}^{+0.056}$  & - & - & - \\
NGC5506 &$=\Gamma_1$& 0.270 $_{-0.153}^{+0.147}$  & - & - & - \\
NGC5548 &1.64& 0.162 $_{-0.162}^{+0.134}$ & - & - & -  \\
2MASXJ15504152 & 1.9$^\dag$ & 13.7 $_{-8.9}^{+59.5}$ &56.7   &   43.0  &    43.6\\
NGC6552 &$=\Gamma_1$& 112 $_{-48}^{+83}$ &1020.0  &    42.9   &   43.6\\
NGC6890 &1.9$^\dag$&52.1 $_{-47.0}^{+43.2}$&76.6     & 42.2   &   42.3\\
MRK0509 &$=\Gamma_1$& 0.309 $_{-0.071}^{+0.062}$  & - & - & - \\
NGC7172 &$=\Gamma_1$& 0.255 $_{-0.256}^{+0.102}$  & - & - & - \\
3C445 &$=\Gamma_1$& 0.317 $_{-0.152}^{+2.74}$  & - & - & - \\
NGC7314 &$=\Gamma_1$& 0.131 $_{-0.117}^{+0.115}$  & - & - & - \\
NGC7469 &$=\Gamma_1$& 0.698 $_{-0.083}^{+0.087}$  & - & - & - \\
NGC7479 &$=\Gamma_1$& 0.293 $_{-0.295}^{+3.71}$  & - & - & - \\
ESO148-IG002 &$=\Gamma_1$& 41.6 $_{-43.1}^{+32.6}$  &45.3 &     43.2   &   43.2\\
NGC7582 &1.49&107&400.0   &   42.6    &  42.5\\
NGC7674 & 1.91 $_{-0.37}^{+0.31}$ & 143 $_{-148}^{+120}$& 185.0    &  43.6  &    43.7 \\

\hline
\end{tabular}
\end{center}
\end{table*}

\begin{table*}
\centering
\caption{X-ray spectral fitting data. Parameters of the warm absorber, if it is present. Column (1) Galaxy name; Column (2) the logarithm of the ionised column density (\cmsq); Column (3) the logarithm of the ionisation parameter, $\xi$. }
\label{table:xrayanalysis_specfit_cwa}
\begin{center}
\begin{tabular}{l l l}
\\
\hline
Name & log$_{10}$\nh & log$_{10}$$\xi$  \\
(1) & (2) & (3) \\

\hline
MRK0335 &22.9&3.88\\
NGC0262 & 22.3 $_{-0.2}^{+0.8}$ & 3.26 $_{-0.25}^{+0.74}$ \\
NGC0526A & 21.4 $_{-0.6}^{+0.4}$ & 2.49 $_{-0.21}^{+0.19}$ \\
NGC0520 & 21.9 $_{-0.2}^{+0.3}$ & -4.00 $_{-0.00}^{+1.28}$ \\
2MASXJ01500266 & 21.4 $_{-0.1}^{+0.3}$ & -0.680 $_{-0.55}^{0.40}$ \\
NGC0695 & 21.9 $_{-0.4}^{+0.3}$ & 0.246 $_{-0.00}^{+0.738}$ \\
NGC1052 &23.0 $_{-0.1}^{+0.2}$&2.71 $_{-0.23}^{+0.21}$\\
NGC1194 & 22.2 $_{-0.4}^{+0.2}$ & 0.884 $_{-0.00}^{+0.226}$ \\
NGC1291 & 21.2 $_{-0.1}^{+0.0}$ & -3.90 $_{-0.10}^{0.25}$ \\
MRK0618 & 20.7 $_{-0.6}^{+0.5}$ & -0.161 $_{-0.26}^{ 0.662}$ \\
ESO362-G018 & 22.3 $_{-0.7}^{+0.3}$ & 3.00 $_{-0.19}^{+0.32}$ \\
IC0450 & 21.9 $_{-0.3}^{+0.0}$ & 0.119 $_{-0.028}^{+0.082}$ \\
UGC03973 & 21.1 $_{-0.1}^{+0.0}$ & -3.64 $_{-0.36}^{+0.04}$ \\
MRK0704 & 23.0 $_{-0.1}^{+0.1}$ & 2.44 $_{-0.16}^{+0.15}$ \\
NGC3227 & 21.4 $_{-0.0}^{+0.1}$ & -1.09 $_{-0.03}^{+0.05}$ \\
MESSIER066 & 21.7 $_{-1.7}^{+1.5}$ & 1.66 $_{-0.00}^{+2.34}$ \\
NGC3976 & 21.6 $_{-0.4}^{+0.4}$ & -3.82 $_{-0.18}^{+0.66}$ \\
NGC4013 & 22.8 $_{-0.3}^{+1.1}$ & -1.37 $_{-2.63}^{+5.08}$ \\
ARP244 & 20.9 $_{-0.1}^{+0.1}$ & -3.20 $_{-0.33}^{+0.08}$ \\
NGC4253 &20.9&-4\\
NGC4414 & 21.1 $_{-0.2}^{+0.2}$ & -2.90 $_{-0.01}^{+0.08}$ \\
NGC4490 & 20.7 $_{-0.7}^{+0.4}$ & -2.40 $_{-1.60}^{+ 6.40}$ \\
NGC4559 & 20.7 $_{-0.2}^{+0.1}$ & -3.64 $_{-0.36}^{+2.44}$ \\
NGC4593 & 21.3 $_{-0.2}^{+0.1}$ & 1.82 $_{-0.05}^{+0.07}$ \\
MESSIER104 & 20.9 $_{-0.4}^{+0.1}$ & -0.285 $_{-0.329}^{+ 0.164}$ \\
NGC4631 & 21.1 $_{-1.1}^{+0.3}$ & 0.836 $_{-0.00}^{+3.16}$ \\
NGC4666 & 21.8 $_{-0.0}^{+0.0}$ & 0.564 $_{-0.082}^{+0.043}$ \\
MESSIER063 & 23.4 $_{-1.2}^{+0.1}$ & 3.84 $_{-1.01}^{+0.16}$ \\
NGC5194 & 23.5 $_{-0.1}^{+0.0}$ & 2.83 $_{-0.20}^{+0.27}$ \\
MESSIER083 & 22.1 $_{-0.3}^{+0.2}$ & 2.40 $_{-0.16}^{+0.11}$ \\
IRASF13349+2438 &21.1 $_{-0.1}^{+0.1}$&-1.62 $_{-0.1}^{+0.05}$\\
NGC5548 &20.8&-4\\
NGC5775 & 21.6 $_{-0.5}^{+0.1}$ & -3.14 $_{-0.08}^{+0.33}$ \\
MRK0509 &20.2&2.06\\
IC5169 & 21.3 $_{-0.0}^{+0.5}$ & -3.94 $_{-0.06}^{+0.55}$ \\
NGC7314 & 20.9 $_{-0.2}^{+0.1}$ & 2.08 $_{-0.08}^{+0.20}$ \\
NGC7469 &20.8&2.2\\
ESO148-IG002 & 21.3 $_{-0.1}^{+0.4}$ & -3.93 $_{-0.07}^{+3.87}$ \\
NGC7771 & 21.1 $_{-0.1}^{+0.2}$ & -3.67 $_{-0.33}^{+ 0.33}$ \\

\hline
\end{tabular}
\end{center}
\end{table*}

\begin{table*}
\centering
\caption{Parameters of the iron K$\alpha$ line, be it narrow or broad, if it is present. Column (1) Galaxy name; Column (2) the EW of the narrow line at 6.4 keV modelled by either {\tt trans} or {\tt pexmon}; Column (3) the energy of the line constrained to be between 6.2 and 6.6 keV, not associated with any other component; Column (4) the gaussian width of this component; Column (5) the EW of this component. }
\label{table:xrayanalysis_specfit_feka}
\begin{center}
\begin{tabular}{l l l l l}
\hline
Name & EW$_{6.4}$ & E$_{\rm Fe K\alpha}$ & $\sigma_{\rm Fe K\alpha}$ & EW$_{\rm Fe K\alpha}$  \\
(1) & (2) & (3) & (4) & (5) \\

\hline
MRK0335 &31$_{-8}^{+8}$&6.51$_{-0.11}^{+0.09}$&0.42$_{-0.13}^{+0.44}$&134$_{-38}^{+42}$\\
NGC0017 &214$_{-214}^{+240}$&-&-&-\\
NGC0262 &28$_{-7}^{+7}$&-&-&-\\
UGC00545 &-&6.6$_{-0.1}^{+0.0}$&0.31$_{-0.2}^{+0.33}$&143$_{-73}^{+90}$\\
NGC0424 &576$_{-187}^{+187}$&6.51$_{-0.31}^{+0.09}$&$<1$&109$_{-109}^{+224}$\\
NGC0526A &26$_{-9}^{+9}$&6.5$_{-0.1}^{+0.1}$&0.34$_{-0.1}^{+0.39}$&99$_{-36}^{+38}$\\
NGC0513 &116$_{-48}^{+48}$&6.2$_{-0.0}^{+0.2}$&$<0.35$&48$_{-39}^{+166}$\\
2MASXJ01500266 &-&6.72$_{-0.23}^{+0.26}$&0.28$_{-0.28}^{+0.43}$&715$_{-448}^{+628}$\\
NGC1052 &124$_{-23}^{+23}$&6.51$_{-0.11}^{+0.09}$&0.05$_{-0.05}^{+0.26}$&23$_{-11}^{+95}$\\
NGC1068 &281&6.6&0.18&396\\
ARP118 &60$_{-21}^{+21}$&6.47$_{-0.08}^{+0.13}$&0.04$_{-0.04}^{+0.16}$&40$_{-34}^{+43}$\\
MCG-02-08-039 &220$_{-140}^{+140}$&-&-&-\\
NGC1194 &194$_{-46}^{+46}$&6.6$_{-0.1}^{+0.0}$&0.26$_{-0.13}^{+0.36}$&239$_{-174}^{+179}$\\
NGC1313 &28$_{-28}^{+62}$&-&-&-\\
NGC1320 &701$_{-163}^{+163}$&6.5$_{-0.1}^{+0.1}$&0.06$_{-0.06}^{+0.09}$&239$_{-138}^{+209}$\\
NGC1365 &78&-&-&-\\
NGC1386 &1710$_{-373}^{+373}$&-&-&-\\
3C120 &19$_{-5}^{+5}$&6.43$_{-0.04}^{+0.08}$&0.15$_{-0.05}^{+0.22}$&47$_{-16}^{+21}$\\
MRK0618 &30$_{-30}^{+37}$&6.2$_{-0.0}^{+0.2}$&$<1.77$&45$_{-41}^{+404}$\\
NGC1808 &10$_{-10}^{+52}$&-&-&-\\
ESO362-G018 &257$_{-63}^{+63}$&6.54$_{-0.08}^{+0.06}$&0.12$_{-0.12}^{+0.51}$&108$_{-86}^{+189}$\\
2MASXJ05210136 &71$_{-56}^{+56}$&-&-&134$_{-72}^{+72}$\\
2MASXJ05580206 &-&6.37$_{-0.17}^{+0.16}$&0.11$_{-0.11}^{+0.48}$&45$_{-40}^{+30}$\\
IC0450 &63$_{-20}^{+20}$&-&-&-\\
UGC03973 &-&6.43$_{-0.06}^{+0.07}$&0.11$_{-0.07}^{+0.07}$&205$_{-81}^{+91}$\\
NGC2655 &140$_{-140}^{+180}$&-&-&-\\
MRK0704 &30$_{-23}^{+23}$&6.29$_{-0.09}^{+0.07}$&0.15$_{-0.15}^{+0.12}$&132$_{-91}^{+74}$\\
UGC05101 &127$_{-127}^{+129}$&6.55$_{-0.04}^{+0.05}$&$<0.29$&170$_{-135}^{+659}$\\
NGC2992 &32$_{-7}^{+7}$&6.33$_{-0.13}^{+0.12}$&0.33$_{-0.16}^{+0.35}$&60$_{-24}^{+26}$\\
MESSIER081 &38$_{-11}^{+11}$&6.6$_{-0.1}^{+0.1}$&$<0.13$&23$_{-14}^{+22}$\\
MESSIER082 &-&6.48&$<0.01$&44\\
NGC3079 &411$_{-173}^{+173}$&6.6$_{-0.1}^{+0.0}$&0.07$_{-0.07}^{+0.19}$&305$_{-183}^{+919}$\\
NGC3147 &17$_{-17}^{+67}$&6.5$_{-0.1}^{+0.1}$&0.05$_{-0.05}^{+0.26}$&162$_{-128}^{+394}$\\
NGC3227 &174$_{-22}^{+22}$&-&-&-\\
NGC3516 &67$_{-8}^{+8}$&6.41$_{-0.01}^{+0.02}$&0.12$_{-0.02}^{+0.12}$&132$_{-22}^{+19}$\\
NGC3690 &-&6.6$_{-0.1}^{+0.0}$&0.31$_{-0.11}^{+0.31}$&913$_{-346}^{+380}$\\
NGC4051 &117$_{-19}^{+19}$&6.55$_{-0.15}^{+0.04}$&0.06$_{-0.06}^{+0.22}$&69$_{-26}^{+117}$\\
NGC4151 &174$_{-20}^{+12}$&6.5$_{-0.1}^{+0.0}$&$<0.07$&27$_{-7}^{+24}$\\
NGC4253 &45$_{-8}^{+8}$&-&-&33$_{-8}^{+8}$\\
NGC4388 &144$_{-21}^{+21}$&-&-&-\\
MESSIER058 &90$_{-44}^{+44}$&6.6$_{-0.1}^{+0.0}$&0.2$_{-0.1}^{+0.2}$&206$_{-103}^{+129}$\\
NGC4593 &108$_{-26}^{+26}$&-&-&-\\
NGC4631 &492$_{-478}^{+431}$&-&-&-\\
NGC4666 &155$_{-100}^{+100}$&6.47$_{-0.21}^{+0.04}$&$<0.4$&234$_{-170}^{+229}$\\
UGC08058 &-&6.49$_{-0.13}^{+0.11}$&0.24$_{-0.11}^{+0.26}$&523$_{-230}^{+967}$\\
NGC4968 &4510$_{-1697}^{+1697}$&-&-&-\\
MCG-03-34-064 &139$_{-29}^{+29}$&6.2$_{-0.0}^{+0.1}$&0.36$_{-0.08}^{+0.37}$&254$_{-87}^{+81}$\\
NGC5194 &470$_{-160}^{+160}$&6.5$_{-0.1}^{+0.1}$&$<0.2$&176$_{-130}^{+795}$\\
ESO383-G035 &23$_{-5}^{+5}$&6.2$_{-0.0}^{+0.1}$&0.51$_{-0.09}^{+0.5}$&185$_{-26}^{+27}$\\
IRASF13349+2438 &-&6.45$_{-0.2}^{+0.15}$&0.61$_{-0.13}^{+0.67}$&774$_{-226}^{+259}$\\
NGC5256 &362$_{-219}^{+219}$&-&-&-\\
MRK0273 &202$_{-105}^{+105}$&-&-&137$_{-92}^{+95}$\\
IC4329A &36$_{-4}^{+4}$&6.42$_{-0.05}^{+0.08}$&0.3$_{-0.1}^{+0.3}$&72$_{-13}^{+16}$\\
UGC08850 &160$_{-74}^{+74}$&6.6$_{-0.1}^{+0.0}$&0.06$_{-0.06}^{+0.11}$&132$_{-78}^{+505}$\\

\hline
\end{tabular}
\end{center}
\end{table*}

\begin{table*}
\centering
\caption{Continued. }
\begin{center}
\begin{tabular}{l l l l l}
\hline
Name & EW$_{6.4}$ & E$_{\rm Fe K\alpha}$ & $\sigma_{\rm Fe K\alpha}$ & EW$_{\rm Fe K\alpha}$  \\

\hline

NGC5506 &31$_{-9}^{+9}$&6.43$_{-0.06}^{+0.11}$&0.28$_{-0.08}^{+0.33}$&116$_{-29}^{+33}$\\
NGC5548 &22$_{-7}^{+7}$&6.4$_{-0.1}^{+0.1}$&0.15$_{-0.05}^{+0.18}$&61$_{-24}^{+28}$\\
2MASXJ15115979 &-&6.2$_{-0.0}^{+0.2}$&0.31$_{-0.14}^{+0.46}$&237$_{-106}^{+160}$\\
NGC6240 &312$_{-67}^{+67}$&-&-&134$_{-50}^{+50}$\\
NGC6552 &1490$_{-510}^{+510}$&-&-&-\\
AM1925-724 &194$_{-194}^{+239}$&-&-&-\\
MRK0509 &39$_{-7}^{+7}$&6.58$_{-0.04}^{+0.02}$&0.00$_{-0.01}^{+0.06}$&17$_{-6}^{+8}$\\
ESO286-IG019 &491$_{-491}^{+718}$&-&-&-\\
NGC7172 &36$_{-8}^{+8}$&6.49$_{-0.02}^{+0.01}$&$<0.01$&36$_{-8}^{+10}$\\
NGC7213 &-&6.39$_{-0.02}^{+0.02}$&0.05$_{-0.05}^{+0.03}$&89$_{-17}^{+18}$\\
3C445 &87$_{-26}^{+26}$&-&-&-\\
NGC7314 &16$_{-9}^{+9}$&6.46$_{-0.1}^{+0.12}$&0.35$_{-0.11}^{+0.38}$&121$_{-35}^{+42}$\\
NGC7469 &84$_{-9}^{+9}$&-&-&-\\
NGC7479 &768$_{-429}^{+429}$&-&-&-\\
NGC7582 &370$_{-29}^{+29}$&6.6$_{-0.1}^{+0.0}$&0.29$_{-0.18}^{+0.19}$&269$_{-173}^{+80}$\\
NGC7674 &272$_{-173}^{+173}$&-&-&-\\
NGC7771 &-&6.38$_{-0.18}^{+0.22}$&0.18$_{-0.18}^{+0.27}$&487$_{-365}^{+876}$\\

\hline
\end{tabular}
\end{center}
\end{table*}

\begin{table*}
\centering
\caption{Details of the Compton thick sources found in this sample. Column (1) Galaxy name; Column (2) the directly measured \nh (\cmsq); Column (3) the EW of the Fe K$\alpha$ line (eV); Column (4) the reflection fraction determined from the {\tt pexmon} model.
}
\label{table:xrayanalysis_cthick}
\begin{center}
\begin{tabular}{l r r r}
\hline
Name & log$_{10}$($N_{\rm H}$) & EW & R  \\
(1) & (2) & (3) & (4) \\

\hline
NGC1068	& -					& 677					&10.3\\
NGC1320		& 24.3$_{-0.3}^{+1.2}$	& 940$_{-585}^{+851}$ &40.4$_{-39}^{+44.9}$\\
NGC1386		& 20.7$_{-0.1}^{+0.1}$	& 1710$_{-373}^{+373}$&145$_{-147}^{+160}$\\
NGC1667		& 24.4$_{-1.1}^{+1.8}$	& -&-\\
NGC3079		& 20.7$_{-0.7}^{+0.3}$	& 716$_{-525}^{+2180}$&16.3$_{-9.7}^{+7.84}$\\
NGC3690		& 20.8$_{-0.8}^{+0.3}$	& 913$_{-346}^{+380}$ & -\\
MESSIER088	& 21.3$_{-0.2}^{+0.4}$	& -&17.6$_{-17.6}^{+24.2}$\\
NGC4968		& 24.5$_{-0.3}^{+0.9}$	& 4510$_{-1700}^{+1700}$&-\\
NGC5194		& 20.6$_{-0.1}^{+0.1}$	& 646$_{-525}^{+2930}$&51.3$_{-41.1}^{+31.6}$\\
2MASXJ15504152& 21.6$_{-0.7}^{+0.4}$	& -&13.7$_{-8.87}^{+59.5}$\\
NGC6552		& -					& 1490$_{-510}^{+510}$	&112$_{-47.7}^{+83.2}$\\
NGC6890		& 21.0$_{-0.7}^{+0.5}$	& -&52.1$_{-47}^{+43.2}$\\
NGC7479		& 24.3$_{-0.4}^{+0.5}$	& 768$_{-429}^{+429}$&0.29$_{-0.3}^{+3.71}$\\
ESO148-IG002	& -					& -&41.6$_{-43.1}^{+32.6}$\\
NGC7582		& 20.6				& 639$_{-414}^{+197}$&107\\
NGC7674		& -					& 272$_{-173}^{+173}$&143$_{-148}^{+120}$\\

\hline
\end{tabular}
\end{center}
\end{table*}
\twocolumn


\label{lastpage}
\end{document}